\documentclass[smallcondensed,natbib]{svjour3} 

\usepackage{amssymb,amsmath,amsfonts}
\usepackage{graphicx}
\usepackage{color}
\usepackage{ulem}             
\usepackage{multicol}
\usepackage{caption}
\usepackage{pgf,tikz}
\usetikzlibrary{arrows}
\usepackage{colortbl}

\usepackage{xcolor}

\usepackage{amssymb}
\usepackage{latexsym}
\usepackage{pstricks}
\usepackage{pst-plot}
\usepackage{tabularx}

\usepackage{soul}
\usepackage{cancel}

	\usetikzlibrary[patterns]

\DeclareMathOperator{\e}{e}

\newcommand\Lam{{\Lambda} }

\newcommand\eps{{\varepsilon} }

\newcommand\Hb{{\overline H} }
\newcommand\cM{{\cal M} }

\newcommand\xt{{\tilde x}}

\newcommand{\qtext}[1]{\quad \text{#1}\quad}

\newcommand\cG{{\cal G} }

\newcommand\gO{{\cal O} }
\newcommand\cS{{\cal S} }

\newcommand\cF{{\cal F} }
\newcommand\cZ{{\cal Z} }
\newcommand\gH{{\cal H} }
\newcommand\cW{{\cal W} }
\newcommand\cR{{\cal R} }
\newcommand\cV{{\cal V} }

\newcommand\cC{\mathcal{C}}

\newcommand\Dv{{\Delta \varpi} }

\newcommand\Fsf{{\cF^{s\!f}}}
\newcommand\Fsc{{\cF^{sc}}}

\newcommand{\be}{\begin{equation}}
\newcommand{\ee}{\end{equation}}

\newcommand*\circled[1]{\tikz[baseline=(char.base)]{
            \node[shape=circle,draw,inner sep=1pt] (char) {#1};}}

\newcommand\cL{{\cal L} }

 {\everymath{\displaystyle\everymath{}}\array}%
 {\endarray}

\begin{document}

\title{On the coplanar eccentric non restricted co-orbital dynamics}

\author{A. Leleu \and  P. Robutel \and  A.C.M. Correia}

\institute{ 
              1. Physics Institute, Space Research and Planetary Sciences, Center for Space and Habitability
 - NCCR PlanetS - University of Bern - Switzerland\\
 				2. IMCCE, Observatoire de Paris, UPMC, CNRS UMR8028, 77 Av. Denfert-Rochereau, 75014 Paris - France.\\
 				3. Departemento de F\`isica, I3N, Universidade de Aviero, Campus de Santiago, 2810-193 Aveiro - Portugal\\         
               \email{adrien.leleu@space.unibe.ch; philippe.robutel@obspm.fr and correia@ua.pt} 
                 }     

\maketitle

\begin{abstract}

We study the phase space of eccentric coplanar co-orbitals in the non-restricted case.  Departing from the quasi-circular case, we describe the evolution of the phase space as the eccentricities increase. We find that over a given value of the eccentricity, around $0.5$ for equal mass co-orbitals, important topological changes occur in the phase space. These changes lead to the emergence of new co-orbital configurations and open a continuous path between the previously distinct trojan domains near the $L_4$ and $L_5$ eccentric Lagrangian equilibria. These topological changes are shown to be linked with the reconnection of families of quasi-periodic orbits of non-maximal dimension.   


\end{abstract}

\section{Introduction}
\label{sec:intro}

Co-orbitals are two bodies $m_1$ and $m_2$ orbiting around a more massive body $m_0$ with the same mean mean-motion. This configuration is also called a $1:1$ mean motion resonance. In the coplanar circular case, the dynamics of this resonance is well known. Out of the $5$ equilibrium points found by Euler and Lagrange, the first $3$ were shown to be unstable by \cite{Liouville1842}, while $L_4$ and $L_5$ are linearly stable when $\mu = \frac{m_1+m_2}{m_0+m_1+m_2} \lesssim 1/27$ \citep{Ga1843}. When the masses satisfy this relation, the bodies can librate around the $L_4$ and $L_5$ equilibrium on stable orbits called Trojan, or tadpole. This libration transcribes by an oscillation of the resonant angle $\zeta=\lambda_1-\lambda_2$, where $\lambda_j$ is the mean longitude of the mass $m_j$. As the quantity $\mu$ decreases, stable orbits with larger amplitude of libration become available. However, the amplitude of libration of $\zeta$ can not increase indefinitely in the trojan domain: at some point a separatrix emanating from the unstable equilibrium $L_3$ is crossed, beyond which the bodies are in a configuration called horseshoe \citep{Ga1977,Ed1977}. In this configuration, $\zeta$ librates around $180^\circ$ with a larger amplitude, the orbits encompassing the $L_3$, $L_4$ and $L_5$ equilibrium points. Horseshoe orbits are stable for $\mu \lesssim 2 \times 10^{-4}$ \citep{Robe2002}. 

1-D models were developed for the averaged coplanar quasi-circular case \citep{Ed1977,RoPo2013}, describing the co-orbital dynamics as long as $m_1$ and $m_2$ are not too close to each other (outside of the Hill's sphere). However, if we consider the inclined and/or eccentric cases, the phase space is significantly more complex. New co-orbital configurations appear, such as quasi-satellites \citep{Namouni1999,MiIn2006,SiArNeZe2013,PoRoVi2017} in the eccentric case and retrograde co-orbitals \citep{MoNa2013} in the inclined one. For no-null eccentricities and/or inclination, secular dynamics kicks in, increasing the number of dimension to consider in order to correctly describe the dynamics.
 
\cite{GiuBeMiFe2010} studied the co-planar eccentric dynamics in the planetary case for $m_1$ and $m_2$ of the order of $10^{-3} m_0$. They noticed that, as the eccentricity of the co-orbitals increases, the stable domain of quasi-satellites configuration increases, and the trojan domains shrink. They also found that, in addition to the eccentric Lagrangian equilibrium $L_4$ and $L_5$, the trojan domains have another periodic solution of the averaged problem: the Anti-Lagrange equilibria. The position of these equilibria evolves in the phase space as the eccentricity increases.\\

In this work, we aim to push further the understanding of the coplanar eccentric co-orbital dynamics. Beside the intrinsic interest of studying the 1:1 mean motion resonance, an understanding of its different configurations is essential to the development of methods of detection adapted to the co-orbital resonance, as well as the estimation of false positives that can be induced to the detection of other orbital configurations \citep[see for example][and references therein]{FoGa2006,GiuBe2012,LeRoCo2015,LeRoCoLi2017}. Although coplanarity seems a strong assumption for any real-life application, the study of this peculiar case is interesting because it is still representative of systems with a small mutual inclination\footnote{The planar dynamics is decoupled from the dynamics of the inclinations at first order in the inclination, see \cite{RoPo2013}.}. 

We know that, at least in the quasi-circular case, some co-orbital configurations are stable only if $\mu$ is smaller than a given value. Less massive co-orbitals may then have a phase space more complex than the one described in \cite{GiuBeMiFe2010}. On the other hand, thorough numerical study of trajectories is increasingly difficult as $\mu$ decreases since the time scales involved in the dynamics are longer. As a compromise, we will consider co-orbitals in the range of rocky planets with respect to the star ($10^{-5}m_0 \sim 10^{-6}m_0$), and see how the co-orbital phase space behaves at high eccentricities. 

After a brief review of the quasi-circular coplanar case in section \ref{sec:QCCD}, we will describe the evolution of the phase space in the case $m_1=m_2$ (which simpler due to an additional symmetry), going from the quasi-circular case up to eccentricities of $0.7$. Although the phase space evolves in a very predictable way for eccentricities lower than $\approx 0.5$, we show that the topology dramatically changes for higher values. In a final section we check that the changes that were observed in the case $m_1=m_2$ occur for different planetary masses as well.

\section{Quasi-circular coplanar dynamics}
\label{sec:QCCD}

The dynamics of the quasi-circular coplanar co-orbitals is well known \citep{Ga1977,Ed1977,RoPo2013}. In this section we give an overview of its main features in the planetary case ($m_1 \leq m_2 \ll m_0$).

\subsection{Hamiltonian of the averaged planetary problem}
\label{sec:H3bp}

We start with the 3-body problem Hamiltonian $\gH$ in canonical cartesian heliocentric coordinates \citep{LaRo1995,RoNiPo2016}:
\begin{equation}
\gH = \gH_K({\bf r}_j) + \eps \gH_P({\bf r}_j,{\bf \tilde{r}}_j)\, ,
\label{eq:Ht}
\end{equation}
where
\begin{equation}
\gH_K = \sum^2_{j=1} \left( \frac{||{\bf \tilde{r}}_j||^2}{2 \beta_j}-\frac{ \mu _j \beta _j}{||{\bf r}_j||} \right)\, , 
\end{equation}
%
%
is the Keplerian part of the hamiltonian, $\eps = {\rm max}(\frac{ m_1}{m_0},\frac{m_2}{m_0})$ is a small parameter such that $m_j =  \eps m'_j$. ${\bf r}_j$ is the position of $m_j$ with respect to $m_0$, and ${\bf \tilde{r}}_j$ is the barycentric linear momentum. $\beta_j$ is the reduced mass ratio $\beta_j= \frac{m_0 m'_j}{m_0+ \eps m'_j}$, and $\mu _j= \cG (m_0+\eps m'_j)$ where $\cG$ is the gravitational constant. The perturbed part of the Hamiltonian reads:
\begin{equation}
\gH_P = \frac{{\bf \tilde{r}}_1 \cdot {\bf \tilde{r}}_2}{m_0} - \cG \frac{m'_1 m'_2}{||{\bf r}_1-{\bf r}_2||}\, .
\end{equation}
In order to get closer to the orbital elements, we rewrite the Hamiltonian in the Poincar\'e set of variables:
\begin{equation}
\begin{aligned}
\Lambda_j &=\beta_j\sqrt{\mu_ja_j}\, , \vspace{2cm} &\lambda_j &=\lambda_j\, , \\ 
x_j & =\sqrt{\Lambda_j}\sqrt{1-\sqrt{1-e_j^2}}\operatorname{e}^{i\varpi_j}\, , \vspace{2cm} &\tilde{x}_j &=-i\bar{x}_j\, , \\
\end{aligned}
\label{eq:poincvar}
\end{equation}
that is:
\begin{equation}
\gH = \gH_K(\Lambda_1,\Lambda_2) + \eps \gH_P(\lambda_1,\lambda_2,\Lambda_1,\Lambda_2,x_1,x_2,\xt_1,\xt_2)\, .
\label{eq:Hpoinc}
\end{equation}
We study here the $1:1$ mean motion resonance. We are hence in the neighbourhood of the exact Keplerian resonance defined by:
\begin{equation}
\frac{\partial \gH_K}{\partial \Lambda_1}(\Lambda_1,\Lambda_2) = \frac{\partial \gH_K}{\partial \Lambda_2}(\Lambda_1,\Lambda_2) \, ,
\label{eq:defeta1}
\end{equation}
where:
\begin{equation}
\gH_K =- \sum^2_{j=1}  \left( \frac{\mu_j^2 \beta_j^3}{2 \Lambda_j^2} \right)\, . 
\label{eq:keppoinc}
\end{equation}
We note $\Lambda_1^0$ and $\Lambda_2^0$ the solution of the equations (\ref{eq:defeta1}) and (\ref{eq:keppoinc}).

Since the mean motions $n_j$ of the two bodies are close at any given time, the quantity $\zeta=\lambda_1-\lambda_2$ evolves slowly with respect to the longitudes. We note $\nu \propto \sqrt{ \eps} n_1$ the fundamental frequency associated with the resonant angle $\zeta$. We process to the following canonical change of variables:
\begin{equation}
 \begin{pmatrix}
  \zeta \\
  \zeta_2
   \end{pmatrix} =
   \begin{pmatrix}
  1 & -1 \\
  0 & 1
   \end{pmatrix} 
      \begin{pmatrix}
  \lambda_1 \\
  \lambda_2
   \end{pmatrix}
      , \hspace{1cm}
       \begin{pmatrix}
  Z \\
  Z_2
   \end{pmatrix} =
   \begin{pmatrix}
  1 & 0 \\
  1 & 1
   \end{pmatrix} 
      \begin{pmatrix}
  \Lambda_1-\Lambda_1^0 \\
  \Lambda_2-\Lambda_2^0
   \end{pmatrix}\, ,
   \label{eq:tl1}
   \end{equation}
to obtain the following Hamiltonian:
\begin{equation}
H = H_K(Z,Z_2) + \eps H_P(\zeta,\zeta_2,Z,Z_2,x_1,x_2,\xt_1,\xt_2)+ \gO(\eps^2), 
\label{eq:Hts}
\end{equation}
with
\begin{equation}
H_K(Z,Z_2) = -\frac{\beta_1^3\mu_1^2}{2(\Lambda_1^0+Z)^2} -\frac{\beta_2^3\mu_2^2}{2(\Lambda_2^0-Z+Z_2)^2}.
\end{equation}
In the Hamiltonian (\ref{eq:Hts}), a third time scale appears. This time scale, called secular, is slow with respect to the orbital period and the resonant motion. It is associated with the orbital precession and then to the variables $x_j$ and $\xt_j$ as $\dot{x}_j=\eps \partial H_P / \partial \tilde{x}_j =  \gO({\eps})$.
 The separation between the fast time scale (associated with the mean motions) and the other time scales allows for the averaging over the fast angle $\zeta_2$. 
 We process this averaging by applying the time-one map of the Hamiltonian flow generated by the auxiliary function $\cW$:
\begin{equation}
 \begin{split}
{\cW}(\zeta,Z,Z_2,x_j,\xt_j) = \eps \frac{1}{2 \pi} \int_0^{\zeta_2} \left[ \Hb_P- H_P \right] d\zeta_2\, ,
\end{split}
\end{equation}
where
\begin{equation}
 \begin{split}
\Hb_P(\zeta,Z,Z_2,x_j,\xt_j) & = \frac{1}{2 \pi} \int_0^{2 \pi}  H_P d\zeta_2 \, .
\end{split}
\end{equation}
We hence obtain the averaged Hamiltonian:
\begin{equation}
\Hb = \cL_{{\cW}} H. 
\label{eq:Hb}
\end{equation}
where $\cL_{{\cW}}$ is the Lie transform:
\begin{equation}
\cL_{\cW}= Id + \{\cW,\cdot\} + \{\cW,\{\cW,\cdot\}\} + ...
\label{eq:Lietrans}
\end{equation}
with $\{\cdot,\cdot\}$ the Lie bracket. We note $\chi_\cM$ the canonical change of variable close to the identity:
\begin{equation}
\chi_\cM = \cL_{-{\cW}}\, .
\label{eq:Hbtrans}
\end{equation}
The previous variables can be written as a function of the new ones:
\begin{equation}
(\zeta,\zeta_2,Z,Z_2,x_j,\xt_j)=\chi_\cM(\zeta',\zeta'_2,Z',Z'_2,x'_j,\xt'_j)\, .
\label{eq:Hbvar}
\end{equation}
$\cW$ is of size $\eps$, the variables of the averaged problem are hence $\eps$-close from the variables of the full 3-body problem. From now on we write the new variables: ($\zeta,\zeta_2,Z,Z_2,x_j,\xt_j)$. We obtain:
\begin{equation}
\Hb = H_K(Z,Z_2) + \eps \Hb_P(\zeta,Z,Z_2,x_j,\xt_j) + \gO(\eps^2)\, . 
\label{eq:Hbp}
\end{equation}
We note that $Z_2$ is a constant of the averaged problem. Without loss of generality, we can take $Z_2=0$. In this case, the equations (\ref{eq:defeta1}) and (\ref{eq:tl1}) give:
\begin{equation}
\frac{\mu^2_1 \beta^3_1}{(\Lambda_1^0)^3}=\frac{\mu^2_2 \beta^3_2}{(\Lambda_2^0)^3}=\left( \frac{ \Lambda_1+\Lambda_2}{\mu_1^{2/3}\beta_1 + \mu_2^{2/3}\beta_2} \right)^3\, .
\label{eq:etaval}
\end{equation}
From this we define the mean mean-motion $\eta$ common to both co-orbitals:
\begin{equation}
\eta=\frac{\mu^2_1 \beta^3_1}{(\Lambda_1^0)^3} = \frac{\mu^2_2 \beta^3_2}{(\Lambda_2^0)^3}\, ,
\label{eq:defeta}
\end{equation}
and the averaged Hamiltonian becomes:
\begin{equation}
\Hb = H_K(Z) + \eps \Hb_P(\zeta,Z,x_j,\xt_j) + \gO(\eps^2). 
\label{eq:Hb}
\end{equation}

\subsection{Invariance of the circular manifold}
\label{sec:ivcp}
We can expand $\Hb_P$ (\ref{eq:Hb}) in Taylor series in the neighbourhood of ($x_1$,$x_2$)=($0$,$0$) \citep[see][]{RoPo2013}:
\begin{equation}
\Hb_P(\zeta,Z,x_j,\xt_j) = \sum_{(p,\tilde{p}) \in \mathbb{N}^4} C_{p,\tilde{p},q,\tilde{q}}(\zeta,Z) x_1^{p_1} x_2^{p_2}\xt_1^{\tilde{p}_1} \xt_2^{\tilde{p}_2},
\label{eq:Hpt}
\end{equation}
where  
$C_{p,\tilde{p}}$ are non-zero if and only if the coefficients $(p,\tilde{p}) \in \mathbb{N}^4$ follow the D'Alembert rule:
\begin{equation}
p_1 + p_2 = \tilde{p}_1 + \tilde{p}_2 .
\label{eq:DAr}
\end{equation}
This previous relation is equivalent to the fact that the total angular momentum is an integral of the problem, that is:
\be
\Lam_1 + \Lam_2 -ix_1\xt_1 -ix_2\xt_2 = cte.
\label{eq:ang_mom}
\ee

Therefore, the expansion (\ref{eq:Hpt}) contains only monomial of even total degree in ( $x_j$, $\tilde{x}_j$). As a consequence, the set $\mathcal{C}_{0}$, defined as: 
\begin{equation}
\mathcal{C}_{0} = \{(\zeta,Z,x_j,\xt_j ) / x_j=\xt_j=0 \} \, ,
\label{eq:circplanvar}
\end{equation}
that we call ``circular invariant manifold", is invariant by the flow of the averaged Hamiltonian (\ref{eq:Hb}).

\subsection{The circular dynamics}
\label{sec:Dyncp}

Restricting the Hamiltonian (\ref{eq:Hb}) to the circular coplanar manifold $\mathcal{C}_0$, \cite{RoNiPo2016} obtained an integrable approximation of $\Hb$  at the order ($Z^2$,$\eps$).
The equation canonically associated with that Hamiltonian can be rewritten as a 2nd order differential equation, generalising the model obtained by \cite{Ed1977}:
\begin{equation}
\ddot{\zeta}=-3 \eps \eta^2 \frac{m'_1+m'_2}{m_0} \left( 1- (2- 2\cos \zeta)^{-3/2} \right) \sin \zeta\, .
\label{eq:eqerdi}
\end{equation}

\subsubsection{The circular  motion}
\label{sec:pperdi}

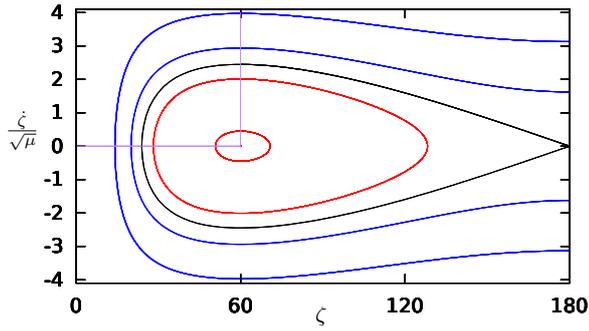
\begin{figure}[h!]
\begin{center}
\includegraphics[width=0.6\linewidth]{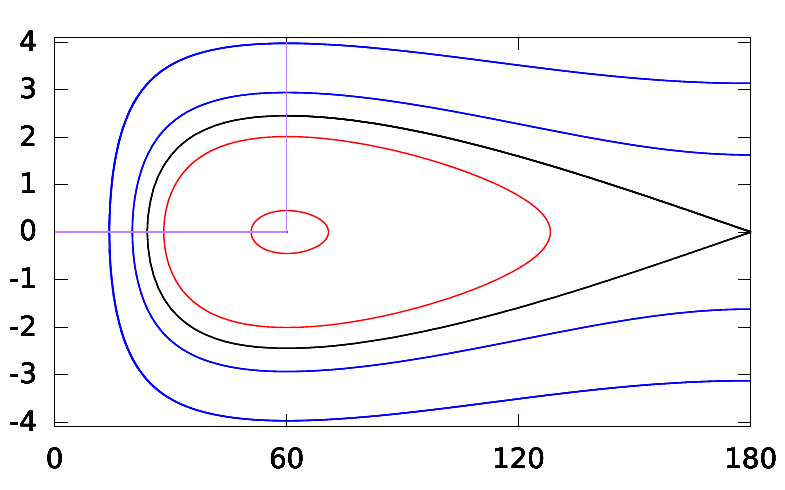}\\
  \setlength{\unitlength}{0.067\linewidth}
\begin{picture}(.001,0.001)
\put(-5,3.5){{$\frac{\dot{\zeta}}{\sqrt{\mu}}$}}
\put(0,0.5){{$\zeta$}}
\end{picture}
\caption{\label{fig:ppH0b} Phase portrait of equation (\ref{eq:eqerdi}). The separatrix (black curve) splits the phase
space in two different domains: inside the separatrix the region associated with the tadpole
orbits (in red) and the horseshoe domain (blue orbits) outside. The phase portrait is
symmetric with respect to $\zeta = 180^\circ$ . The horizontal purple segment indicates the range
of variation of $\zeta_0$ while the vertical one shows the section used as initial condition to draw
Fig. \ref{fig:stabzr}. See the text for more details. 
}
\end{center}
\end{figure}

The phase portrait of the 1-D model (eq. \ref{eq:eqerdi}) is given figure~\ref{fig:ppH0b} in the ($\zeta,\dot{\zeta}/\sqrt{\mu}$) plane, where
\begin{equation}
\mu =  \frac{m_1+m_2}{m_0+m_1+m_2}\, 
\label{eq:mu}
\end{equation}
is of size $\eps$. The phase portrait was plotted for a given value of the masses, but the topology of the phase space does not depend on their value.\\

Tadpole orbits (in red) librate around $L_4$ or $L_5$, while horseshoe orbits librate with large amplitude, encompassing the $L_4$, $L_3$ and $L_5$ equilibria. This libration of the resonant angle $\zeta$ is associated with the fundamental frequency $\nu$, which is small with respect to the mean mean-motion: $\nu \propto \eta \sqrt{\eps}$. In the vicinity of the $L_4$ and $L_5$ equilibrium, we have \citep{Charlier1906}:
\be
\nu_0 = \eta \sqrt{\frac{27}{4}\mu}.
\label{eq:nu_L}
\ee

Note that any trajectory in this phase space can be identified by its initial conditions $(t_0,\zeta_0)$ such that $\zeta(t_0)=\zeta_0$ and $\dot\zeta(t_0) =0$, where $\zeta_0$ is the minimal value of $\zeta$ on its trajectory, and $t_0$ is the first positive instant when $\zeta_0$ is reached. $\zeta_0$ sets the shape of the orbit, and $t_0$ gives the position of the bodies at a given time. Finally, the parameter $\eta\sqrt{\mu}$ gives the time scale of the resonant motion and the scale in the $Z$ direction \citep[$Z$ being proportional to $\dot \zeta$, see][]{RoPo2013}. \\

\subsection{Stability of quasi-circular co-orbitals}
\label{sec:stabcirc}

\begin{figure}[h!]
\begin{center}
\includegraphics[width=.49\linewidth]{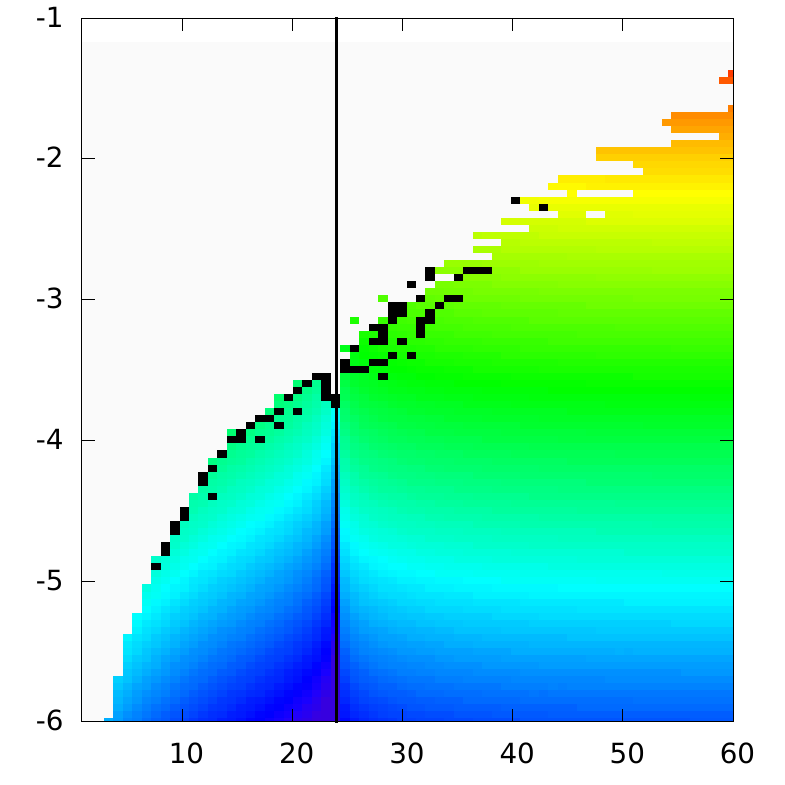}
\includegraphics[width=.49\linewidth]{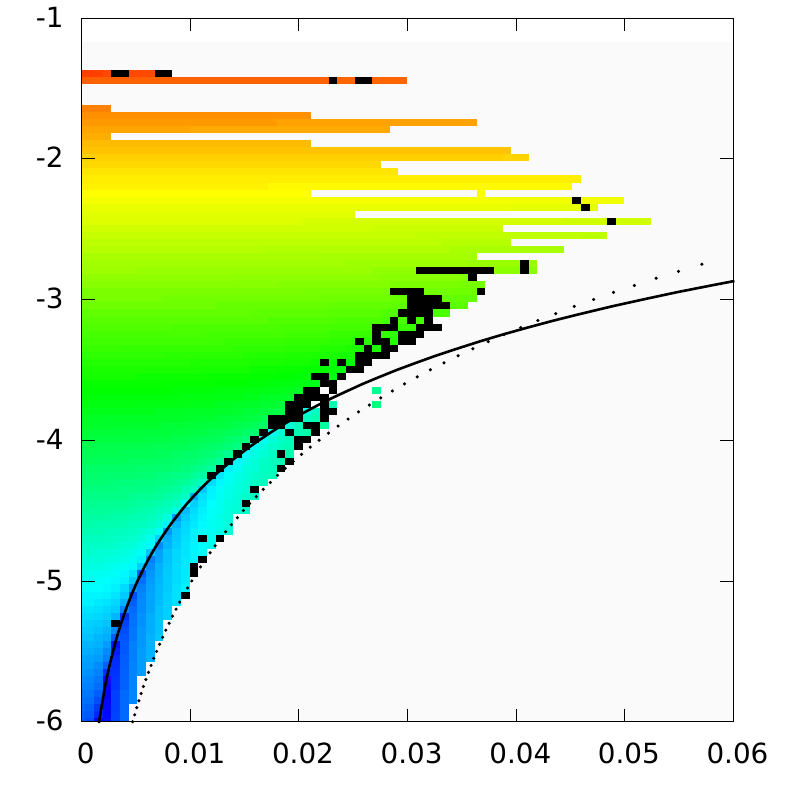}\\
\vspace{0.3cm}
\includegraphics[width=.49\linewidth]{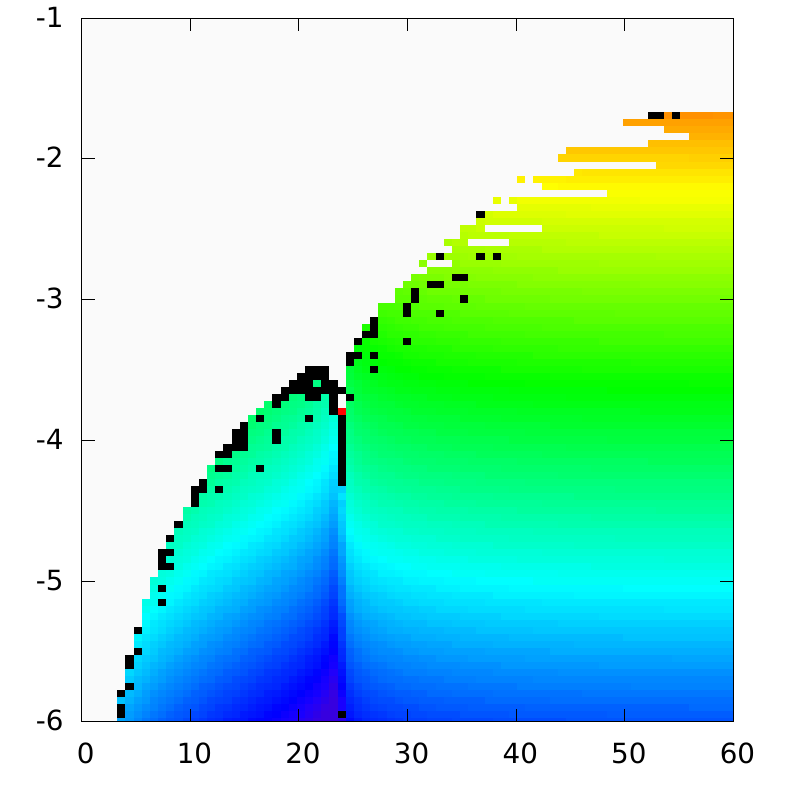}
\includegraphics[width=.49\linewidth]{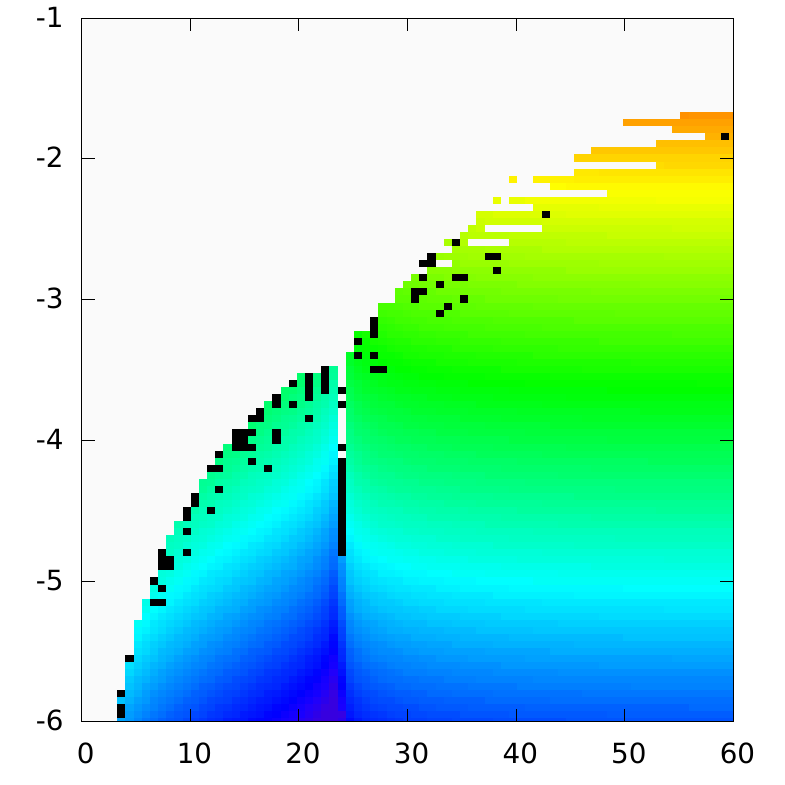}\\
\vspace{0.3cm}
\includegraphics[width=.49\linewidth]{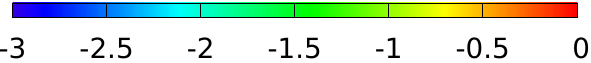}\\
  \setlength{\unitlength}{0.1\linewidth}
\begin{picture}(.001,0.001)
\put(-5,8.5){\rotatebox{90}{$\log_{10}(\mu)$}}
\put(0,8.5){\rotatebox{90}{$\log_{10}(\mu)$}}
\put(-5,3.5){\rotatebox{90}{$\log_{10}(\mu)$}}
\put(0,3.5){\rotatebox{90}{$\log_{10}(\mu)$}}
\put(-2.5,1){{$\zeta_0$}}
\put(2.5,1){{$\zeta_0$}}
\put(-2.5,6){{$\zeta_0$}}
\put(2.5,6){{$\Delta a/a$}}
\put(-4.25,10.75){{$(a)$}}
\put(0.75,10.75){{$(b)$}}
\put(-4.25,5.5){{$(c)$}}
\put(0.75,5.5){{$(d)$}}
\put(-0.5,0){{$\log_{10}(\nu/\eta)$}}
\end{picture}
\vspace{0.5cm}
\caption{\label{fig:stabzr} Stability of coplanar quasi-circular co-orbitals as a function of $\log_{10}(\mu)$ and $\zeta_0$ for the graphs (a), (c) and (d), and $\Delta a/a$ for the graph (b). For (a) and (b) $m_2=m_1$, for (c) $m_2=10 m_1$ and for (d) $m_2=100 m_1$. The black line in (a) and (b) shows the position of the separatrix between the tadpole and horseshoe domains. The color code gives the value of the libration frequency. See the text for more details.
}
\end{center}
\end{figure}
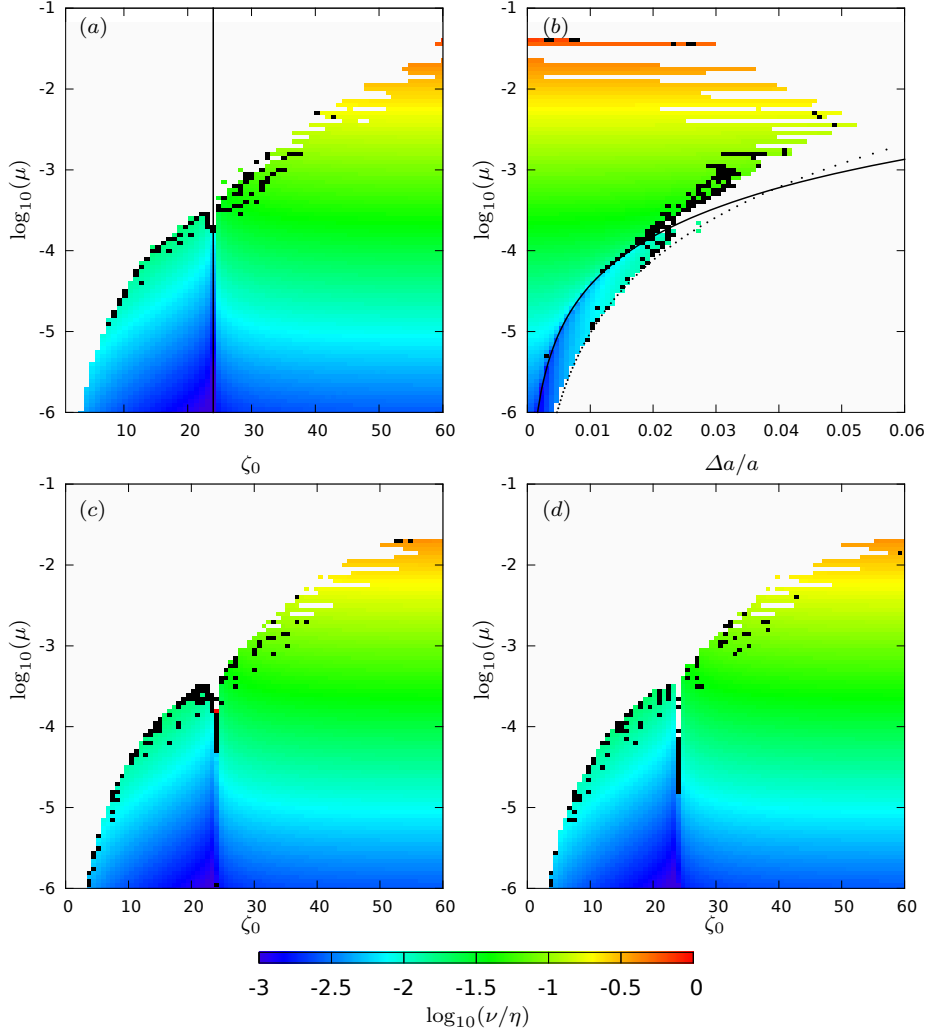

To study the stability of quasi-circular coplanar co-orbitals, we integrate the 3-body problem for a grid of initial conditions. As we saw in section \ref{sec:pperdi}, taking initial conditions in the $\zeta_0$ direction while taking $t_0=0$ allows to study all the possible co-orbital configurations in the coplanar circular case. We hence take $\zeta_0 \in [ 0, 60^\circ ] $ and $\mu= \frac{m_1+m_2}{m_0+m_1+m_2} \in [10^{-6 },10^{-1}]$ for our grid of initial conditions for the graphs (a), (c) and (d) in figure \ref{fig:stabzr}. In the graph (b), we check the width of the stability domain in the direction $Z$. We set $m_0=1\,M_\odot$, $a_1 = a_2 =1$~au, $e_1 = e_2 =0.05$, $\varpi_2=\lambda_2$, and $\lambda_1=\varpi_1=0^\circ$. The mass of each planet is given by the $y$ coordinate (the value of $\mu$) and the relation between $m_1$ and $m_2$: for graphs (a) and (b) $m_2=m_1$, for (c) $m_2=10 m_1$, and for (d) $m_2=100 m_1$. 

For each set of initial conditions, the system is integrated over $5 \times 10^{6}$~orbital periods using the symplectic integrator SABA4 \citep{LaRo2001} with a time step of $0.01001$ orbital periods.  Trajectories with a relative variation of the total energy above $10^{-6}$ are considered unstable. Note that the integrator is not especially well suited to handle close encounters. As a result, some stable trajectories might be labeled as unstable, and is that sense, the results presented here are conservative. Unstable trajectories, along with those ejected from the resonance before the end of the integration, are identified with white pixels. These short term instabilities are generally due to the overlap of secondary resonances \citep{RoGa2006,PaEf2015}. The black pixels identify the initial condition for which the diffusion of the libration frequency $\nu$ between the first and second half of the integration is higher than  $10^{-6}$ \citep[the stability check, along with the other numerical studies in this work, were performed using TRIP,][]{Gastineau2011}. Most of the black pixels are close to the stability boundary or the separatrix. The remaining trajectories are expected to be stable for a duration longer than $10^7$ orbital periods \citep{Laskar1990,RoGa2006}. For these trajectories, the color code gives the value of $log_{10}(\nu/\eta)$.

For $\mu$ close to the Gascheau's criterion value ($\mu\approx 0.037$), orbits are stable only in the vicinity of the Lagrangian equilibrium, confined by the chaos induced by the resonances $\nu = \eta/2$, $\nu = \eta/3$, and $\nu = \eta/4$. As $\mu$ decreases, orbits with larger amplitude of libration become stable, until stable horseshoe configurations appear for $\mu \approx 3\times 10^{-4}$ or lower \citep{Robe2002}. For these small $\mu$ values, the instability induced by the resonances is significant only near the stability border \citep[see][in the restricted case]{PaEf2015,RoGa2006,ErNa2007}. The stability domain of the horseshoe configuration in the $\zeta_0$ direction is bound by the Hill sphere around the collision, of width $\mu^{1/3}$ \citep[see][]{RoPo2013}. \\     

The graph (b) represents another section of the same phase space as graph (a): the initial conditions are taken along the purple vertical line in figure~\ref{fig:ppH0b}. The black curves delimit the trojan and horseshoe domains \citep{RoPo2013}. Combining the information of the graphs (a) and (b), we find that the co-orbital domain is at its largest for $10^{-3} < \mu < 10^{-2}$, and that the horseshoe domain ($\propto \mu^{1/3}$) becomes larger than the tadpole one ($\propto \mu^{1/2}$) as $\mu$ tends to $0$ \citep{DeMu1981a}.   \\

The graphs (a), (c) and (d) show that the mass repartition between co-orbitals does not impact much the stability, excepted in the vicinity of the separatrix. \\

\subsection{Periodic orbits' families in the neighbourhood of the circular Lagrangian and Eulerian equilibria}
\label{sec:LE}

The average problem, as defined in section  \ref{sec:H3bp}, possesses three fixed points\footnote{It is proven in \cite{RoNiPo2016} that the averaging process is not convergent in a neighbourhood of the collision between the two planets including the Hill sphere associated with this collision. The two Eulerian configurations that correspond to $L_1$ and $L_2$ are consequently excluded from the present study.}: two correspond to the Lagrange (circular) equilateral configurations,  $L_4$ and $L_5$ for $\zeta = \pm\pi/3$, $Z=x_1=x_2=0$, and the third one to the Euler configuration $L_3$ where the two planets are in the both sides of the more massive body for $\zeta =\pi$, $Z=x_1=x_2=0$. From these two equilibria (for symmetry reasons, $L_4$ and $L_5$ are dynamically equivalent), emanate several remarkable families of periodic orbits. These families being extensively described in \cite{RoPo2013}, only their main features will be discussed in this section.

The circular Lagrangian configuration ($L_4$) corresponding to an elliptic (stable) equilibrium\footnote{As long as the Gascheau criterion is fulfilled.}, gives rise to three periodic orbit families, according to the Lyapunov central theorem \cite[see][]{MeHa1992}. The first one is included entirely in the circular invariant manifold $\cC_0$. These orbits are those presented in the section \ref{sec:pperdi} in the neighbourhood of $L_4$. Their frequency tends to $\eta\sqrt{ 27\mu}/2$ as they approach the fixed point.

The second Lyapunov family, denoted by $\cF^1_{4}$, corresponds to a one-parameter family which is tangent, at its origin, to the orbits satisfying the relations 
\be
a_1 = a_2,  \quad  m_1e_1= m_2 e_2, \quad \zeta=\pi/3  \qtext{and}  \varpi_1 -\varpi_2 = \zeta + \pi.
\label{eq:relat_AL}
\ee
 This particular configuration is conserved over time while precessing at the secular frequency $g$ close to $27 \eta\mu/8$. $\cF^1_{4}$ is nothing but the beigining of the anti-lagrange family described by \cite{GiuBeMiFe2010} in the case of the reduced problem (see Sect. \ref{sec:RoP}). Let us mention that, although the relations (\ref{eq:relat_AL}) provide a good approximation of the $\cF^1_{4}$'s orbits for small eccentricities, they are no longer valid for high eccentricities \citep[see][]{GiuBeMiFe2010, HaVo2011}.

The last family, which is not strictly speaking a Lyapunov family, since it is only made of fixed points, is the one containing the eccentric Lagrange configurations that will be denoted by $\cF^2_{4}$. Indeed, these orbits, that fulfil the relations 
\be
a_1 = a_2,  \quad e_2= e_1  \qtext{and}  \quad \zeta = \varpi_1 -\varpi_2 =\pi/3
\label{eq:relat_L}
\ee
for all eccentricities, do not precess. In other words, the frequency associated with this last family is equal to zero, which corresponds to the fact that two eigenvalues of the linearised averaged system at the circular Lagrangian configurations vanish.

For the Eulerian point $L_3$, the situation is quite different. Its corresponding averaged linearised system has a pair of real eigenvalues, a pair of purely imaginary eigenvalues and two others equal to zero \cite[see][]{RoPo2013}. Only two Lyapunov-like families emanate from this point: the anti-Lagrange family  $\cF^1_{3}$ highlighted by \cite{HaPsyVo2009} and the eccentric Euler family $\cF^2_{3}$. The family $\cF^1_{3}$ is tangent, at its origine, to the orbits that satisfy the relations 
\be
a_1 = a_2,  \quad  m_1e_1= m_2 e_2, \quad \zeta=\pi \qtext{and}  \varpi_1 -\varpi_2 =0
\label{eq:relat_AE}
\ee
This condition is broken as the family moves away from $L_3$ \citep{HaPsyVo2009}.  As for the configurations belonging to $\cF^1_{4}$, the two ellipses, which are aligned in this case, precess at a frequency close to $27 \eta\mu/8$.

The last family, $\cF^2_{3}$, is obviously the one that corresponds to the elliptic Eulerian equilibria. For a given eccentricity, the associated ellipses pair satisfies the relations 
\be
a_1 = a_2,  \quad e_1=  e_2, \quad \zeta=0\qtext{and}  \varpi_1 -\varpi_2 =  \pi.
\label{eq:relat_E}
\ee

\section{Reduction of the problem in the eccentric case}

\begin{figure}
\begin{minipage}{0.49\linewidth}
 \begin{center}
 \includegraphics[width=1.8\linewidth, height=.7\linewidth]{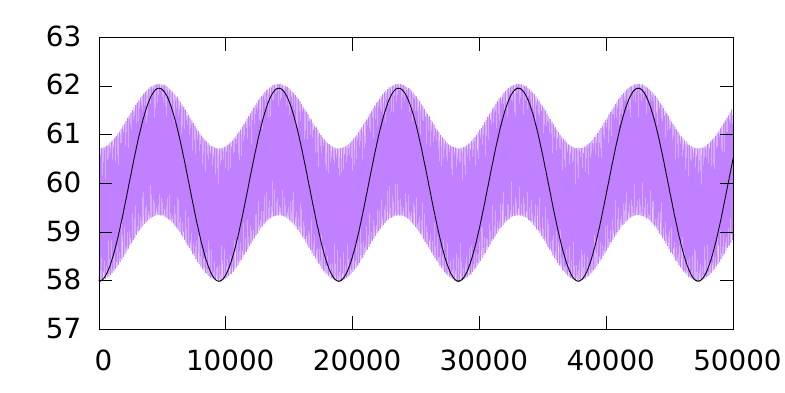} \\
   \setlength{\unitlength}{0.074\linewidth}
\begin{picture}(.001,0.001)
\put(-6.8,5){\rotatebox{90}{$\Dv$, $\mathbin{\color{purple}\zeta}$}}
\end{picture}
\end{center}
\vspace{-1.3cm}
 \begin{center}
 \includegraphics[width=1.8\linewidth,height=.7\linewidth]{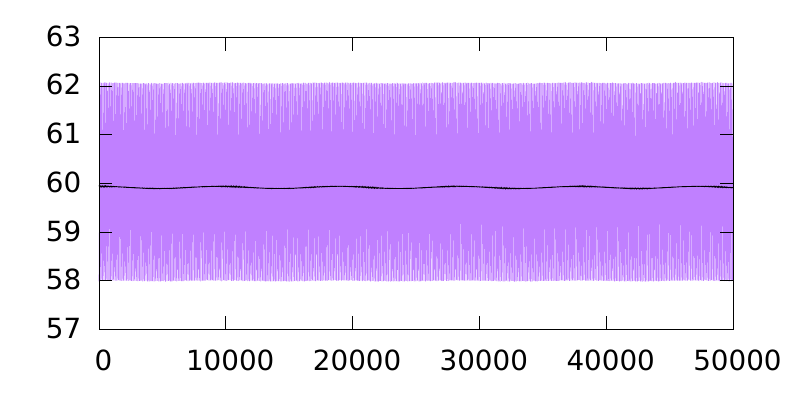} \\
   \setlength{\unitlength}{0.074\linewidth}
\begin{picture}(.001,0.001)
\put(-6.8,5){\rotatebox{90}{$\Dv$, $\mathbin{\color{purple}\zeta}$}}
\end{picture}
\end{center}
\vspace{-1.3cm}
 \begin{center}
 \includegraphics[width=1.8\linewidth,height=.7\linewidth]{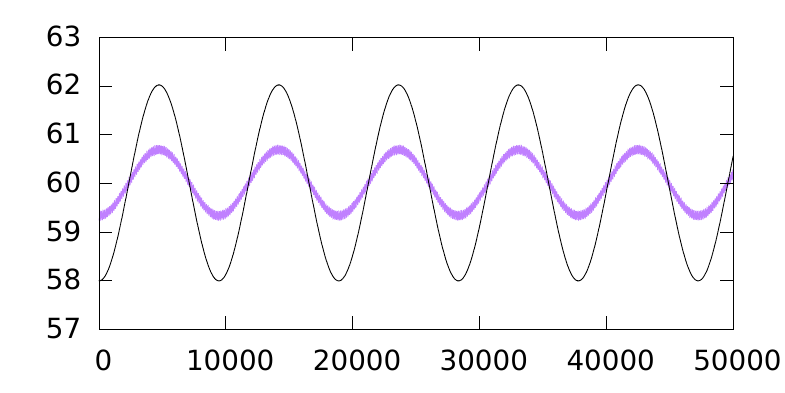} \\
   \setlength{\unitlength}{0.074\linewidth}
\begin{picture}(.001,0.001)
\put(-6.8,5){\rotatebox{90}{$\Dv$, $\mathbin{\color{purple}\zeta}$}}
\put(4,0){t[orbital period]}
\end{picture}
\end{center}
\end{minipage}
\caption{\label{fig:Flk} 
Temporal variations of the angles $\zeta=\lambda_1- \lambda_2$ (purple) and $\Dv=\varpi_1-\varpi_2$ (black) for three different initial conditions in the full three body problem (i.e. non-averaged, the $\lambda_j$ and $\varpi_j$ are osculating astrocentric elements), for $m_1=m_2=10^{-4}$ and $e_1=e_2=0.4$. Top: generic trajectory; Middle: trajectory near the $\Fsf$ family; Bottom: trajectory near the $\Fsc$ family, see the text for the definition of these families.
}
 \end{figure}
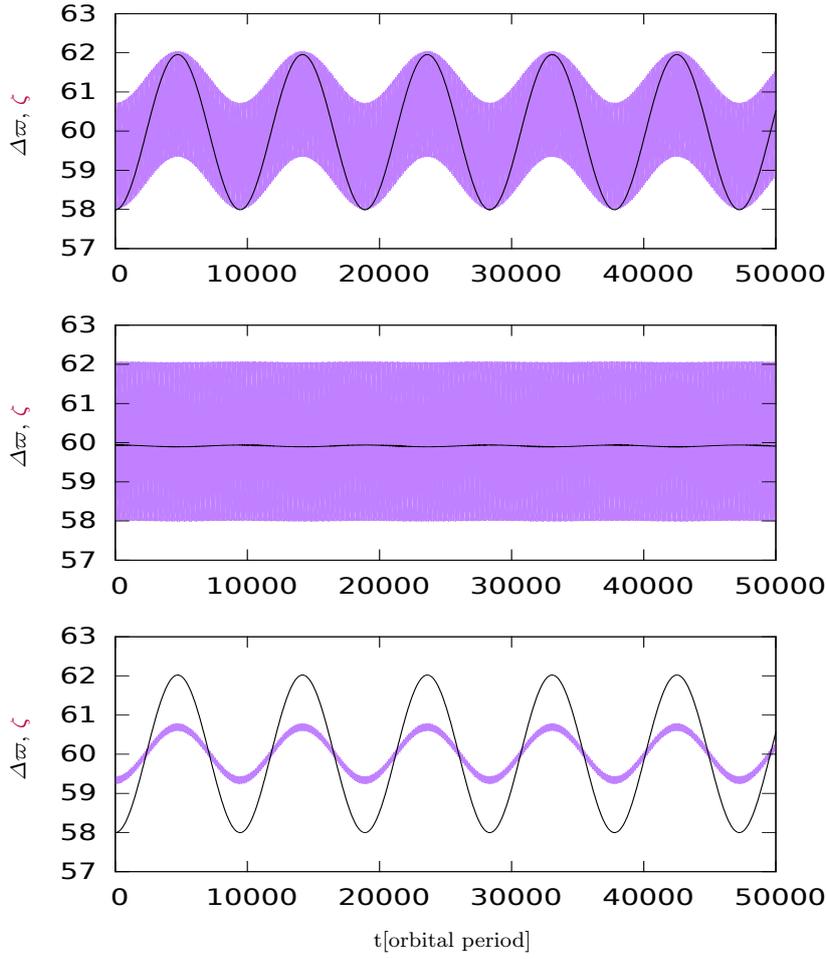

\subsection{Conservation of the total angular momentum}
\label{sec:RoP}

When the eccentricities are different from zero, it is possible to eliminate one more degree of freedom by using the conservation of the total angular momentum. Starting from the averaged Hamiltonian   (\ref{eq:Hpoinc}) and  following \cite{GiuBeMiFe2010}, 
we introduce the canonical coordinate system $(\zeta,\Dv,q,Q,\cZ,\Pi, J_1,J_2)$ given by: 
\begin{equation}
\begin{aligned}
\zeta & =\lambda_1 - \lambda_2; & \cZ &=(\Lambda_1-\Lambda_2)/2\\
\Dv & =\varpi_1 - \varpi_2; & \varPi &= i(x_2 \xt_2 - x_1 \xt_1)/2\\
q &= \varpi_1+\varpi_2; & J_1 &= (\Lambda_1+\Lambda_2-i(x_1 \xt_1+x_2 \xt_2))/2\\
 Q &= \lambda_1+ \lambda_2 - q; & J_2 &=(\Lambda_1+\Lambda_2)/2\, .
\end{aligned}
\label{eq:CIred}
\end{equation}
Since the action $J_1$ is an integral of the motion (half the total angular momentum, equation \ref{eq:ang_mom}), the angle  $q = \varpi_1+\varpi_2$ can be ignored and the system associated with the reduced Hamiltonian $\gH_\cR$ possesses only three degrees of freedom and depends on the parameter $J_1$. This Hamiltonian can additionally be averaged over the fast angle $Q$ to become the averaged reduced Hamiltonian, denoted $\gH_{\cR\cM}$. This new function has only two degrees of freedom and depends on the two parameters $J_1$ and $J_2$. This last integral can be considered as a scaling factor associated with the mean semi-major axis (hence the mean mean-motion) and will be omitted it the subsequent sections. 
As a consequence, for a given value of $J_1$,  the coordinates $(\zeta,\Dv,\cZ,\Pi)$ are adapted to the averaged reduced system in study.

Before going further, let us interpret the remarkable periodic orbits described in Sect. \ref{sec:LE}. Since the coordinate system (\ref{eq:CIred}), has a singularity when $e_1=e_2=0$, the  circular manifold $\cC_0$ does not belong to any averaged reduced phase  space. Regarding the other Lyapunov families $\cF^j_k$, the intersection of one of them with the surface  $J_1=cte$ is reduced to a single point.  As for a given periodic orbit of these families, the angle $\Dv$ does not depend on the time, these intersection points are equilibrium points of the averaged reduced problem. 
We call $L_k=\cF^2_k \cap \{J_1= \text{cte} \}$ and $AL_k=\cF^1_k \cap \{J_1= \text{cte} \}$ these fixed points. From now on, the equilibrium point $L_k$ refers to the given eccentric equilibrium point except if `circular' is mentioned.

More generally, a generic quasi-periodic solution of the averaged reduced problem depends on two fundamental frequencies: The frequency $\nu$, which is of order $\sqrt\mu$ and mainly associated with the semi-fast component $(\zeta,\cZ)$, and the secular frequency $g = \gO(\mu)$ related to the slow variations of $(\Dv, \Pi)$. 

Some of these quasi-periodic orbits have only one frequency and are consequently periodic. Let us denote by $\Fsf$ the semi-fast periodic orbit family, defined by:
\begin{equation}
\frac{\partial }{\partial \varPi}\gH_{\cR\cM} = \frac{\partial }{\partial \Dv} \gH_{\cR\cM}= 0\, ,
\label{eq:defFsf}
\end{equation}
 and $\Fsc$ the secular one, defined by\footnote{The $\Fsc$ and $\Fsf$ correspond respectively to the $\sigma-family$ and the $\Dv-family$ studied in \cite{GiuBeMiFe2010}.}:
\begin{equation}
\frac{\partial }{\partial \zeta}\gH_{\cR\cM} = \frac{\partial }{\partial \cZ} \gH_{\cR\cM}= 0\, .
\label{eq:defFsf}
\end{equation}
In the neighbourhood of the fixed points $L_k$ and $AL_k$, the set $\Fsc$ coincides with the secular Lyapunov family of periodic orbits originated at these points, while $\Fsf$ merges with the semi-fast Lyapunov family connected to $L_4$ an $AL_4$ ($L_3$ and $AL_3$ being hyperbolic fixed points, they possesses only one Lyapunov family).

Examples of these trajectories are displayed in Fig. \ref{fig:Flk}. The top graph  shows the variation of $\zeta$ (purple) and $\Dv$ (black) for a generic quasi-periodic orbit: both the semi-fast evolution (here $2\pi/\nu \approx 100$ orbital periods) and the secular one ($ \approx 10000$ orbital periods) are visible on $\zeta$, while $\Dv$ evolves mainly on the secular time scale. In the no-averaged and no-reduced problem, this trajectory possesses an additional precession frequency (which would leave the chosen angles $\zeta$ and $\Dv$ invariants) and small short time variations, which leads to a  quasi-periodic trajectory possessing  4 fundamental frequencies in the full 3-body problem. 
 
The middle graph represents a trajectory with its initial conditions close to the $\Fsf$ family: the secular time scale associated with the frequency $g$ does not impact the orbit, it is hence a periodic orbit of semi-fast frequency $\nu$ in the averaged reduced problem and a  quasi-periodic orbit with  two 2 frequencies in the averaged problem and three in the full problem. Finally, the bottom graph represents a trajectory with its initial conditions close to the $\Fsc$ family: the semi-fast time scale associated with the frequency $\nu$ does not impact the orbit, which is quasi-periodic with two frequencies in the averaged problem.

\subsection{Reference manifold $\cV$ in the case $m_1=m_2$}
\label{sec:RM}

In the circular coplanar case, we saw in section \ref{sec:Dyncp} that the initial conditions of the system were equivalent to a couple ($\zeta_0,t_0$) where $\zeta_0$ defines the orbit and $t_0$ defines a trajectory on this orbit. We could hence explore the characteristics of all the trajectories of the phase space by studying only the trajectories having for initial condition ($\zeta_0$, $t_0=0$). This reduces the relevant space of initial conditions to a 1-dimensional space.

In the eccentric case, the $4$ dimensions of the reduced restricted phase space require 4 initial conditions $(\zeta,\Dv,\cZ,\varPi)$ to define a given trajectory. Following the circular case, we want to define a 2-dimensional manifold $\cV$ of initial conditions which would be representative of the 4-dimensional phase space of the averaged reduced problem \citep{MiFeBe2006}. We consider that $\cV$ is a representative manifold of the averaged reduced phase space if the trajectories emanating from this surface explore a significant part of the entire phase space. 

In the case $m_1=m_2$, the manifold

\be
\cV=\left\{(\zeta,\Dv) \in [0, 2\pi]^2 \qtext{with} \cZ = \Pi = 0\right\} ,
\label{eq:RM}
\ee

that is, $a_1=a_2$ and $e_1=e_2$, is a good candidate for a given value of the masses and the total angular momentum, as it contains the  $L_k$, $AL_k$ equilibria and the $\Fsf$ and $\Fsc$  families, at least for low eccentricities (see section \ref{sec:LE}). We want that the trajectories emanating from $\cV$ explore the entire phase space. Since the Hamiltonian flow is continuous, it is equivalent to show that any trajectory of the phase space goes as close as we want to $\cV$ in a finite time. We demonstrate this result at first order in eccentricity in section \ref{sec:foe}, and numerically for higher eccentricities ($e_1=e_2=0.4$) in section \ref{sec:fle}.
 
  In the case $m_1 \neq m_2$, the definition of a reference manifold is significantly more complicated. An algorithm to obtain such manifold is proposed in \cite{these}, section 2.6.2.

\section{Phase space of eccentric co-orbitals in the case $m_1=m_2$}
\label{sec:ECC}

In this section we study the impact of the total angular momentum $J_1$ (which is equivalent to the value of the eccentricities) on the dynamics and the stability of the co-orbital configuration.

\subsection{Position of the $\Fsc$ on the reference manifold $\cV$}
\label{sec:recon}

The separation between the semi-fast and the secular time scales (appendix \ref{sec:tsai}) allows us to determine the position of the intersection between $\cV$ and the $\Fsc$ families by studying the critical points of the averaged Hamiltonian (appendix \ref{sec:SAIF}). Note that the determination of the position of the $\Fsc$ with this method is independent from $\eps$. As long as $m_1=m_2$, it is hence independent of the value of the planetary masses.\\

\begin{figure}[h!]
\begin{center} 
\includegraphics[width=0.4\linewidth]{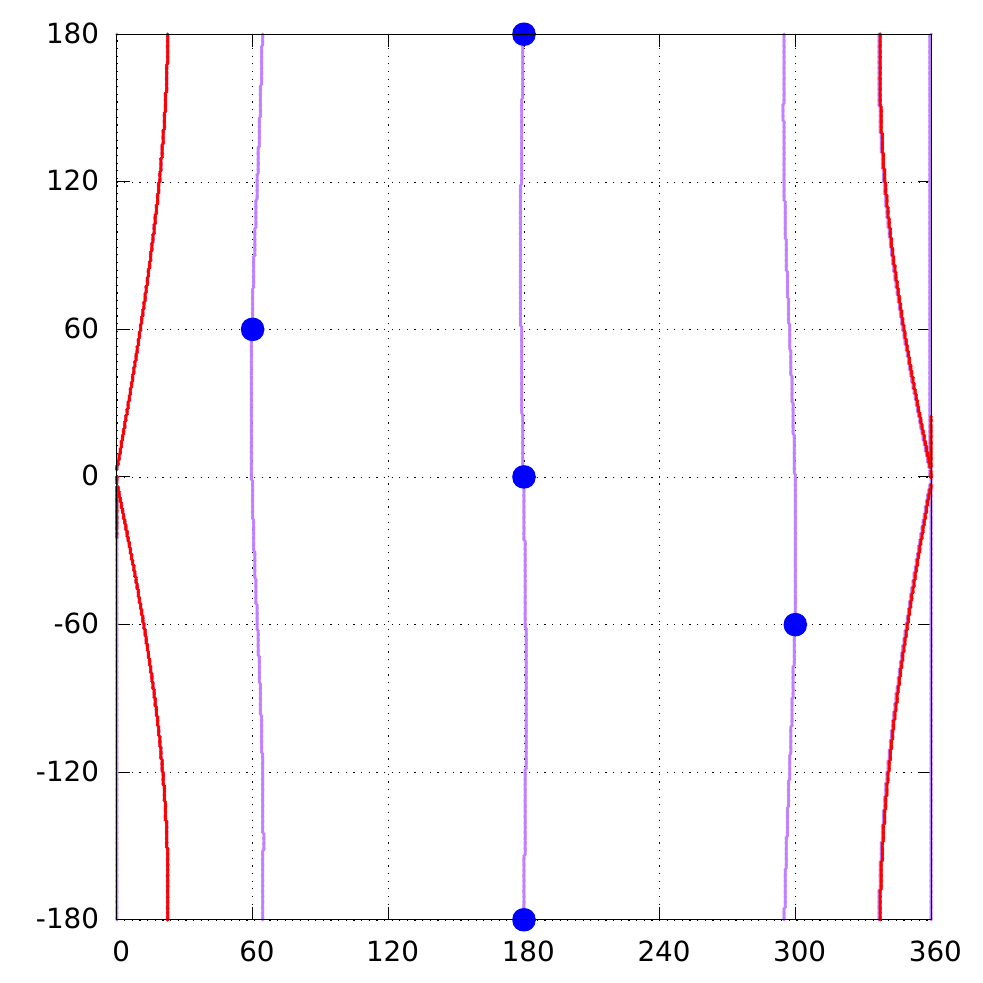}
\includegraphics[width=0.4\linewidth]{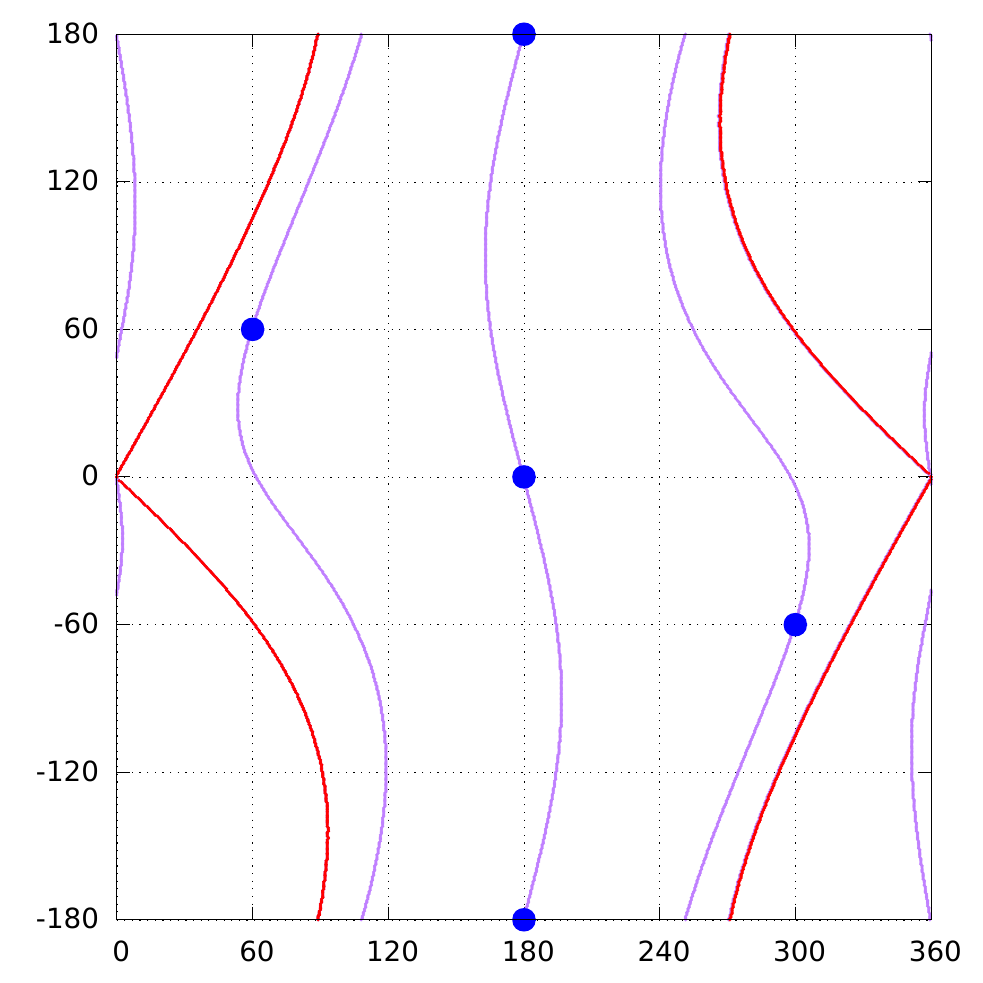}\\
\includegraphics[width=0.4\linewidth]{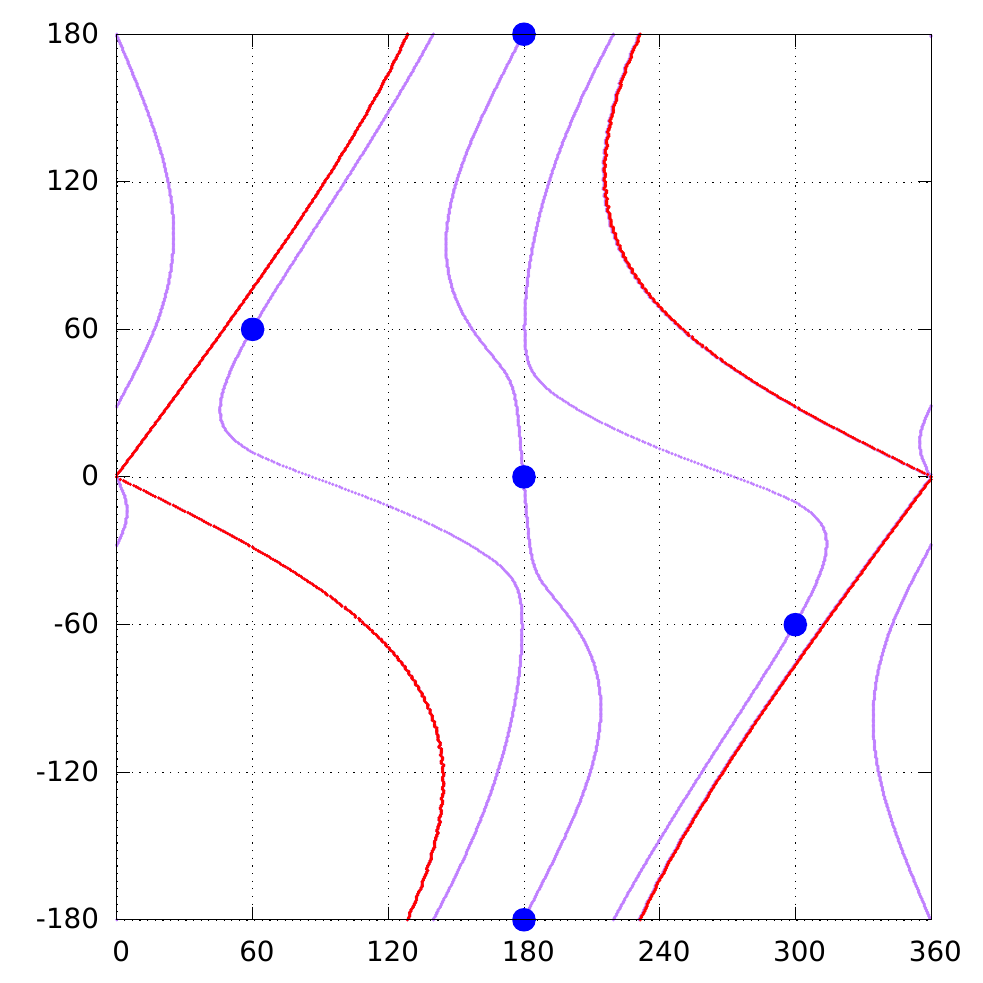}
\includegraphics[width=0.4\linewidth]{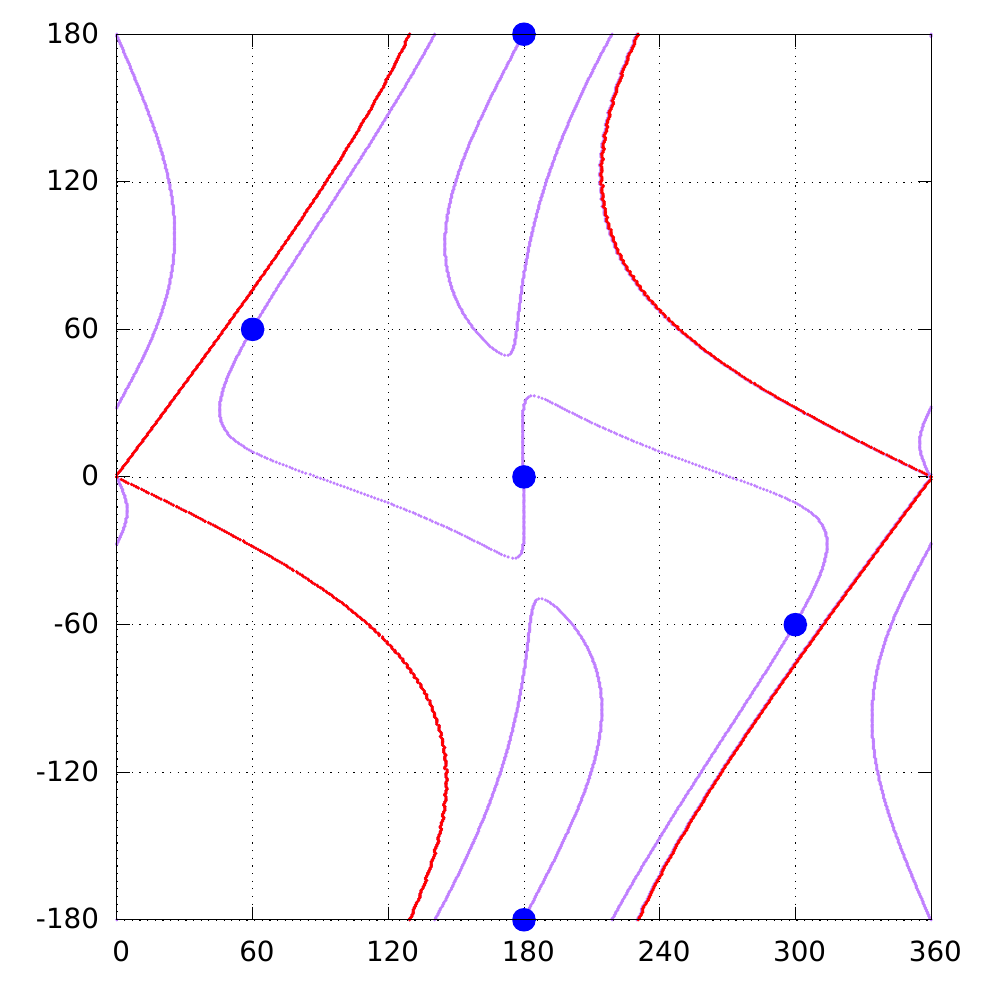}\\
\includegraphics[width=0.4\linewidth]{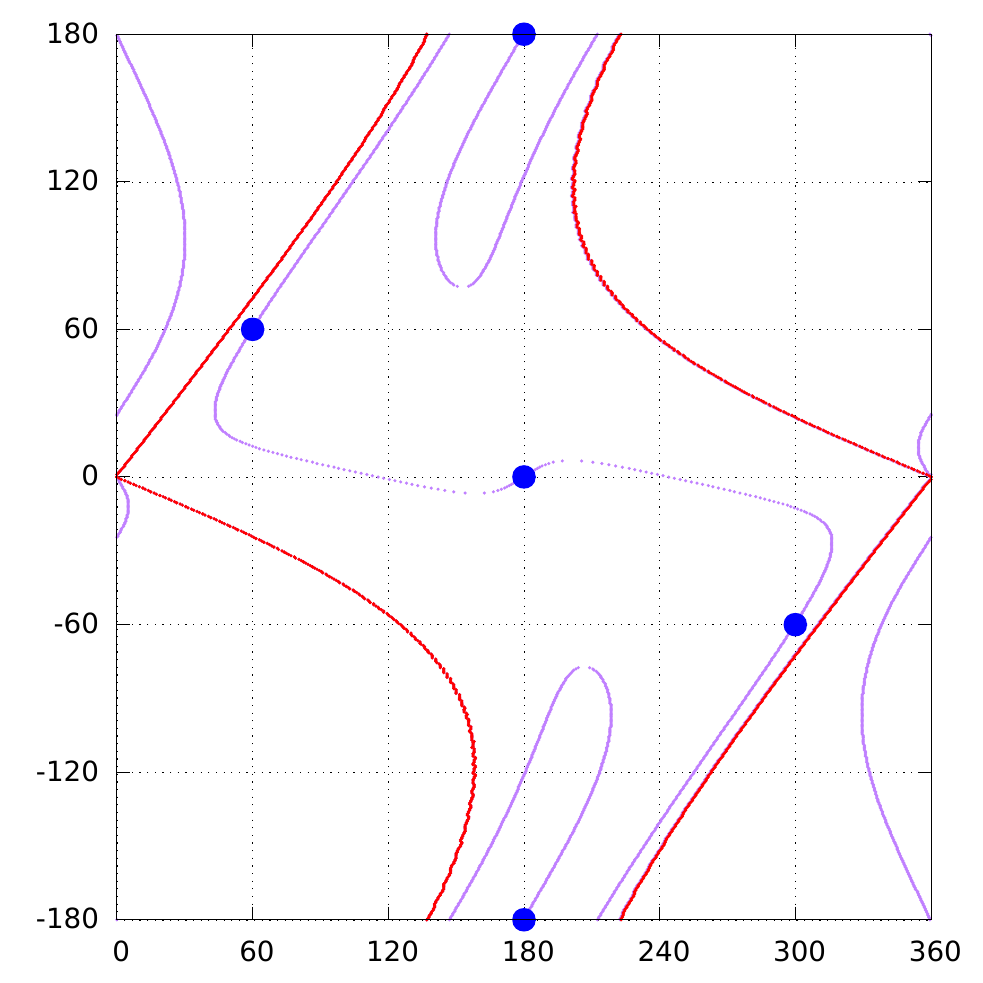}
\includegraphics[width=0.4\linewidth]{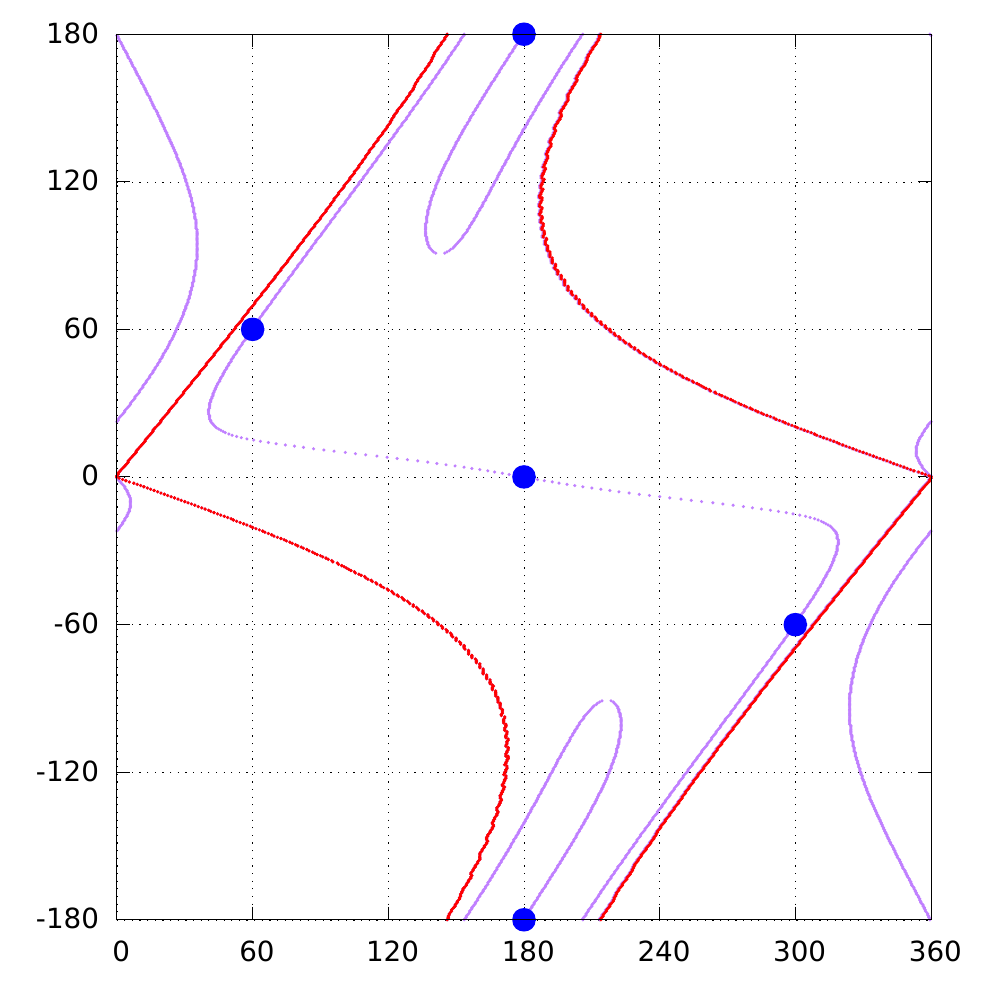}\\
 \setlength{\unitlength}{0.1\linewidth}
\begin{picture}(.001,0.001)
\put(-4.1,2.1){\rotatebox{90}{$\Dv$}}
\put(-4.1,6.1){\rotatebox{90}{$\Dv$}}
\put(-4.1,10.2){\rotatebox{90}{$\Dv$}}
\put(-2,0){{$\zeta$}}
\put(2.1,0){{$\zeta$}}
\put(-.6,3.8){{(e)}}
\put(-.6,7.8){{(c)}}
\put(-.6,11.8){{(a)}}
\put(3.5,3.8){{(f)}}
\put(3.5,7.8){{(d)}}
\put(3.5,11.8){{(b)}}
\put(-2.95,9.5){{$\cF^{sc}_4$}}
\put(-1.85,9.5){{$\cF^{sc}_3$}}
\put(-1.25,11.2){{$\cF^{sc}_5$}}
\end{picture}
\caption{\label{fig:Fb0meq}  $\Fsc$ (in purple), and the collision manifold (in red), on $\cV=\{a_1=a_2,e_1=e_2\}$. (a) $e_j=0.1$; (b) $e_j=0.4$; (c) $e_j=0.6$; (d) $e_j=0.605$; (e) $e_j=0.65$ et (f) $e_j=0.7$. The blue dots show the position of the} $L_k$ and $AL_3$ \citep[the position of $AL_4$ and $AL_5$ evolve with $e_j$, see][]{GiuBeMiFe2010}. See the text for more details.  
\end{center}
\end{figure}

  In figure \ref{fig:Fb0meq} we show the $\Fsc$ and the collision manifold on $\cV$ (that is, the points of $\cV$ that verify the condition (\ref{eq:condFb0n}), see appendix \ref{sec:appmm} for more details). Each graph corresponds to a different value of the total angular momentum (hence a different value of $e_1=e_2$). The curve $\cV \cap \Fsc$ is represented in purple, the blue circles represent the $L_k$ and $AL_k$ that have a fixed position on the ($\zeta$,$\Dv$) plane ($AL_4$ and $AL_5$ are hence excluded), and the supposed intersection between $\cV$ and the collision manifold\footnote{
The red curves satisfy the relation (\ref{eq:condFb0n}), contain the collision point $(0,0)$, and are located at a relevant position for the collision manifold (see the purple curves in the unstable areas in the figures \ref{fig:glob_e01} to \ref{fig:glob_e7_m6}). The collision manifold satisfies equation (\ref{eq:condFb0n}) if $\frac{\partial}{\partial \zeta} \gH_{\cR \cM}$ tends to $- \infty$ when we get close to the collision from one side and $+ \infty$ from the other side.} is represented in red.\\

For small eccentricities ($\leq 0.1$), we are in the neighbourhood of the circular case and the direction of $\Dv$ does not impact much the position of the $\Fsc$. Note that the same branch of the $\Fsc$ family contains the $L_3$ and $AL_3$ equilibria. We will call this branch $\cF_3^{sc}$. Similarly, we call $\cF_4^{sc}$ (resp. $\cF_5^{sc}$) the branch going through $L_4$ and $AL_4$ (resp. $L_5$ and $AL_5$). For $e_1=e_2=0.1$, a new curve appears for $\zeta \approx 0^\circ$ (the curve is mingled with the axis $\zeta=0$ in the figure (a)). This branch of $\Fsc$ intersects the domain of the quasi-satellite configuration.

When we increase the eccentricity, there is a growing dependence on the direction of $\Dv$ for the position of the $\Fsc$ family. Until $e_1=e_2\approx 0.6$, the sole effect of the increasing eccentricity is to twist the existing branches of the $\Fsc$.

Between $e_j=0.6$ and $e_j=0.605$, an important topological change occurs: in the averaged problem, the $\cF^{sc}_k$ reconnect in order to create a single continuous family of periodic orbits that goes through all the $L_k$ and $AL_k$ for $k \in \{1,2,3\}$. As we will see in the coming sections, this reconnection leads to a modification of the whole phase space of the eccentric co-orbital resonance.\\

Note that we identify here the families of periodic orbits of the averaged reduced problem. To verify equation (\ref{eq:condFb0}) is only a necessary condition for the associated orbit of the full planar 3-body problem to be a quasi-periodic orbit with $3$ fundamental frequencies, but we still need to check if the orbit is indeed quasi-periodic (not unstable/chaotic).

\subsection{Trajectories emanating from the reference manifold $\cV$}

\label{sec:Vmeq}


In order to represent most of the planar co-orbital dynamics for a given value of $m_1=m_2$ and $J_1(e_1,e_2)$, we take initial conditions on the reference manifold $\cV$ that was defined in section \ref{sec:RM}: $a_1=a_2$ ($=1$~au, the value of the semi-major axis is a scale factor), and $e_1=e_2$. We also chose $m_0$ equal to one solar mass, and $\lambda_1=\varpi_1=0^\circ$. The other initial conditions are given by the coordinate of the point on the grid of initial conditions.\\

We perform here numerical integrations of the full 3-body problem. However, as stated in section \ref{sec:ins}, the result of these integrations (in the case of quasi-periodic orbits) can be interpreted as the trajectories of the averaged reduced problem. As for the quasi-circular case (see figure \ref{fig:stabzr}), we expect that $\eps$ does not change the shape of the orbits, but it only impact the size of the stability domains and the time scale.


For each set of initial conditions, the system is integrated over $10/\eps$ orbital periods using the symplectic integrator SABA4 \citep{LaRo2001} with a time step of $0.01001$ orbital period (eccentricities larger that $0.6$ may require to take a smaller time step in order to avoid to eject stable orbits for numerical reasons). The initial conditions that lead to highly chaotic orbits, or that quit the resonance before the end of the integration are identified by a white pixel in the figure. Moreover, in order to identify the orbits that are not stable on a time scale that is long with respect to $10/\eps$, we compute the variation of the average value of the semi-major axis of the planet $m_1$ between the first and the second half of the integration. The grey pixels identify the initial conditions for which this diffusion is higher than a given small parameter $\epsilon_a$. Since the phase space is symmetric with respect to the point ($\zeta=0,\Dv=0$), we compute and describe only half of the phase space ($\zeta \in [0,180]$). The other half is also displayed for a better understanding of the whole phase space.\\

In the figures \ref{fig:glob_e01} and \ref{fig:glob_e6} we show the integration of the grid of initial conditions of $\cV$ for $e_j=0.01$, $0.4$, $0.65$ and $0.7$. In each case, $m_1=m_2=10^{-5} m_0$. The left graphs represent the mean value of $\zeta$ over the whole integration and the right ones represent the mean value of $\Dv$. When markers such as $\times$ or $+$ are displayed on the left graphs, they indicate the point of the manifold in the neighbourhood of which a given orbit (examples plotted in figures \ref{fig:orbe01}, \ref{fig:orbe4} and \ref{fig:orbe7}) crosses the plane quasi-periodically. Note that a generic trajectory crosses in the neighbourhood of 4 distinct points of the the reference manifold \citep[see][for more details]{MiFeBe2006,these}. On the right plot, the numerical criteria (\ref{eq:condFb02}) and (\ref{eq:condFb1}) (developed in appendix \ref{sec:SAIF}) are used to identify the position of the intersection between $\cV$ and $\Fsc$ in brown and $\Fsf$ in black. For comparison, we also plot in these graphs the result of the research of critical points of the Hamiltonian (figure \ref{fig:Fb0meq}), to identify the position of the $\Fsc$ families (in purple in the figures \ref{fig:glob_e01} to \ref{fig:glob_e7_m6}).\\


\subsubsection{Quasi-circular case}

\begin{figure}[h!]
\begin{center}
\includegraphics[width=0.5\linewidth]{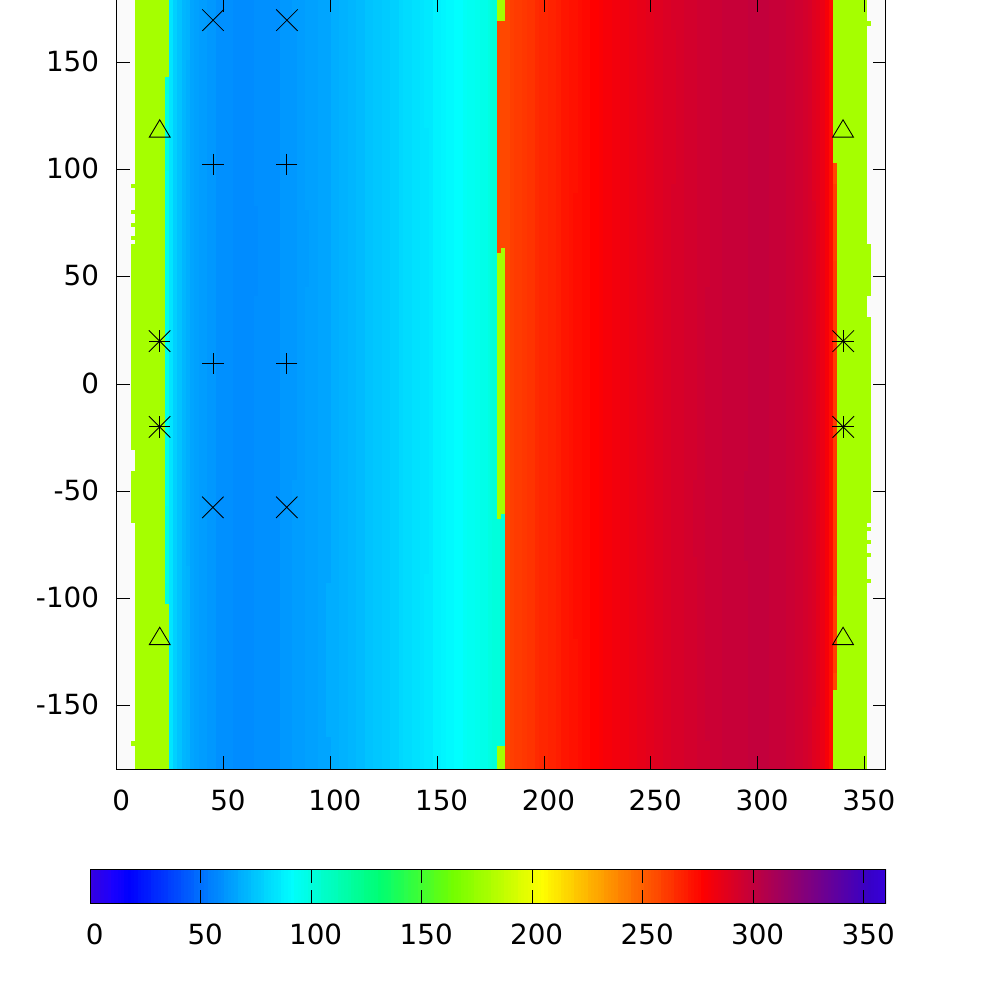}\includegraphics[width=0.5\linewidth]{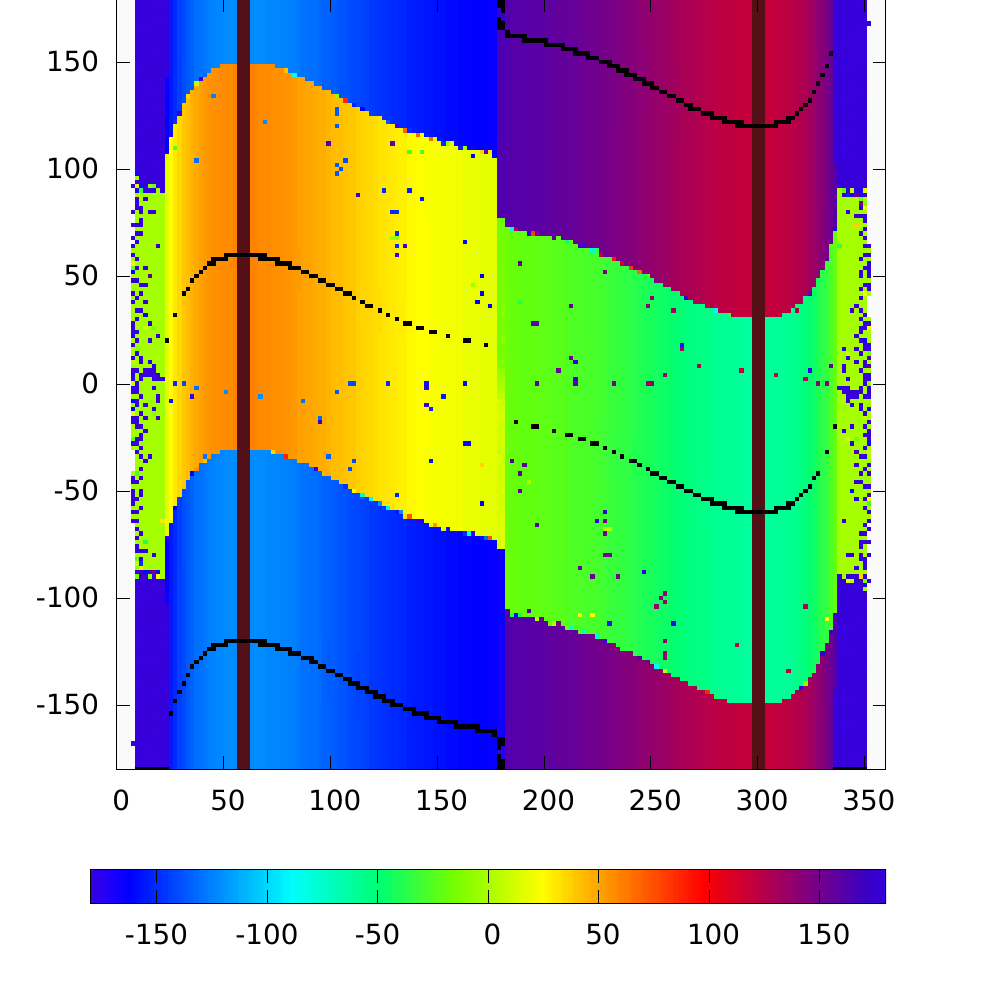}\\
  \setlength{\unitlength}{0.067\linewidth}
\begin{picture}(.001,0.001)
\put(-7.5,4.5){\rotatebox{90}{$\Dv$}}
\put(0,4.5){\rotatebox{90}{$\Dv$}}
\put(-3.8,1.4){{$\zeta$}}
\put(3.8,1.4){{$\zeta$}}
\put(-4,0.2){{moy($\zeta$)}}
\put(3.3,0.2){{moy($\Dv$)}}
\end{picture}\\
\includegraphics[width=0.5\linewidth]{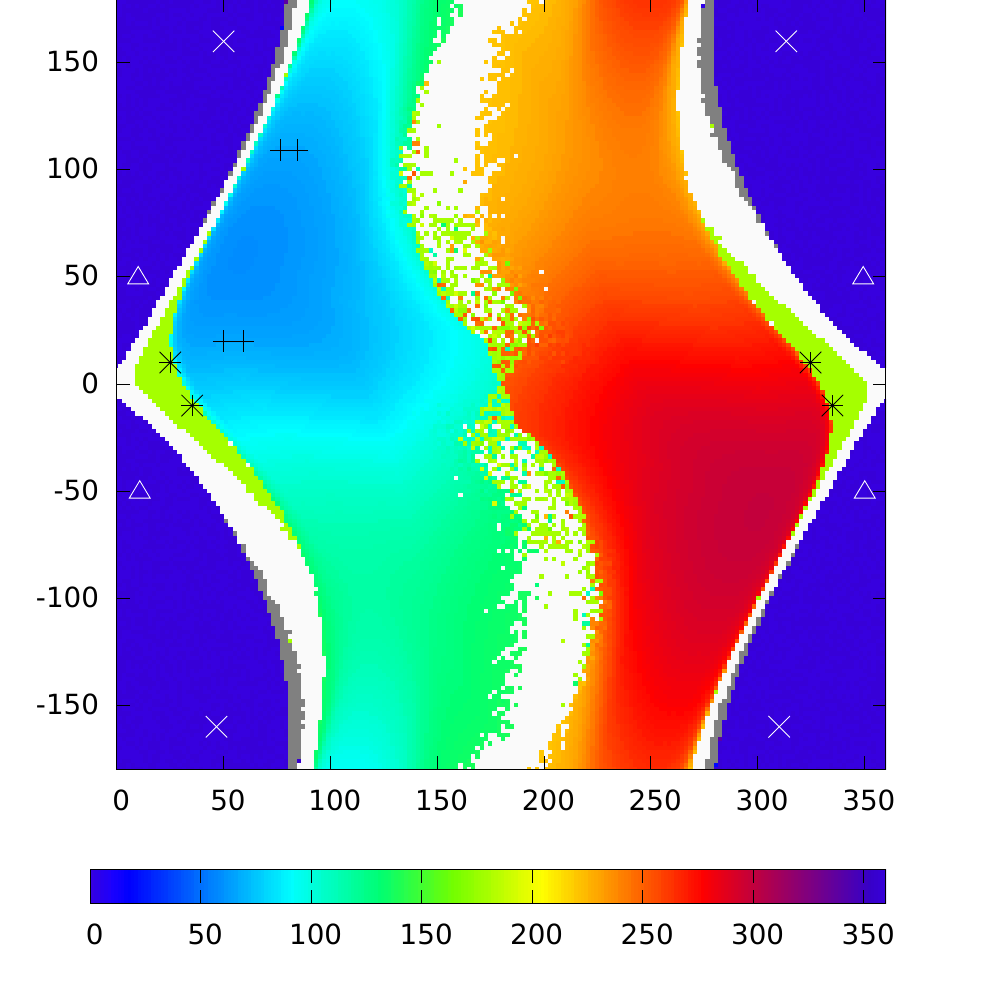}\includegraphics[width=0.5\linewidth]{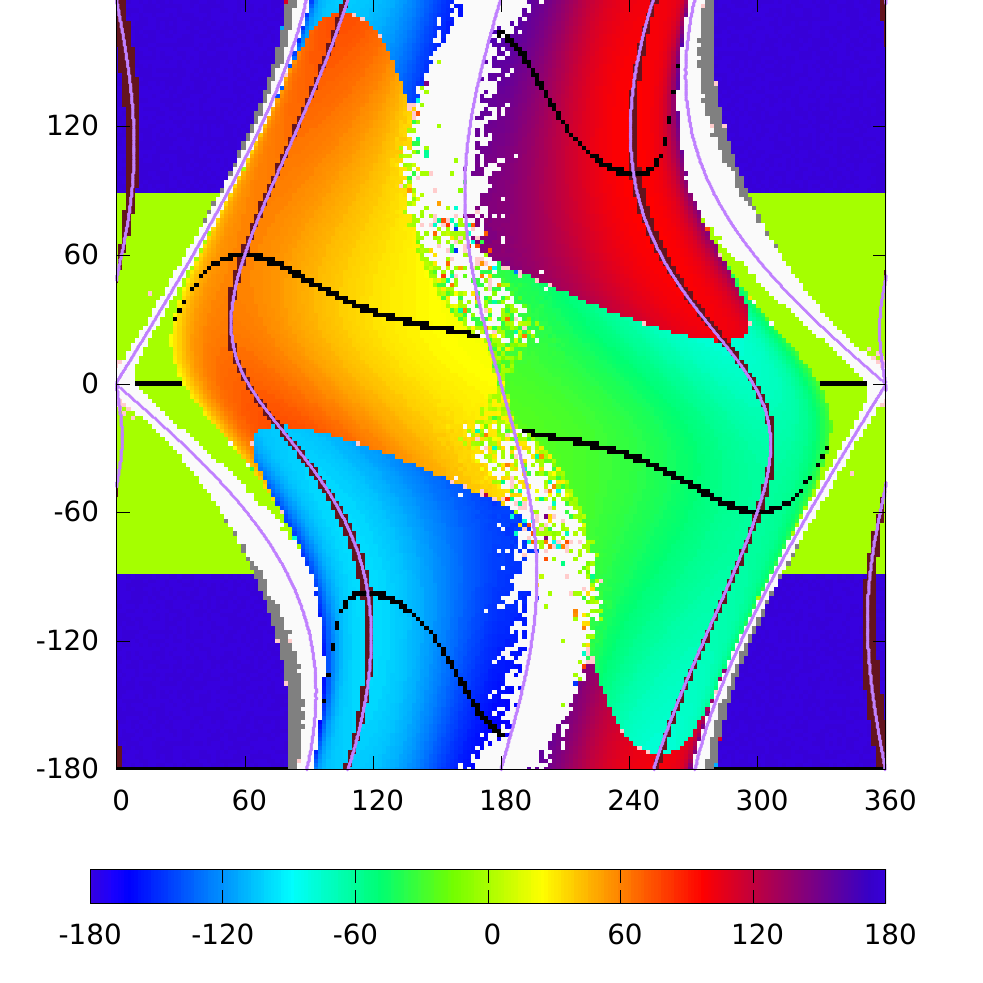}\\
  \setlength{\unitlength}{0.067\linewidth}
\begin{picture}(.001,0.001)
\put(-7.5,4.5){\rotatebox{90}{$\Dv$}}
\put(0,4.5){\rotatebox{90}{$\Dv$}}
\put(-3.8,1.4){{$\zeta$}}
\put(3.8,1.4){{$\zeta$}}
\put(-4,0.2){{moy($\zeta$)}}
\put(3.3,0.2){{moy($\Dv$)}}
\end{picture}
\caption{\label{fig:glob_e01} Grid of initial conditions for $m_1/m_0=m_2/m_0=10^{-5}$, $a_1=a_2=1$~au, and $e_1=e_2=0.01$ (top), $e_1=e_2=0.4$ (bottom). The color code on the left hand graphs gives the mean value of $\zeta$ on the orbit emanating from each initial condition. The markers shows the points of the manifold near which the orbits of the figures~\ref{fig:orbe01} and ~\ref{fig:orbe4} cross. On the right hand graphs the color code indicates the mean value of $\Dv$. In the top right graph, the eccentricities are low and might vanish, $\Dv$ is thus difficult to determine. The orbits in the neighbourhood of $\Fsc$, hence those verifying (\ref{eq:condFb02}) with $\epsilon_\nu=10^{-3.5}$, are represented by brown pixels. The purple curves show the result of the semi-analytical method (eq. \ref{eq:condFb0n}). The orbits close to $\Fsf$, hence those verifying (\ref{eq:condFb1}) with $\epsilon_g= 3^\circ$, are represented by black pixels. The initial conditions that lead to a diffusion of the mean semi-major axis over $\epsilon_a=10^{-5.5}$ are displayed in grey.}
\end{center}
\end{figure}

 \begin{figure}[h!]
 \begin{minipage}{0.49\linewidth}
\includegraphics[width=1\linewidth]{epsfigs/sim20160517a_mesoDlam2.pdf}\\
  \setlength{\unitlength}{0.1\linewidth}
\begin{picture}(.001,0.001)
\put(0,6){\rotatebox{90}{$\Dv$}}
\put(5,1.85){{$\zeta$}}
\put(4.2,0.5){{moy($\zeta$)}}
\end{picture}\\
\vspace{-0.4cm}
\includegraphics[width=1\linewidth]{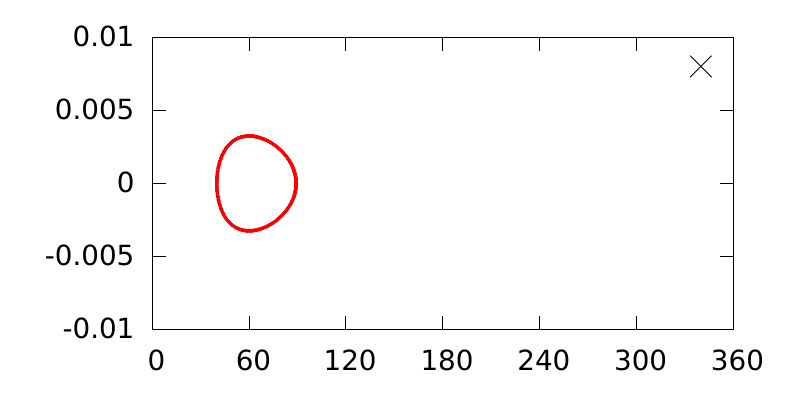}\\
  \setlength{\unitlength}{0.1\linewidth}
\begin{picture}(.001,0.001)
\put(0,1.6){\rotatebox{90}{$a_1-a_2$}}
\put(5.6,-0.3){{$\zeta$}}
\end{picture}\\
\vspace{-0.4cm}
\includegraphics[width=1\linewidth]{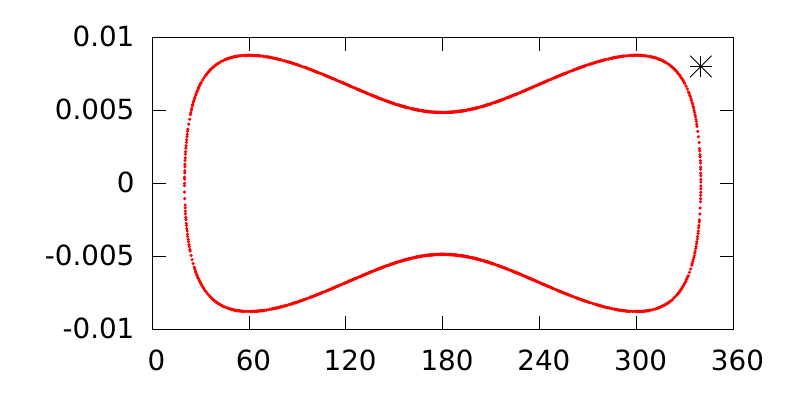}\\
  \setlength{\unitlength}{0.1\linewidth}
\begin{picture}(.001,0.001)
\put(0,1.6){\rotatebox{90}{$a_1-a_2$}}
\put(5.6,-0.3){{$\zeta$}}
\end{picture}\\
\vspace{-0.4cm}
\includegraphics[width=1\linewidth]{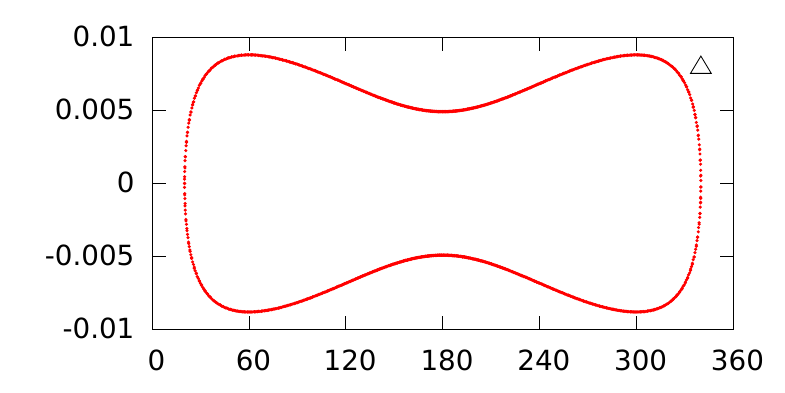}\\
  \setlength{\unitlength}{0.1\linewidth}
\begin{picture}(.001,0.001)
\put(0,1.6){\rotatebox{90}{$a_1-a_2$}}
\put(5.6,-0.3){{$\zeta$}}
\end{picture}\\
\end{minipage}
%
%
%
%
\begin{minipage}{0.49\linewidth}
\includegraphics[width=1\linewidth]{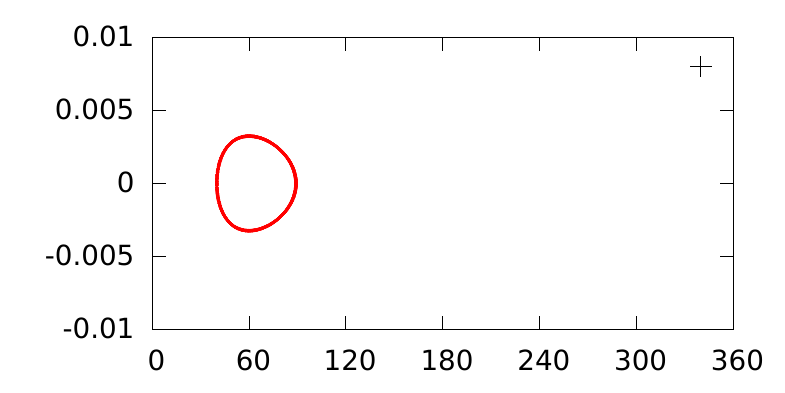}\\
  \setlength{\unitlength}{0.1\linewidth}
\begin{picture}(.001,0.001)
\put(0,2.4){\rotatebox{90}{$a_1-a_2$}}
\put(5.6,0.3){{$\zeta$}}
\end{picture}\\
\vspace{-0.4cm}
\includegraphics[width=1\linewidth]{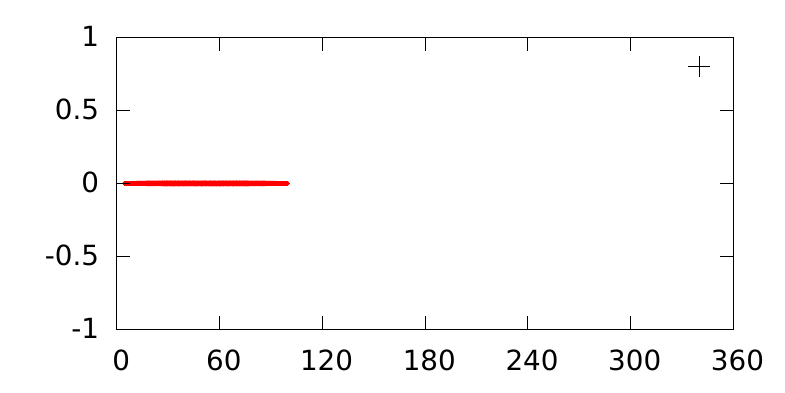}\\
  \setlength{\unitlength}{0.1\linewidth}
\begin{picture}(.001,0.001)
\put(0,1.6){\rotatebox{90}{$e_1-e_2$}}
\put(5.1,-0.3){{$\Dv$}}
\end{picture}\\
\vspace{-0.4cm}
\includegraphics[width=1\linewidth]{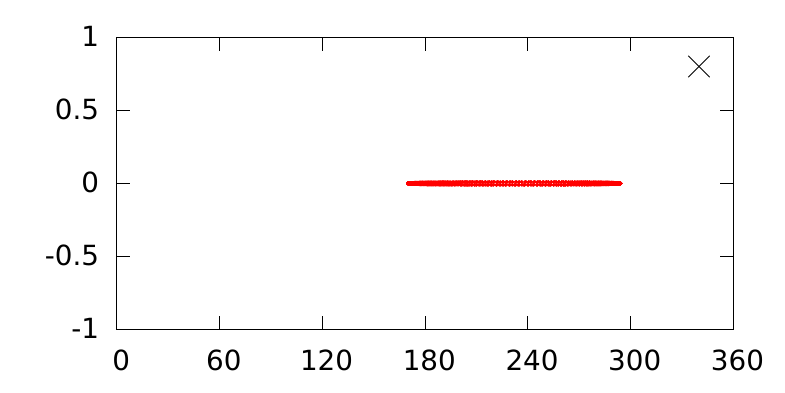}\\
  \setlength{\unitlength}{0.1\linewidth}
\begin{picture}(.001,0.001)
\put(0,1.6){\rotatebox{90}{$e_1-e_2$}}
\put(5.1,-0.3){{$\Dv$}}
\end{picture}\\
\vspace{-0.4cm}
\includegraphics[width=1\linewidth]{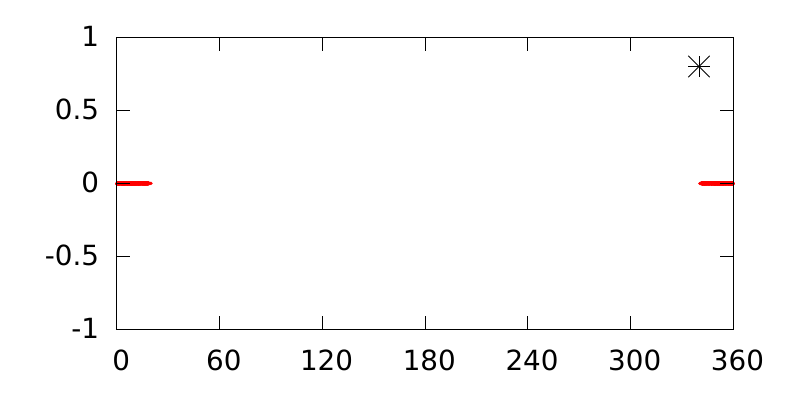}\\
  \setlength{\unitlength}{0.1\linewidth}
\begin{picture}(.001,0.001)
\put(0,1.6){\rotatebox{90}{$e_1-e_2$}}
\put(5.1,-0.3){{$\Dv$}}
\end{picture}\\
\vspace{-0.4cm}
\includegraphics[width=1\linewidth]{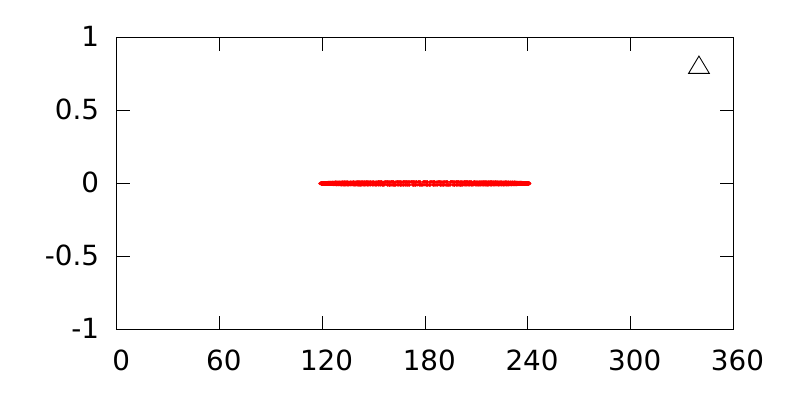}\\
  \setlength{\unitlength}{0.1\linewidth}
\begin{picture}(.001,0.001)
\put(0,1.6){\rotatebox{90}{$e_1-e_2$}}
\put(5.1,-0.3){{$\Dv$}}
\end{picture}\\
 \end{minipage}
 \caption{\label{fig:orbe01} Projections of generic trajectories emanating from the reference manifold for $e_1=e_2=0.01$, $\eps=m_1/m_0=m_2/m_0=10^{-5}$. The plotted orbital elements are the osculating ones (non-averaged, see \ref{sec:ins}). The trajectories were integrated over $5/\eps$ years.These trajectories pass near 4 distinct points of $\cV$ which are represented by symbols in the left-top graph (identical to the left-top graph of figure~\ref{fig:glob_e01}). 
  }
\end{figure}

  \begin{figure}[h!]
 \begin{minipage}{0.49\linewidth}
\includegraphics[width=1\linewidth]{epsfigs/sim20160517b_mesoDlam2.pdf}\\
  \setlength{\unitlength}{0.1\linewidth}
\begin{picture}(.001,0.001)
\put(0,6){\rotatebox{90}{$\Dv$}}
\put(5,1.85){{$\zeta$}}
\put(4.2,0.5){{moy($\zeta$)}}
\end{picture}\\
\vspace{-0.4cm}
\includegraphics[width=1\linewidth]{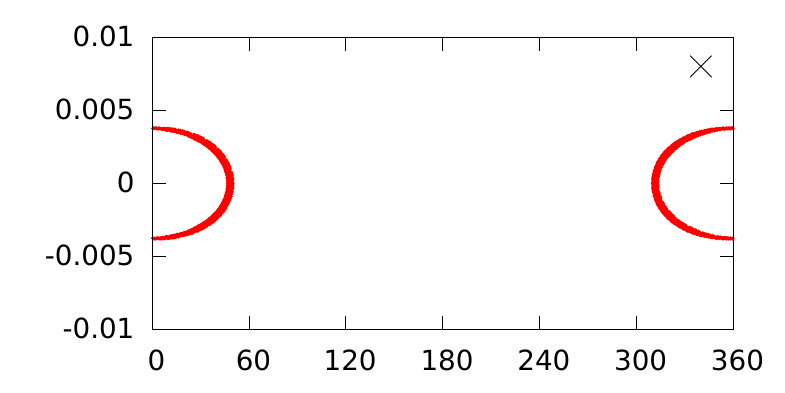}\\
  \setlength{\unitlength}{0.1\linewidth}
\begin{picture}(.001,0.001)
\put(0,1.6){\rotatebox{90}{$a_1-a_2$}}
\put(5.6,-0.3){{$\zeta$}}
\end{picture}\\
\vspace{-0.4cm}
\includegraphics[width=1\linewidth]{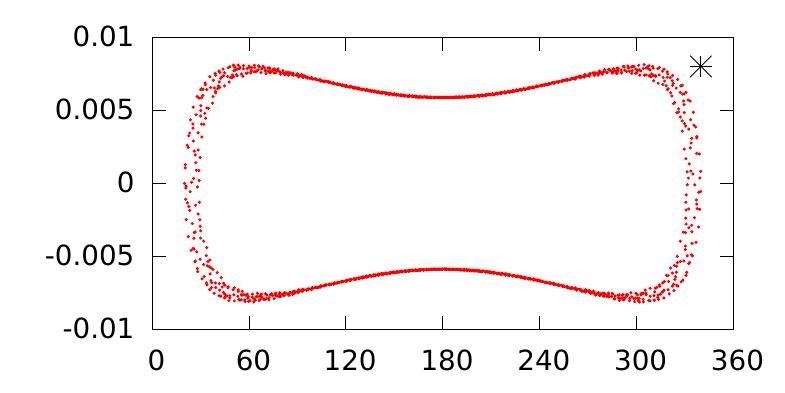}\\
  \setlength{\unitlength}{0.1\linewidth}
\begin{picture}(.001,0.001)
\put(0,1.6){\rotatebox{90}{$a_1-a_2$}}
\put(5.6,-0.3){{$\zeta$}}
\end{picture}\\
\vspace{-0.4cm}
\includegraphics[width=1\linewidth]{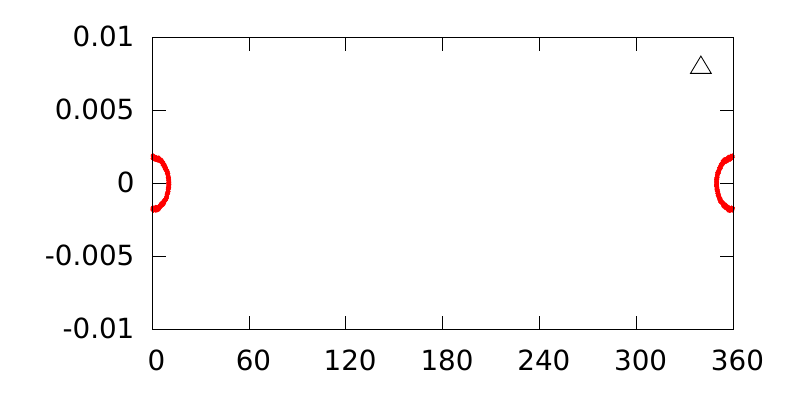}\\
  \setlength{\unitlength}{0.1\linewidth}
\begin{picture}(.001,0.001)
\put(0,1.6){\rotatebox{90}{$a_1-a_2$}}
\put(5.6,-0.3){{$\zeta$}}
\end{picture}\\
\end{minipage}
%
%
%
%
\begin{minipage}{0.49\linewidth}
\includegraphics[width=1\linewidth]{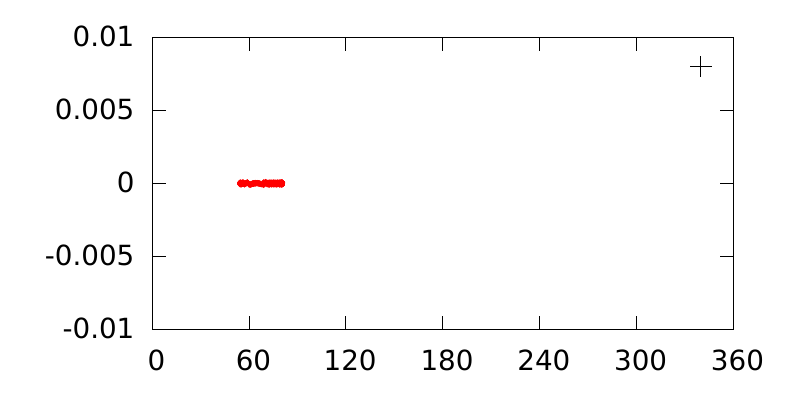}\\
  \setlength{\unitlength}{0.1\linewidth}
\begin{picture}(.001,0.001)
\put(0,2.4){\rotatebox{90}{$a_1-a_2$}}
\put(5.6,0.3){{$\zeta$}}
\end{picture}\\
\vspace{-0.4cm}
\includegraphics[width=1\linewidth]{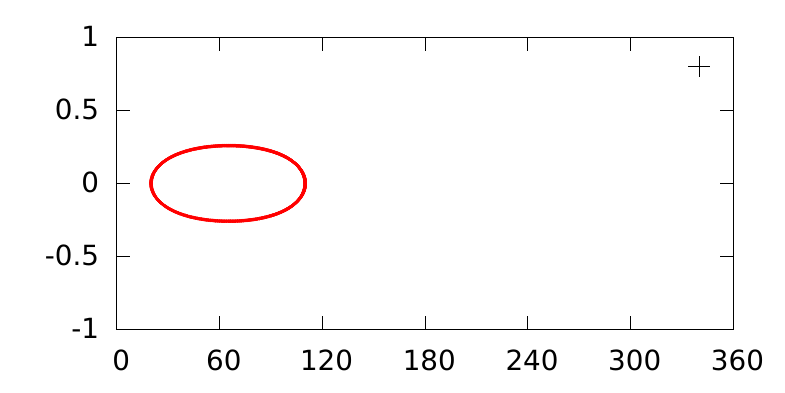}\\
  \setlength{\unitlength}{0.1\linewidth}
\begin{picture}(.001,0.001)
\put(0,1.6){\rotatebox{90}{$e_1-e_2$}}
\put(5.1,-0.3){{$\Dv$}}
\end{picture}\\
\vspace{-0.4cm}
\includegraphics[width=1\linewidth]{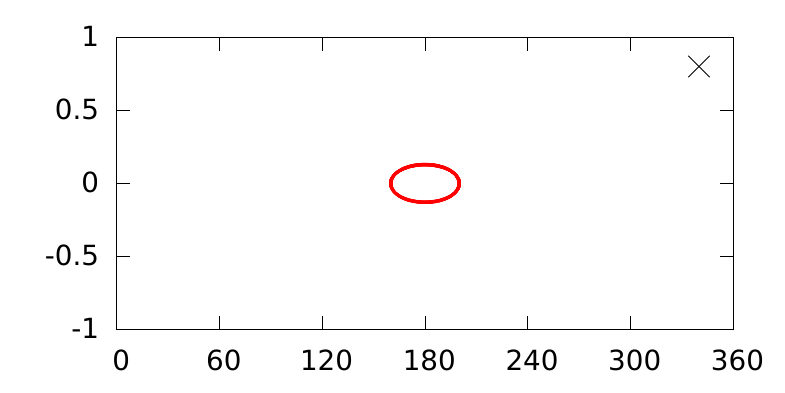}\\
  \setlength{\unitlength}{0.1\linewidth}
\begin{picture}(.001,0.001)
\put(0,1.6){\rotatebox{90}{$e_1-e_2$}}
\put(5.1,-0.3){{$\Dv$}}
\end{picture}\\
\vspace{-0.4cm}
\includegraphics[width=1\linewidth]{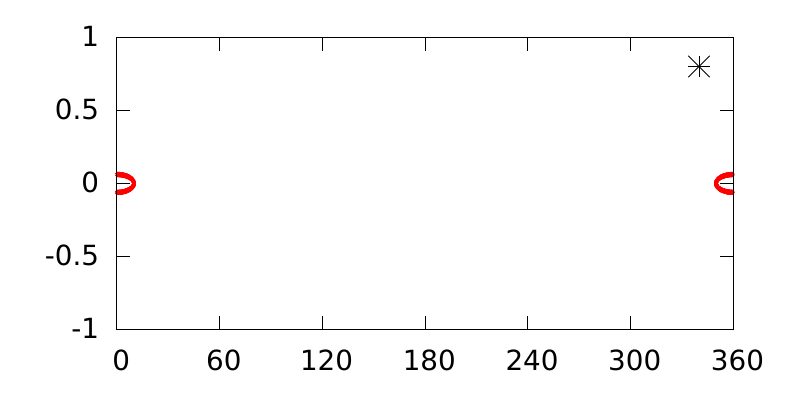}\\
  \setlength{\unitlength}{0.1\linewidth}
\begin{picture}(.001,0.001)
\put(0,1.6){\rotatebox{90}{$e_1-e_2$}}
\put(5.1,-0.3){{$\Dv$}}
\end{picture}\\
\vspace{-0.4cm}
\includegraphics[width=1\linewidth]{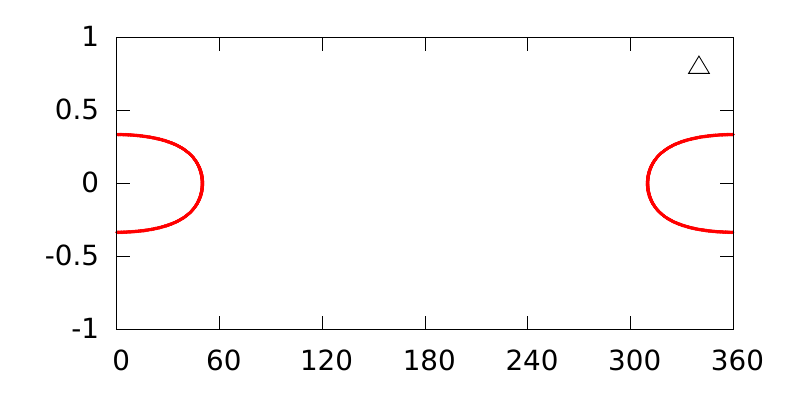}\\
  \setlength{\unitlength}{0.1\linewidth}
\begin{picture}(.001,0.001)
\put(0,1.6){\rotatebox{90}{$e_1-e_2$}}
\put(5.1,-0.3){{$\Dv$}}
\end{picture}\\
 \end{minipage}
 \caption{
\label{fig:orbe4} Projections of generic trajectories emanating from the reference manifold for $e_1=e_2=0.4$, $\eps=m_1/m_0=m_2/m_0=10^{-5}$. The plotted orbital elements are the osculating ones (non-averaged, see \ref{sec:ins}). The trajectories were integrated over $5/\eps$ years. These trajectories pass near 4 distinct points of $\cV$ which are represented by symbols in the left-top graph (identical to the left-bottom graph of figure~\ref{fig:glob_e01}). 
 }
\end{figure}

In the quasi-circular case (figure~\ref{fig:glob_e01}, top), the dynamics of the degree of freedom ($Z,\zeta$) (left graph) is very close to the circular case (equation (\ref{eq:eqerdi}) is relevant at the order one in the eccentricities): we still have a tadpole and a horseshoe domain, with the separatrix located in $\zeta \approx 24^\circ$ and $\approx 336^\circ$, and the initial value of $\Dv$ does not impact much the average value of $\zeta$ on the orbit (left graph). The positions of the families $\Fsc$ and $\Fsf$ are represented on the right graph. Note that the families emanating from the $L_3$ circular equilibrium cannot be identified by the criterion that we developed in appendix \ref{sec:SAIF} because they are in an unstable area (near the separatrix emanating from $L_3$). In addition to the families that we defined in the neighbourhood of a circular equilibrium, there are branches of the reunion $\Fsf$ in the horseshoe domain. We name $\cF^{s\!f}_{\gH \cS}$ the family located at $\Dv=0^\circ$, around which librate the orbits of the green area (right graph). Note that another family is located at $\Dv=180^\circ$ around which librate the horseshoe orbits of the blue area. Both the trojan and horseshoe domains are hence split in two parts: for the trojan orbit, $\Dv$ oscillates either around the branch of $\Fsf$ emanating from $L_k$, or the one emanating from ${AL_k}$. In the horseshoe domain, it oscillates either near $\Dv=0^\circ$ or $\Dv=180^\circ$. Note that this ``split" results from our choice of variables: there are no separatrix between these domains. As we move from $L_4$, the minimum eccentricity that is reached on a given orbit decreases. Eventually, this minimal eccentricity reaches $0$ before it increases again for orbits librating around $AL_4$, hence the discontinuity in the value of $\Dv$ between $L_4$ and $AL_4$. 

The markers on the left graph indicate the points of $\cV$ near which pass the $4$ orbits whose projection on the ($Z,\zeta$) and ($e_1-e_2,\Dv$) plane is represented in figure~\ref{fig:orbe01}. We show orbits in the neighbourhood of $L_4$, $AL_4$, and the two types of horseshoe orbits. Each of these generic orbits passes near $4$ different points of $\cV$, and these points can be divided in 2 pairs which have the same value of $\Dv$. Note that the 4 points representing a given orbit are always positioned in a different quadrant (quadrants that are delimited by the $\Fsc$ and $\Fsf$ families).

\subsubsection{Moderate eccentricities}

We now increase the total angular momentum of the system, assuming $e_1=e_2= 0.4$. The results are displayed in the botom graphs of the figure~\ref{fig:glob_e01}. As we move away from the circular case, the phase space evolves. The quasi-satellite domains appears \citep{Namouni1999,GiuBeMiFe2010,PoRoVi2017}, centred on a fixed point of the averaged reduced problem located at $\zeta=0^\circ$, $\Dv=180^\circ$, which is also the intersection of the families $\Fsf$ and $\Fsc$. We can observe on the bottom right-hand graph of figure~\ref{fig:glob_e01} that the quasi-satellite domain is also splitted in two kinds of quasi-satellites: those for which $\Dv$ librates around $180^\circ$ and those for which it librates around $0^\circ$. As it is the case between the orbit librating around $L_4$ and $AL_4$ (see previous section), the discontinuity between the two domains is due to a non-definition of $\Dv$ when one the eccentricity reaches $0$. The orbits located at the border between these two kind of quasi-satellites are discussed in \cite{Nauenberg2002}. \\

 The dynamics in the Trojan and horseshoe domains remain similar to the quasi-circular case, but the domain where the horseshoe orbits librate around $\Dv=180^\circ$ shrinks on this plane\footnote{When $m_1=m_2$ we suppose that the reference manifold represents all the co-orbital configurations reaching $a_1=a_2$ on their orbit, for a given value of the total angular momentum. However, the relative size of the section of two stability domains by the reference manifold is not necessarily representative of the relative volume of these two configurations in the phase space. For example figure~\ref{fig:stabzr} shows that depending on the chosen section, the horseshoe domain may appear larger or smaller than the tadpole one (for $\mu \approx 10^{-6}$).}. This is due to the increase of the unstable area near the $\cF^{sc}_3$ family and the position of the collision manifold. Indeed, the collision manifold, as well as all the $\cF$ branches, is twisted as the total angular momentum increases (see figure \ref{fig:Fb0meq}). On this plane of initial conditions, this leads to the reduction of the stability domain for trojan and horseshoe configurations and the increase of the stability domain for quasi-satellites.


\subsubsection{Emergence of the asymmetric horseshoe orbits}

\begin{figure}[h!]
\begin{center}
\includegraphics[width=0.7\linewidth]{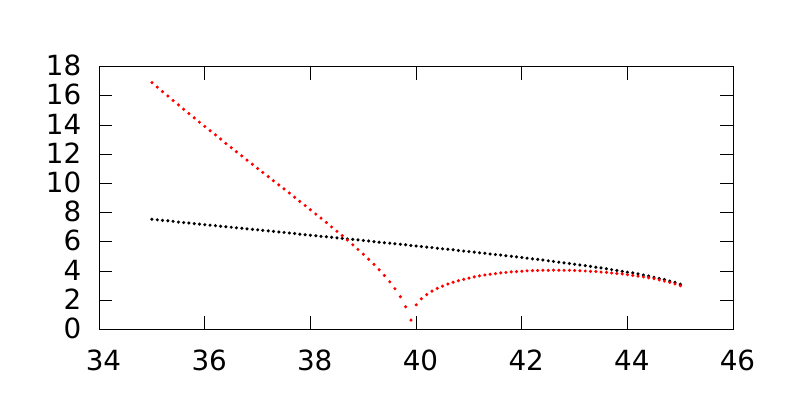}\\
 \setlength{\unitlength}{0.067\linewidth}
\begin{picture}(.001,0.001)
\put(-5.1,3){\rotatebox{90}{$g/\eps$}}
\put(0,0.2){{$\zeta$}}
\end{picture}\\
\includegraphics[width=0.7\linewidth]{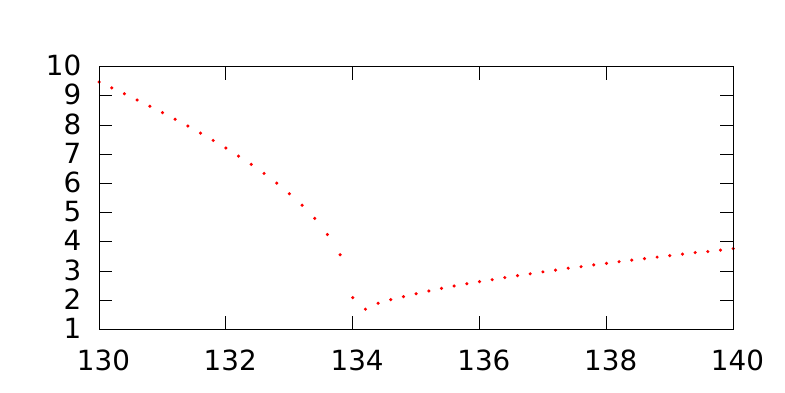}\\
 \setlength{\unitlength}{0.067\linewidth}
\begin{picture}(.001,0.001)
\put(-5.1,3){\rotatebox{90}{$g/\eps$}}
\put(0,0.2){{$\zeta$}}
\end{picture}
\caption{\label{fig:freqge5} Top: evolution of the normalised frequencies $g/(n\eps)$ in red and $\nu/(n\sqrt \eps)$ in black for $\Dv=1^\circ$ and $e_1=e_2=0.55$. Bottom: evolution of the normalised frequency $g/(n\eps)$ along the $\cF^{sc}_{4}$ family for $e_1=e_2=0.7$ (figure~\ref{fig:glob_e6}). These curves are obtained by numerical integration of the 3-body problem and a frequency analysis of the angle $\zeta$ (resp. $\Dv$) to obtain $\nu$ (resp. $g$).
}
\end{center}
\end{figure}
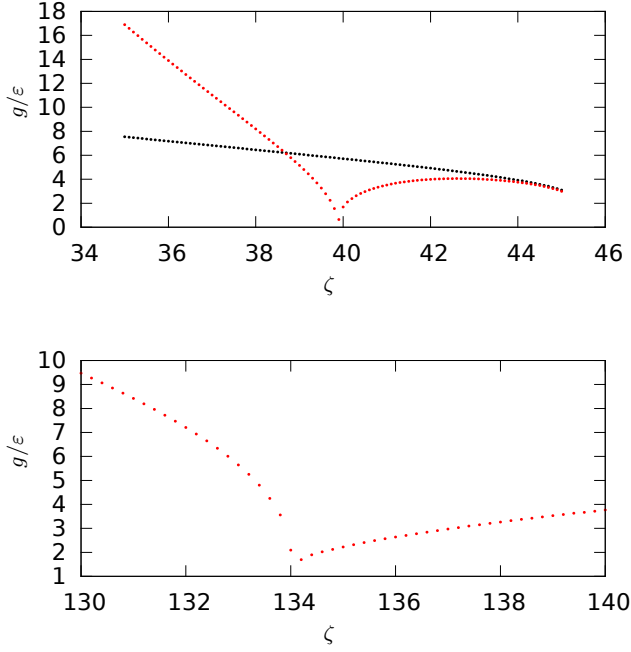

\begin{figure}[h!]
\begin{center}
\includegraphics[width=0.5\linewidth]{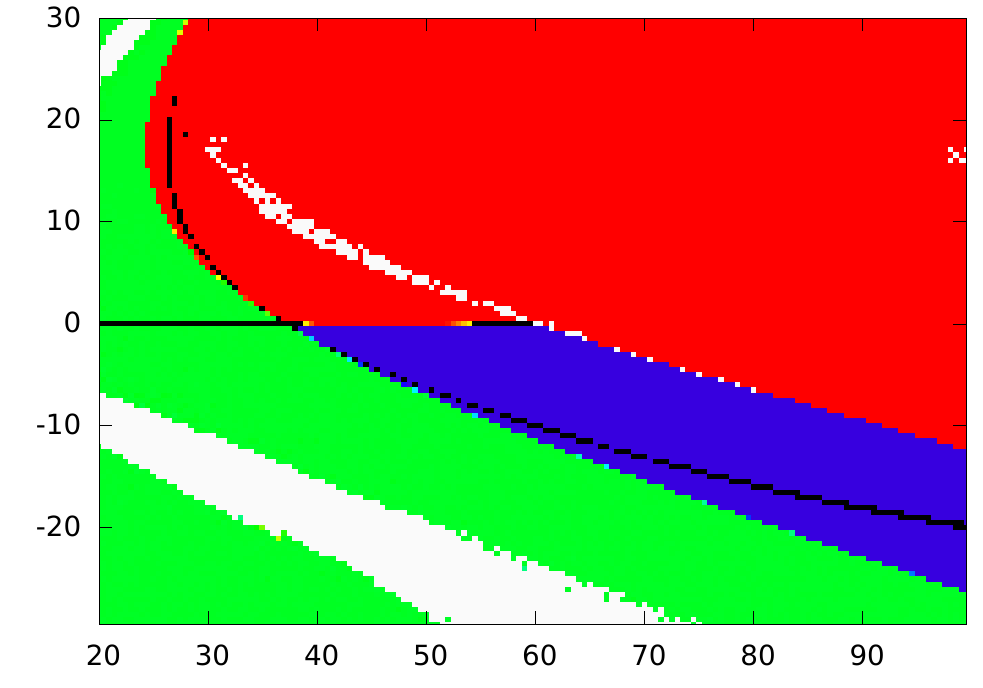}\\
  \setlength{\unitlength}{0.067\linewidth}
\begin{picture}(.001,0.001)
\put(-4,3){\rotatebox{90}{$\Dv$}}
\put(-2.6,2.8){$\cF^{sf}_{\gH \cS, elli.}$}
\put(-2.35,4.2){$\cF^{sf}_{\gH \cS, elli.}$}
\put(0,2.5){$\mathbin{\color{white}\cF^{sf}_{\gH \cS, elli.}}$}
\put(-0.8,3.3){$\cF^{sf}_{\gH \cS, hyp.}$}
\put(0,0.2){{$\zeta$}}
\end{picture}
\caption{\label{fig:zoomAHS} Zoom on the horseshoe area of the reference manifold for $\eps=10^{-6}$ and $e_1=e_2=0.6$, $m_1=m_2$. Each point of the grid is an initial condition. The color code gives an indication of the mean value of $\Dv$ for the trajectory emanating from each initial condition: green when the mean value is $0^\circ$, red when positive, and blue when negative. Note that the position of the separatrix between the horseshoe and trojan domain can be identified by the transition from blue to red on the right-hand side of the horseshoe domain, then by the unstable (white) orbits in its neighbourhood (its position for higher values of $\Dv$ is not visible due to the choice of the color code). The trajectories close to a branch of $\Fsf$, i.e. those verifying equation (\ref{eq:condFb1}) with $\epsilon_g= 3^\circ$, are shown with black pixels. The elliptic branch of the $\cF^{sf}_{\gH \cS}$ prior to the bifurcation ($\zeta\lessapprox 40^\circ,\, \Dv=0^\circ$), and the elliptic branches after the bifurcation (in the red and blue areas) are labelled by $\cF^{sf}_{\gH \cS, elli.}$. The hyperbolic branch ($\zeta\gtrapprox 40^\circ,\, \Dv=0^\circ$) is labelled by $\cF^{sf}_{\gH \cS, hyp.}$.
}
\end{center}
\end{figure}
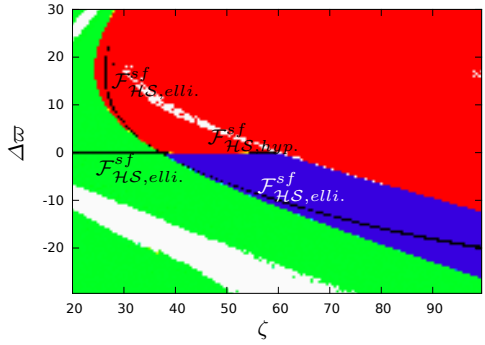

We recall that the horseshoe domain is located between the manifold defined by $\nu=0$ (separatrix emanating from the unstable family $\cF^{sc}_3$) and the unstable area around the collision manifold. For $e_1=e_2 \lesssim 0.5$, it is made only of `symmetric' orbits: as shown by the examples in figures \ref{fig:orbe01} and \ref{fig:orbe4}, these orbits are symmetric with respect to $\zeta=180^\circ$.\\ 

However, for $e_1=e_2  \gtrsim 0.5$, the first notable modification of the phase space appears: the previously elliptic (or normally stable) family of periodic orbits $\cF^{s\!f}_{\gH \cS}$ bifurcates into two elliptic families of periodic orbits (one with $\Dv >0^\circ$ and another with $\Dv <0^\circ$), and one hyperbolic (or normally unstable) family of periodic orbits located at $\Dv =0^\circ$. This bifurcation is due to the encounter of the $\cF^{sf}_{\gH \cS}$ family with the $g=0$ manifold: for $e_1=e_2 \lesssim 0.5$, $|g|$ was monotonously decreasing along $\Dv=0^\circ$ as $\zeta$ increases, but never reaching $0$. However, as $e_1=e_2$ increases, the border of the horseshoe domain ($\{\nu=0\}\cap \cV$) shifts towards larger values of $\zeta$. For $e_1=e_2 \gtrsim 0.5$ the frequency $g$ reaches zero before the separatrix $\nu=0$ is reached\footnote{Interestingly, while the position of $\{\nu=0\}\cap \cV$ for $\Dv=0$ depends strongly of the value of $e_1=e_2$, the position of $\{g=0\}\cap \cV$ for $\Dv=0$ seems to occur around $\zeta=40^\circ$ for any value of $e_1=e_2 \gtrsim 0.5$, see figures \ref{fig:freqge5}, \ref{fig:zoomAHS}, \ref{fig:glob_e6} and \ref{fig:glob_e7_m6}.} (see figure \ref{fig:freqge5}).\\



The effect of this bifurcation is represented in figure \ref{fig:zoomAHS}, which is a section of the reference manifold with $e_1=e_2=0.6$. The green area centred on $\Dv=0^\circ$ is the symmetric horseshoe domain we had for lower eccentricities, and the horizontal black line in the middle of this domain is the family $\cF^{s\!f}_{\gH \cS}$. On the right-hand side of the bifurcation (occurring at $\Dv=0,\zeta\approx40^\circ$), the stable branches of the $\Fsf$ family are identified by black pixels\footnote{Orbits in the close neighbourhood of the unstable family can also verify the condition (\ref{eq:condFb1}) when integrated over a duration of the order of $1/\eps$ because $g$ tends to $0$ for this family.}. The orbits librating around $\Dv >0^\circ$ are represented in red and those librating around $\Dv<0^\circ$ are represented in blue (the tadpole orbits beyond the separatrix $\nu=0$ are also represented in red). In these two domains, the projection of a given orbit in the ($Z,\zeta$) plane is not symmetric with respect to $\zeta=180^\circ$, see figure \ref{fig:orbe7}. We thus call these domains asymmetric horseshoe. The red/green interface and the blue/green interface mark the separatrix $g=0$, while the position of the hyperbolic family can be identified by the transition from red to blue.
  \\

\subsubsection{Reconnection of the $\Fsc$ and $\Fsf$ families}

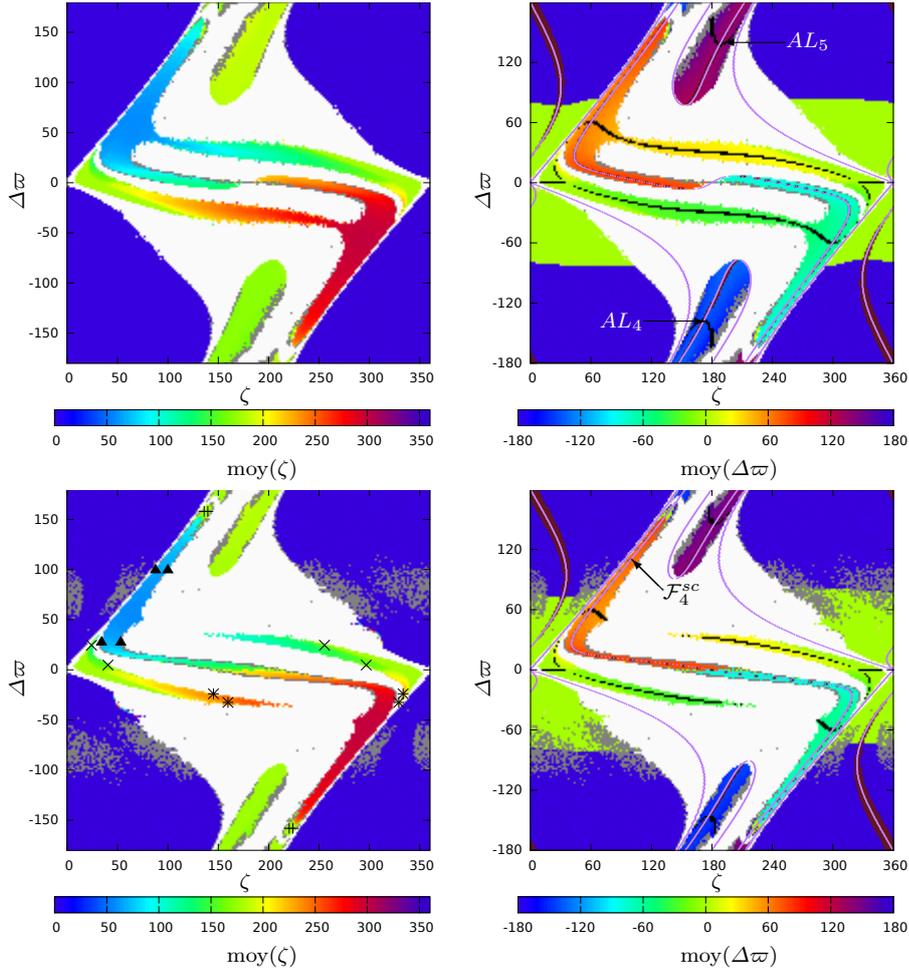
\begin{figure}[h!]
\begin{center}
\includegraphics[width=0.5\linewidth]{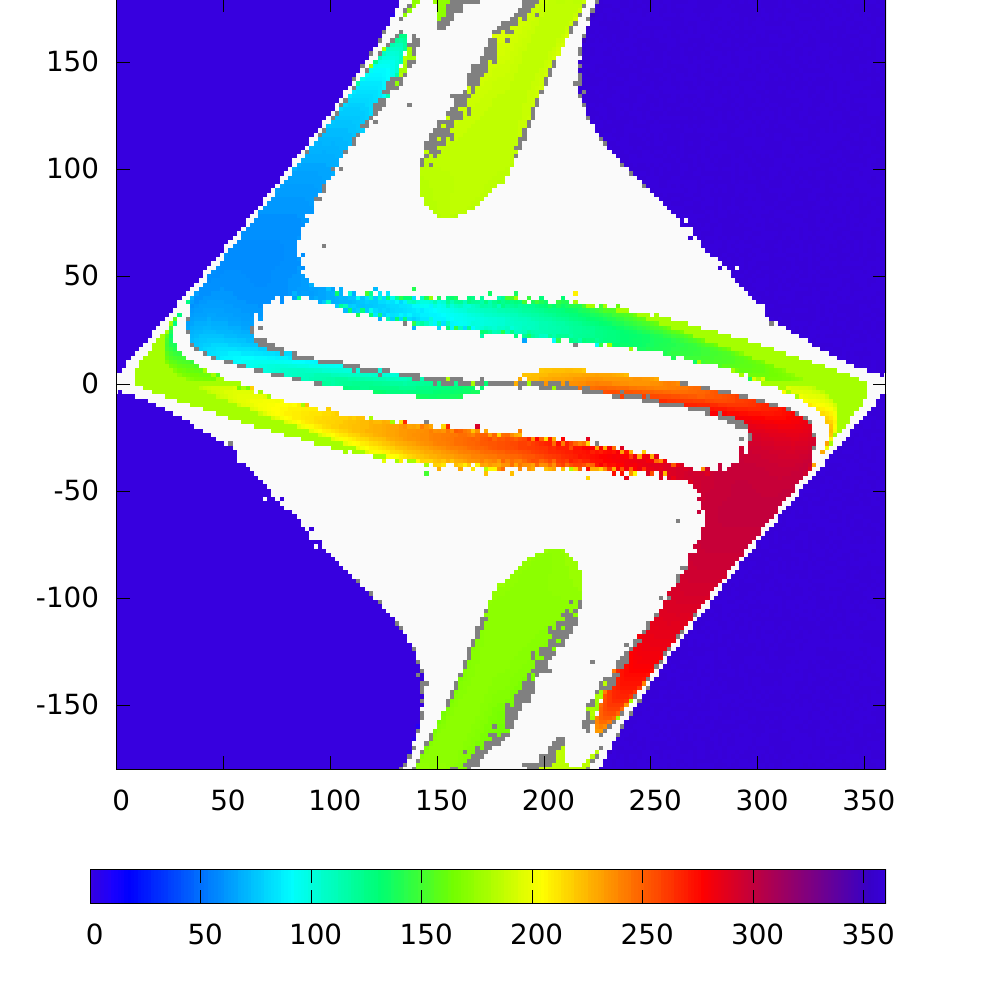}\includegraphics[width=0.5\linewidth]{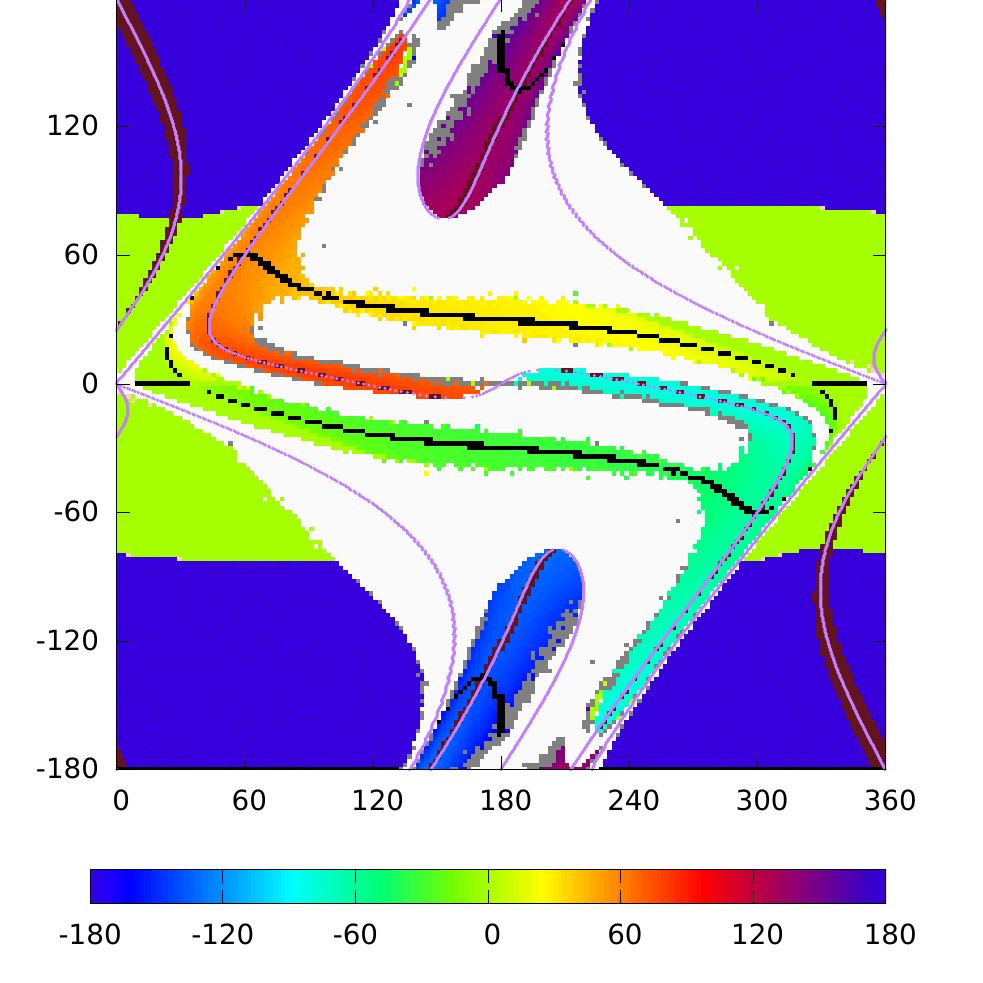}\\
  \setlength{\unitlength}{0.067\linewidth}
\begin{picture}(.001,0.001)
\put(-7.5,4.5){\rotatebox{90}{$\Dv$}}
\put(0,4.5){\rotatebox{90}{$\Dv$}}
\put(-3.8,1.4){{$\zeta$}}
\put(3.8,1.4){{$\zeta$}}
\put(-4,0.2){{moy($\zeta$)}}
\put(3.3,0.2){{moy($\Dv$)}}
\put(2.7,2.7){\vector(1,0){1}}
\put(5,7.22){\vector(-1,0){1}}
\put(2,2.6){$\mathbin{\color{white}AL_4}$}
\put(5,7.15){$\mathbin{\color{white}AL_5}$}
\end{picture}\\
\includegraphics[width=0.5\linewidth]{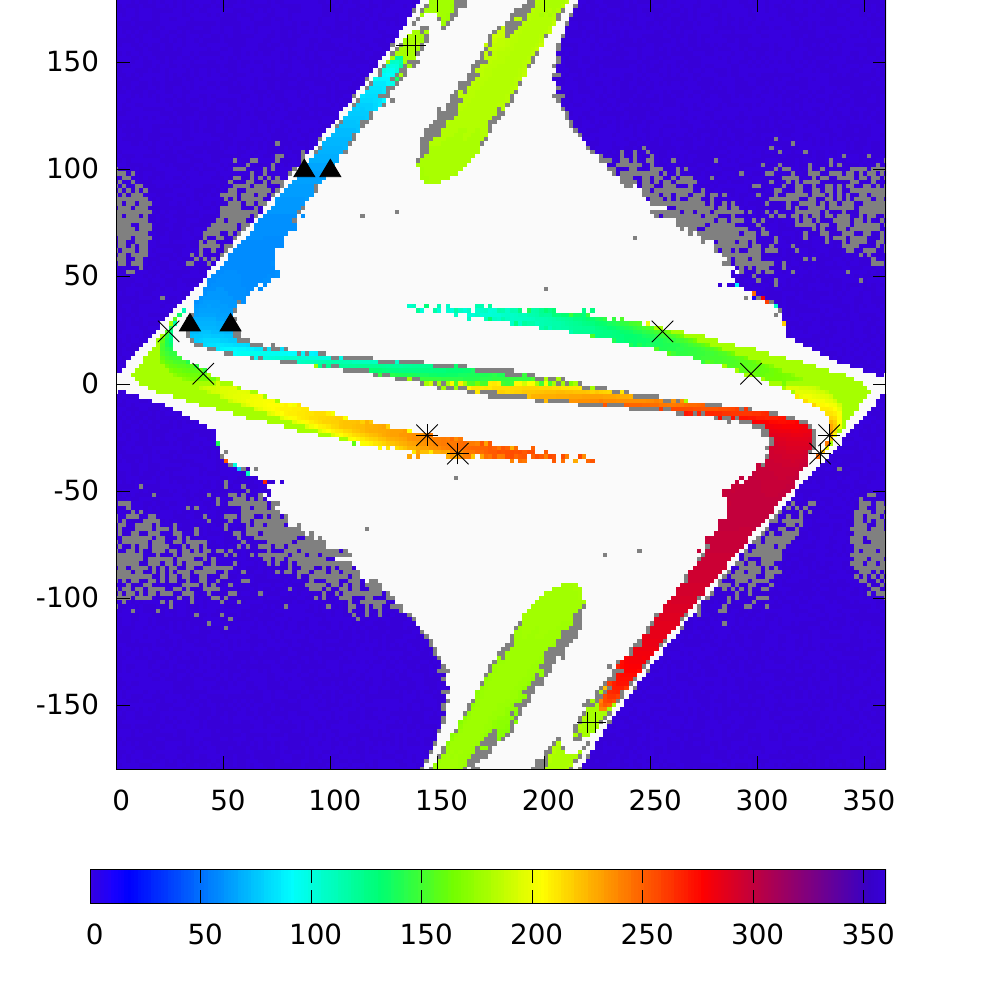}\includegraphics[width=0.5\linewidth]{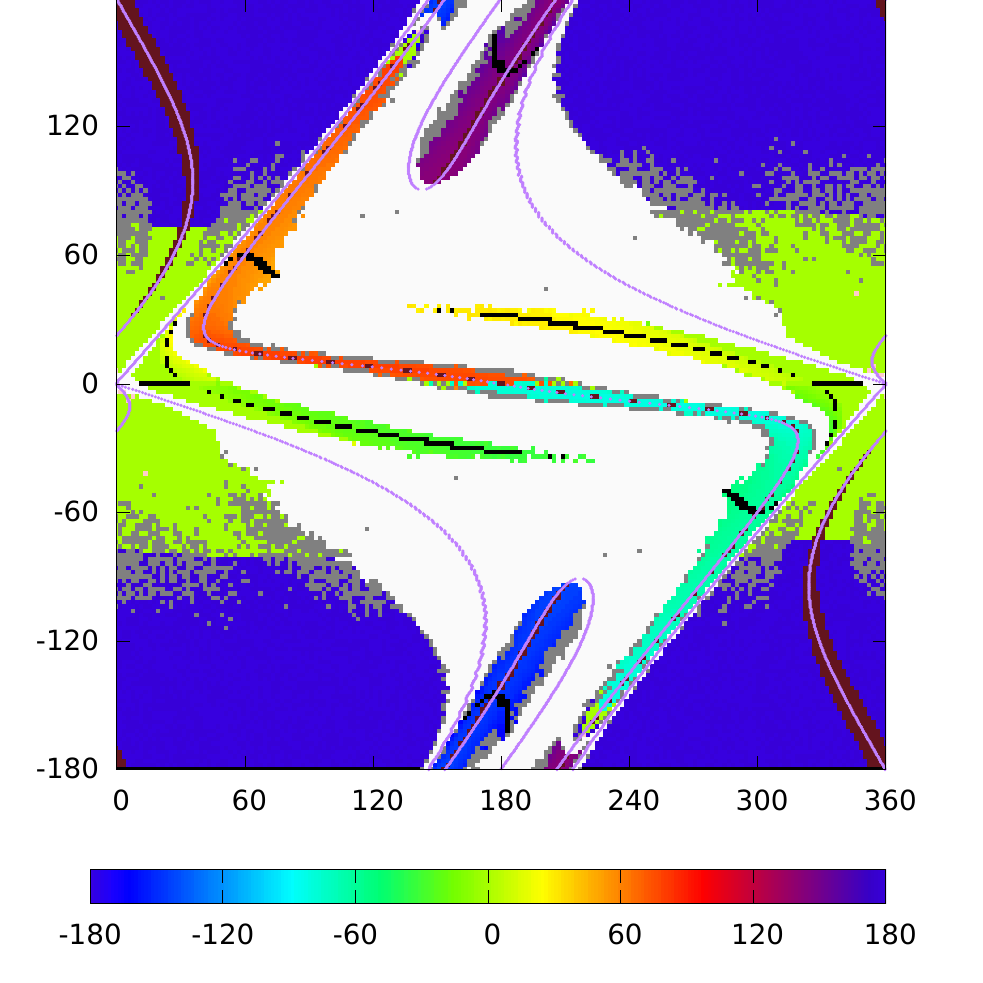}\\
  \setlength{\unitlength}{0.067\linewidth}
\begin{picture}(.001,0.001)
\put(-7.5,4.5){\rotatebox{90}{$\Dv$}}
\put(0,4.5){\rotatebox{90}{$\Dv$}}
\put(-3.8,1.4){{$\zeta$}}
\put(3.8,1.4){{$\zeta$}}
\put(-4,0.2){{moy($\zeta$)}}
\put(3.3,0.2){{moy($\Dv$)}}
\put(3.02,6.24){\vector(-1,1){.5}}
\put(3.02,6.1){$\cF^{sc}_4$}
\end{picture}
\caption{\label{fig:glob_e6} Grid of initial conditions for $m_1/m_0=m_2/m_0=10^{-5}$, $a_1=a_2=1$~au, and $e_1=e_2=0.65$ (top) and $e_1=e_2=0.7$ (bottom). The color code on the left hand graphs gives the mean value of $\zeta$ on the orbit emanating from each initial condition. The orbits in the neighbourhood of $\Fsc$, hence those verifying (\ref{eq:condFb02}) with $\epsilon_\nu=10^{-3.5}$, are represented by brown pixels. The purple curves show the result of the semi-analytical method (eq. \ref{eq:condFb0n}). The branch of the $\Fsc$ family along which the $g$ frequency was computed, figure \ref{fig:freqge5}, is labelled by $\cF^{sc}_4$. The orbits close to $\Fsf$, hence those verifying (\ref{eq:condFb1}) with $\epsilon_g= 3^\circ$, are represented by black pixels. The $AL_k$ equilibrium are fixed points of the averaged reduced problem and are hence located at the intersection of the $\Fsf$ and $\Fsc$ families. The initial conditions that lead to a diffusion of the mean semi-major axis over $\epsilon_a=10^{-5.5}$ are displayed in grey.}
\end{center}
\end{figure}

\begin{figure}[h!]
\begin{center}
\includegraphics[width=0.5\linewidth]{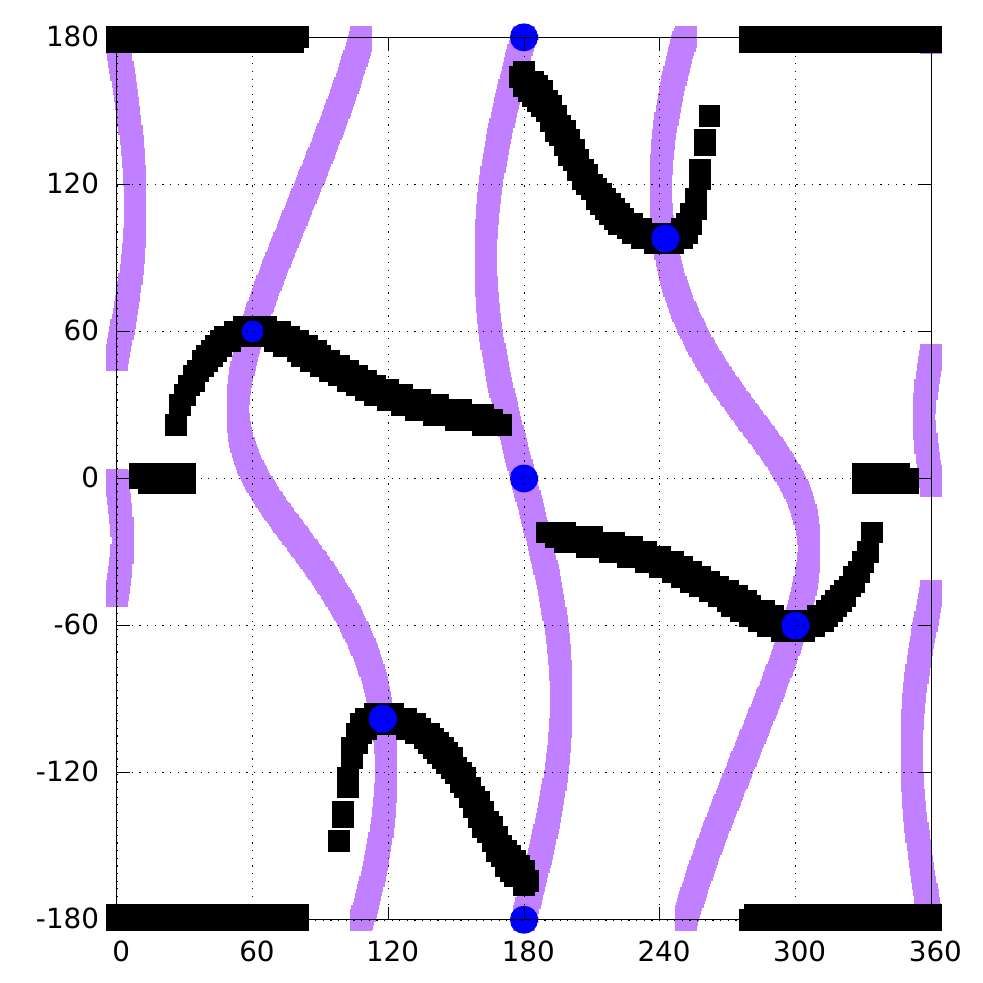}\includegraphics[width=0.5\linewidth]{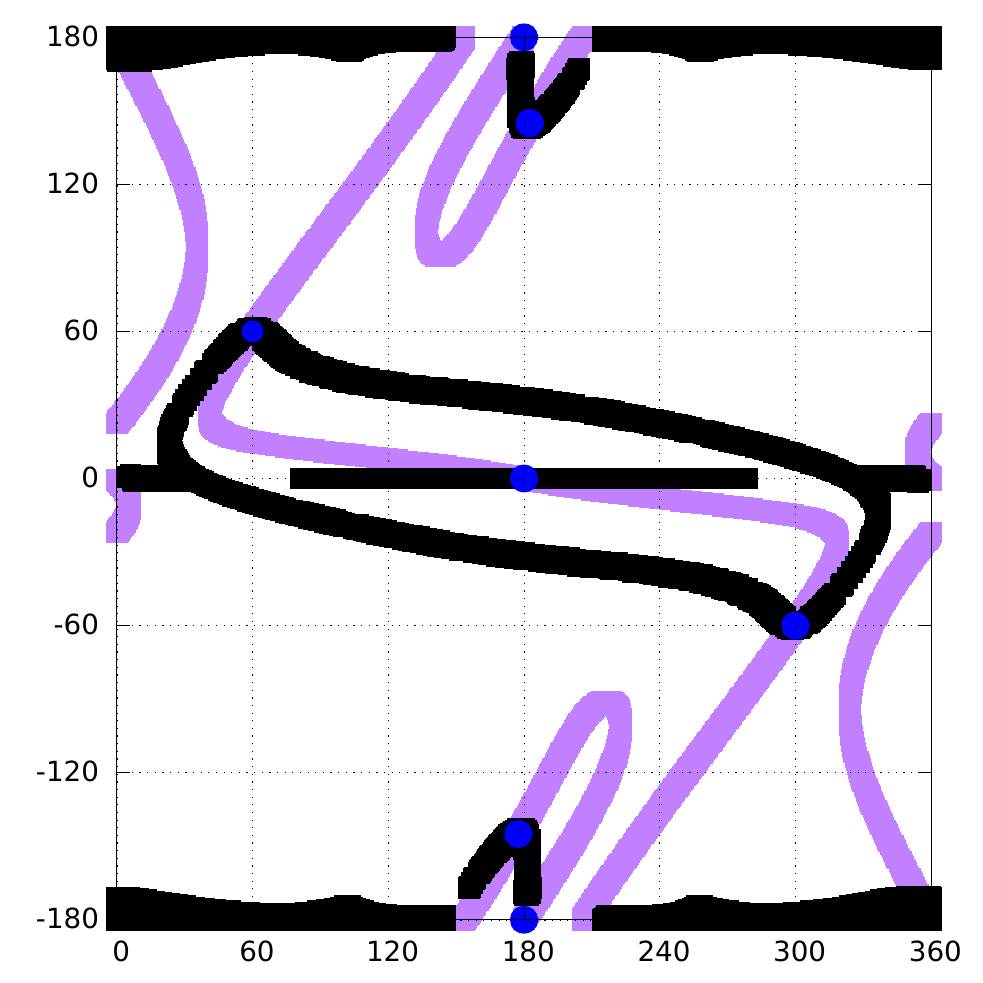}\\
  \setlength{\unitlength}{0.067\linewidth}
\begin{picture}(.001,0.001)
\put(-7.5,4){\rotatebox{90}{$\Dv$}}
\put(0,4){\rotatebox{90}{$\Dv$}}
\put(-3.6,0){{$\zeta$}}
\put(3.85,0){{$\zeta$}}
\put(-6.1,5.6){$L_4$}
\put(-1.3,2.6){$L_5$}
\put(-4.4,2.6){$AL_4$}
\put(-3.3,4.3){$AL_3$}
\put(-3.3,1){$L_3$}
\put(-2,6){$AL_5$}
\end{picture}\\
\caption{\label{fig:recon_schema} Schema of the position of the $\Fsf$ and $\Fsc$ families in the reference plane for $e_1=e_2=0.4$ (left) and $e_1=e_2=0.7$ (right). The position of the $\Fsc$, shown in purple, were computed using the method described in section \ref{sec:recon} (position independent of $\eps=m_1/m_0=m_2/m_0$). The position of the $\Fsf$, shown in black, was computed using the numerical criterion (\ref{eq:condFb1}) applied on orbits integrated in the full (non-averaged) 3-body problem and are hence shown only in the areas where the trajectories are stable at least for a duration comparable to $1/\eps$. The position of the $\Fsf$ families were computed with $\eps=10^{-5}$ for  $e_1=e_2=0.4$ (see also the bottom panels of figure \ref{fig:glob_e01}), and with $\eps=10^{-6}$ for  $e_1=e_2=0.7$ (see also figure \ref{fig:glob_e7_m6}). The blue dots show the position of the $L_k$ and $AL_k$. Since they are fixed points of the reduced averaged problem, they are located at the intersection of the $\Fsf$ and $\Fsc$ families. We recall that the reference plane is a 2-dimensional torus, and we have $\zeta\equiv \zeta+360^\circ$ and  $\Dv\equiv \Dv+360^\circ$. On the right-hand panel, we hence have a continuous branch of $\Fsc$ (purple) going through $L_4$-$AL_4$-$L_3$-$AL_5$-$L_5$-$AL_3$-$L_4$, while a continuous branch of $\Fsf$ (black) links directly $L_4$ to $L_5$. }
\end{center}
\end{figure}

\begin{figure}[h!]
 \begin{minipage}{0.49\linewidth}
\includegraphics[width=1\linewidth]{epsfigs/sim20160517f_mesoDomm2.pdf}\\
  \setlength{\unitlength}{0.1\linewidth}
\begin{picture}(.001,0.001)
\put(0,6){\rotatebox{90}{$\Dv$}}
\put(5,1.85){{$\zeta$}}
\put(4.2,0.5){{moy($\Dv$)}}
\put(2.5,7.6){\red{\circled{1}}}
\put(5.2,7.1){\red{\circled{2}}}
\put(7,6.8){\red{\circled{3}}}
\put(8,6.5){\red{\circled{4}}}
\put(8.3,6){\red{\circled{5}}}
\put(8,5.7){\red{\circled{6}}}
\put(1.9,7.6){\blue{\circled{1}}}
\put(1.6,7.1){\blue{\circled{2}}}
\put(1.4,6.8){\blue{\circled{3}}}
\put(1.6,6.5){\blue{\circled{4}}}
\put(3,6){\blue{\circled{5}}}
\put(6,5.7){\blue{\circled{6}}}
\end{picture}\\
\includegraphics[width=1\linewidth]{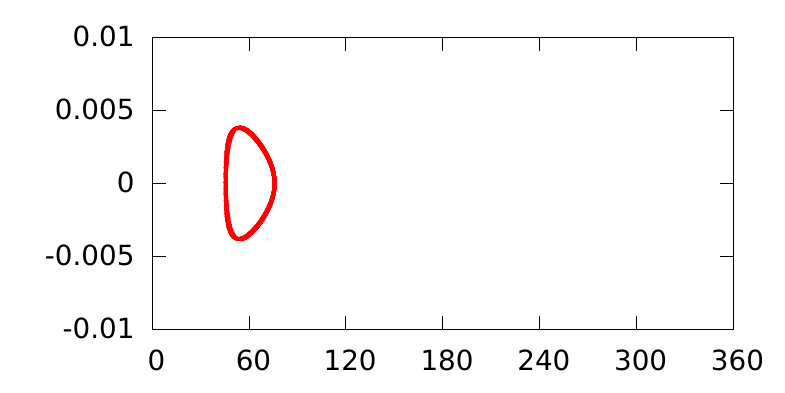}\\
  \setlength{\unitlength}{0.1\linewidth}
\begin{picture}(.001,0.001)
\put(0,2.5){\rotatebox{90}{$a_1-a_2$}}
\put(6,0.5){{$\zeta$}}
\put(8.5,4.5){\red{\circled{1}}}
\end{picture}\\
\includegraphics[width=1\linewidth]{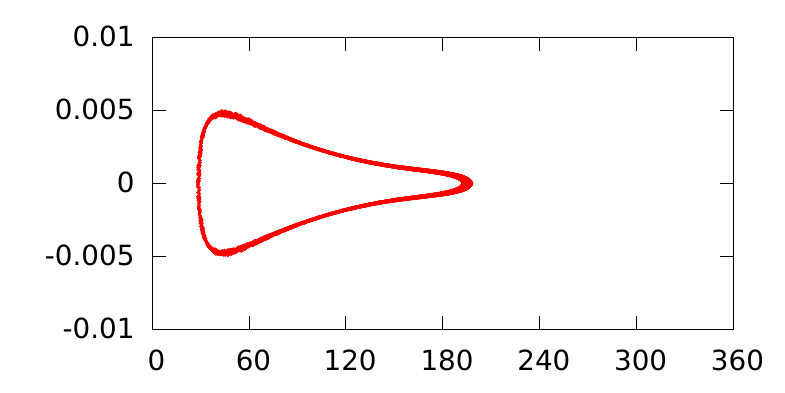}\\
  \setlength{\unitlength}{0.1\linewidth}
\begin{picture}(.001,0.001)
\put(0,2.5){\rotatebox{90}{$a_1-a_2$}}
\put(6,0.5){{$\zeta$}}
\put(8.5,4.5){\red{\circled{2}}}
\end{picture}\\
\end{minipage}
\begin{minipage}{0.49\linewidth}
\includegraphics[width=1\linewidth]{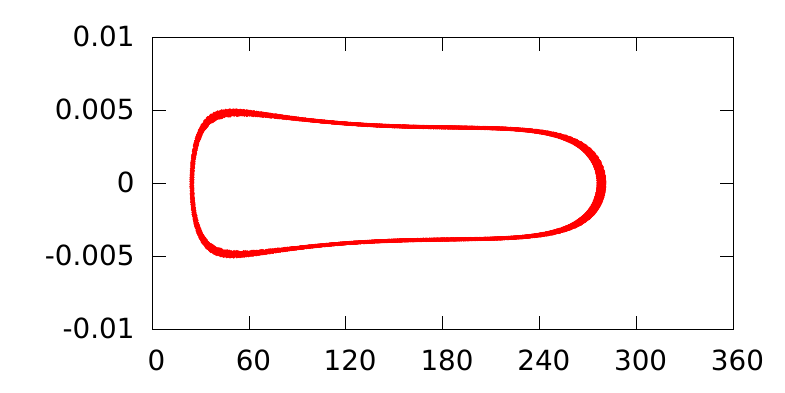}\\
  \setlength{\unitlength}{0.1\linewidth}
\begin{picture}(.001,0.001)
\put(0,2.5){\rotatebox{90}{$a_1-a_2$}}
\put(6,0.5){{$\zeta$}}
\put(8.5,4.5){\red{\circled{3}}}
\end{picture}\\
\includegraphics[width=1\linewidth]{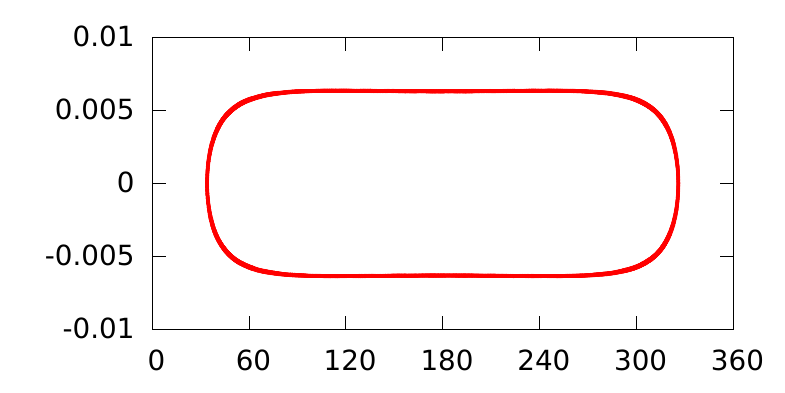}\\
  \setlength{\unitlength}{0.1\linewidth}
\begin{picture}(.001,0.001)
\put(0,2.5){\rotatebox{90}{$a_1-a_2$}}
\put(6,0.5){{$\zeta$}}
\put(8.5,4.5){\red{\circled{4}}}
\end{picture}\\
\includegraphics[width=1\linewidth]{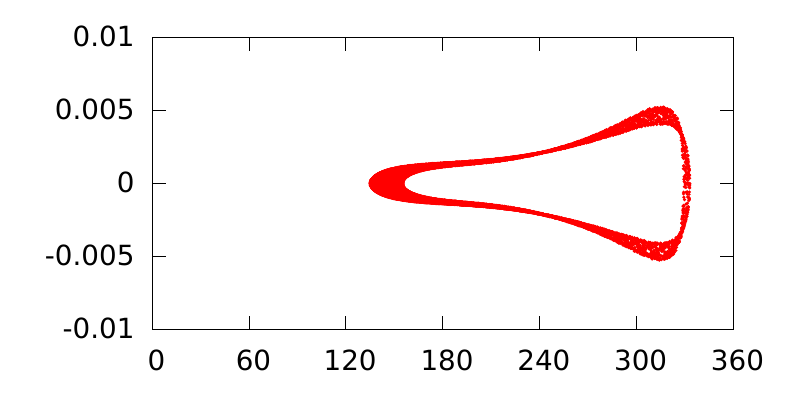}\\
  \setlength{\unitlength}{0.1\linewidth}
\begin{picture}(.001,0.001)
\put(0,2.5){\rotatebox{90}{$a_1-a_2$}}
\put(6,0.5){{$\zeta$}}
\put(8.5,4.5){\red{\circled{5}}}
\end{picture}\\
\includegraphics[width=1\linewidth]{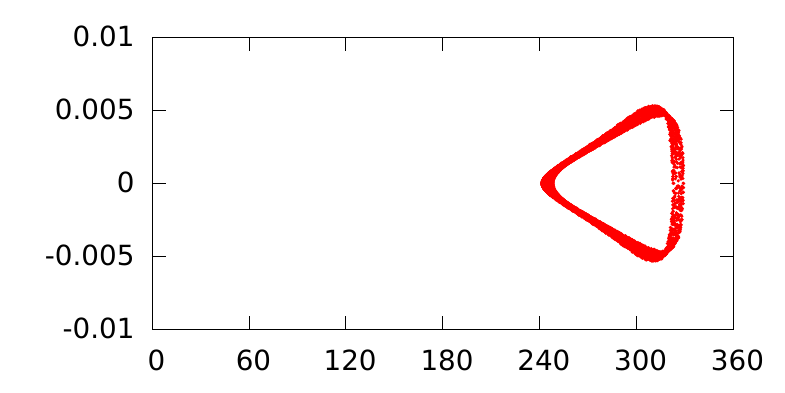}\\
  \setlength{\unitlength}{0.1\linewidth}
\begin{picture}(.001,0.001)
\put(0,2.5){\rotatebox{90}{$a_1-a_2$}}
\put(6,0.5){{$\zeta$}}
\put(8.5,4.5){\red{\circled{6}}}
\end{picture}\\
 \end{minipage}
 \caption{\label{fig:path} Continuous path of quasi-periodic orbits from the $L_4$ eccentric equilibrium to the $L_5$  eccentric equilibrium for $e_1=e_2=0.65$ and $m_1=m_2=10^{-5}m_0$. The plotted orbital elements are the osculating ones (non-averaged, see \ref{sec:ins}). The trajectories were integrated over $5/\eps$ years. On the top-left graph (identical to the top-right graph of figure \ref{fig:glob_e6}) we selected six trajectories of the continuous path from $L_4$ to $L_5$ whose projection are represented in the other graphs. As these trajectories are near the $\Fsf$ family, each of them pass near two distinct points of $\cV$ (in contrast to four distinct points of $\cV$  for generic orbits) that are represented by the blue and red circled numbers on the top-left graph.}
\end{figure}

 \begin{figure}[h!]
 \begin{minipage}{0.49\linewidth}
\includegraphics[width=1\linewidth]{epsfigs/sim20160517g_mesoDlam2.pdf}\\
  \setlength{\unitlength}{0.1\linewidth}
\begin{picture}(.001,0.001)
\put(0,6){\rotatebox{90}{$\Dv$}}
\put(5,1.85){{$\zeta$}}
\put(4.2,0.5){{moy($\zeta$)}}
\end{picture}\\
\vspace{-0.4cm}
\includegraphics[width=1\linewidth]{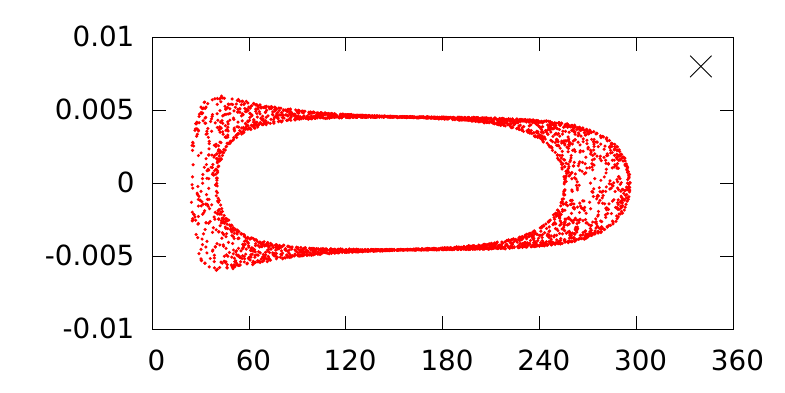}\\
  \setlength{\unitlength}{0.1\linewidth}
\begin{picture}(.001,0.001)
\put(0,1.6){\rotatebox{90}{$a_1-a_2$}}
\put(5.6,-0.3){{$\zeta$}}
\end{picture}\\
\vspace{-0.4cm}
\includegraphics[width=1\linewidth]{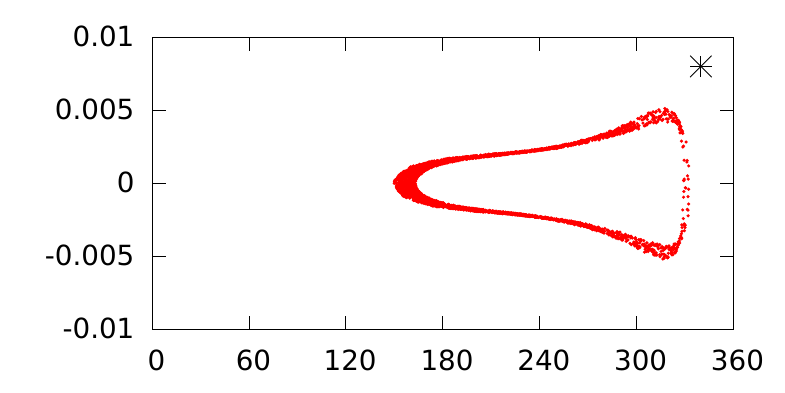}\\
  \setlength{\unitlength}{0.1\linewidth}
\begin{picture}(.001,0.001)
\put(0,1.6){\rotatebox{90}{$a_1-a_2$}}
\put(5.6,-0.3){{$\zeta$}}
\end{picture}\\
\vspace{-0.4cm}
\includegraphics[width=1\linewidth]{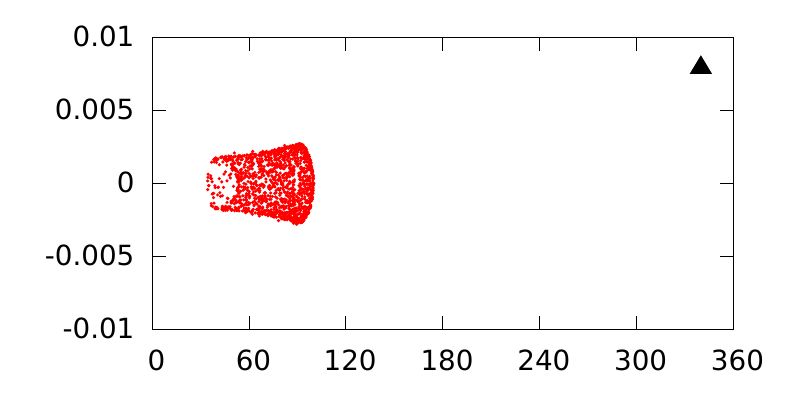}\\
  \setlength{\unitlength}{0.1\linewidth}
\begin{picture}(.001,0.001)
\put(0,1.6){\rotatebox{90}{$a_1-a_2$}}
\put(5.6,-0.3){{$\zeta$}}
\end{picture}\\
\end{minipage}
%
%
%
%
\begin{minipage}{0.49\linewidth}
\includegraphics[width=1\linewidth]{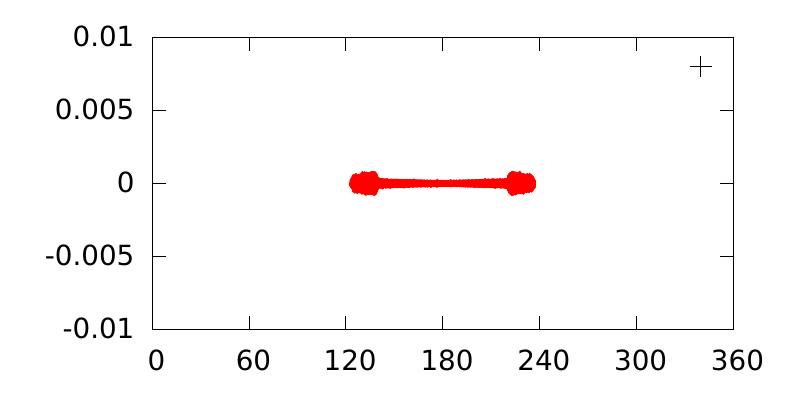}\\
  \setlength{\unitlength}{0.1\linewidth}
\begin{picture}(.001,0.001)
\put(0,2.4){\rotatebox{90}{$a_1-a_2$}}
\put(5.6,0.3){{$\zeta$}}
\end{picture}\\
\vspace{-0.4cm}
\includegraphics[width=1\linewidth]{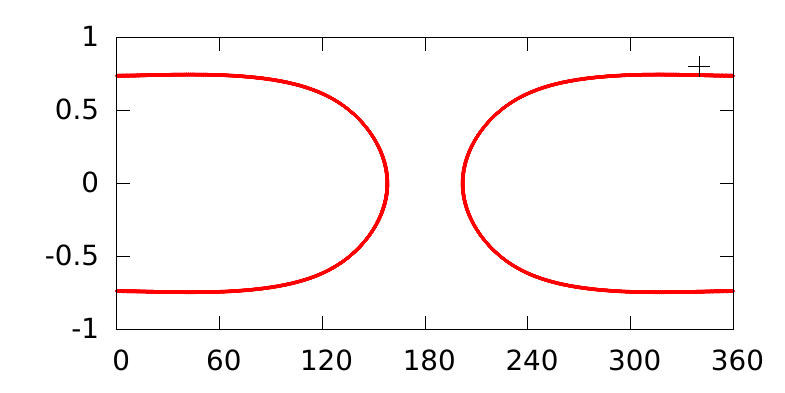}\\
  \setlength{\unitlength}{0.1\linewidth}
\begin{picture}(.001,0.001)
\put(0,1.6){\rotatebox{90}{$e_1-e_2$}}
\put(5.1,-0.3){{$\Dv$}}
\end{picture}\\
\vspace{-0.4cm}
\includegraphics[width=1\linewidth]{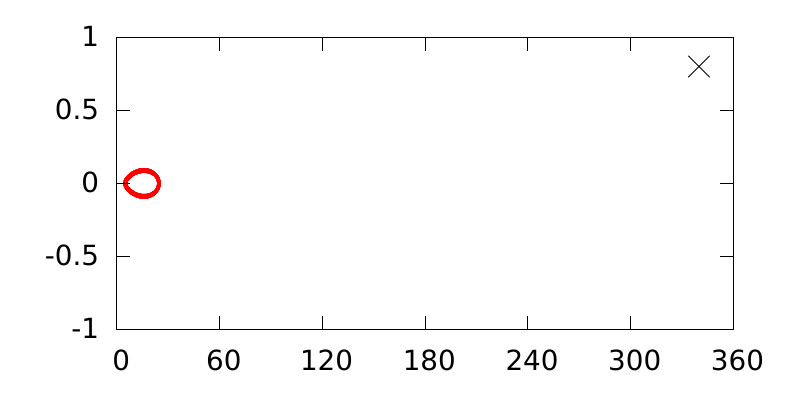}\\
  \setlength{\unitlength}{0.1\linewidth}
\begin{picture}(.001,0.001)
\put(0,1.6){\rotatebox{90}{$e_1-e_2$}}
\put(5.1,-0.3){{$\Dv$}}
\end{picture}\\
\vspace{-0.4cm}
\includegraphics[width=1\linewidth]{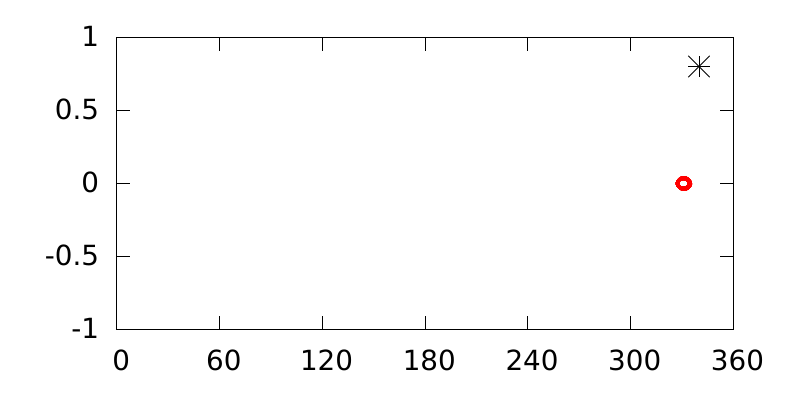}\\
  \setlength{\unitlength}{0.1\linewidth}
\begin{picture}(.001,0.001)
\put(0,1.6){\rotatebox{90}{$e_1-e_2$}}
\put(5.1,-0.3){{$\Dv$}}
\end{picture}\\
\vspace{-0.4cm}
\includegraphics[width=1\linewidth]{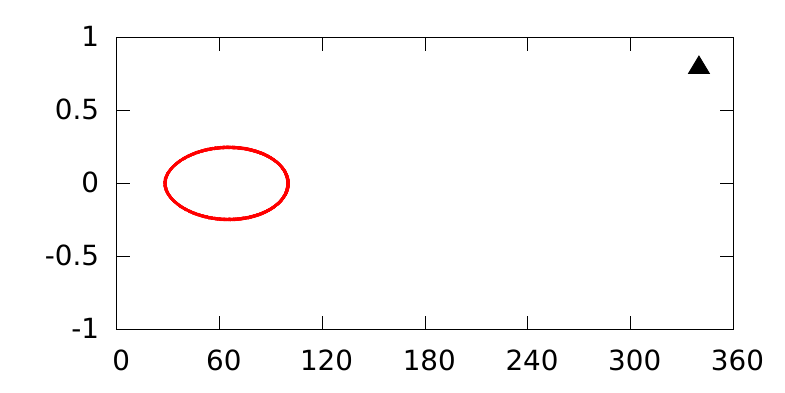}\\
  \setlength{\unitlength}{0.1\linewidth}
\begin{picture}(.001,0.001)
\put(0,1.6){\rotatebox{90}{$e_1-e_2$}}
\put(5.1,-0.3){{$\Dv$}}
\end{picture}\\
 \end{minipage}
 \caption{
\label{fig:orbe7} Projections of generic trajectories emanating from the reference manifold for $e_1=e_2=0.7$, $\eps=m_1/m_0=m_2/m_0=10^{-5}$. The plotted orbital elements are the osculating ones (non-averaged, see \ref{sec:ins}). The trajectories were integrated over $5/\eps$ years. These trajectories pass near 4 distinct points of $\cV$ which are represented by symbols in the top-left graph (identical to the bottom-left graph in figure~\ref{fig:glob_e6}). 
 }
\end{figure}


For $e_1=e_2\leq 0.6$, the $\cF^{sc}_k$ correspond to three separated branches of the $\Fsc$ family, each containing one $L_k$ and one $AL_k$ equilibria. In section \ref{sec:recon}, we showed that these 3 branches reconnected in a continuous $\Fsc$ family for $\e_1=e_2>0.6$. This reconnection leads to a complete restructuring of the whole phase space. \\


The top part of figure \ref{fig:glob_e6} represents the case with $e_1=e_2=0.65$. On the right-hand graph, we clearly see that the stable areas are centred on the $\cF$ families, generating a phase space completely different from the one prior to the reconnection (see figure \ref{fig:glob_e01}). An unstable area appears in the Trojan domain between the orbits librating around $AL_4$ and those librating around $L_4$, clearly differentiating the anti-Lagrange domains (made of orbits librating around $AL_4$ or $AL_5$) from the Trojan domain made of orbits librating around $L_4$ or $L_5$. \\

Although the stability domain of the Trojan and Horseshoe configurations is overall shrinking, new stable areas appear: in addition to the reconnection of the $\Fsc$ families, the $\Fsf$ reconnect as well (see figure~\ref{fig:recon_schema}): the branches of the $\Fsf$ family that contain the equilibria $L_4$ and $L_5$ reconnect to the branches that emanate from the bifurcation of $\cF^{s\!f}_{\gH\cS}$ in the horseshoe domain. This second reconnection link the Trojan domains of $L_4$ and $L_5$ together by the means of what we previously called the the asymmetric horseshoe domains. The consequence of this reconnection is illustrated figure~\ref{fig:path}: it represents $6$ trajectories for $e_1=e_2=0.65$ and $m_1=m_2=10^{-5}$ with initial conditions all taken close to the $\Fsf$ family (the projection of these trajectories on the ($a_1-a_2$,$\zeta$) plane is almost periodic of frequency $\nu$). These six trajectories illustrate that, for high eccentricities we can pass continuously (without crossing separatrix/unstable areas) from a Trojan orbit in the neighbourhood of $L_4$, to an horseshoe orbit, or to a Trojan orbit in the neighbourhood of $L_5$.


\subsubsection{High eccentricities}

\begin{figure}[h!]
\begin{center}
\includegraphics[width=0.7\linewidth]{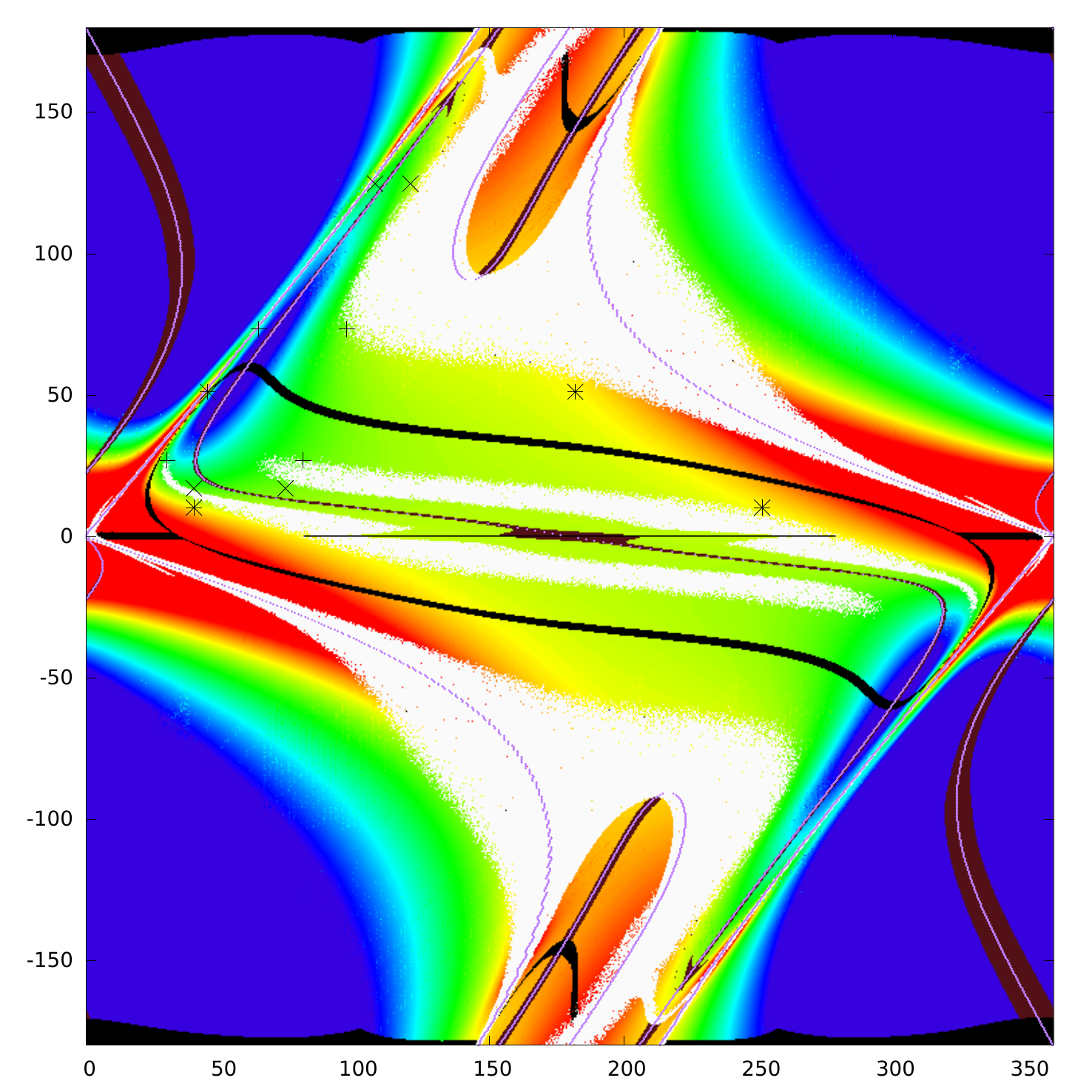}\\
  \setlength{\unitlength}{0.07\linewidth}
\begin{picture}(.001,0.001)
\put(-5,5){\rotatebox{90}{$\Dv$}}
\put(0,0.3){$\zeta$}
\end{picture}
\caption{\label{fig:glob_e7_m6} Variations of the value of the averaged Hamitonian on $\cV$ for $\eps=10^{-6}$ and $e_1=e_2=0.7$. Initial conditions that led to the ejection of the trajectory from the co-orbital resonance before $1/\eps=10^{6}$ orbital periods are displayed in white. Red and blue represent the extrema of the value of Hamiltonian (the scale is not represented because of its non-linearity). The orbits in the neighbourhood of $\Fsc$, i.e. those verifying (\ref{eq:condFb02}) with $\epsilon_\nu=10^{-4}$, are represented by brown pixels (the large amount of brown pixels when we get close to the separatrix is due to a slow variation of $Z$ in these areas with respect to the duration of the integration). The purple curves show the result of the semi-analytical method (eq. \ref{eq:condFb0n}). The orbits close to $\Fsf$, i.e. those verifying (\ref{eq:condFb1}) with $\epsilon_g= 3^\circ$, are represented by black pixels.
}
\end{center}
\end{figure}

The bottom graphs of figure~\ref{fig:glob_e6} show the results of the integrations of $\cV$ for larger eccentricities\footnote{The large amount of grey pixels in the quasi-satellite domain is due to a numerical instabilities: between the blue and the green domain on the right hand graph, each eccentricity vanishes periodically, while the other gets close to $0.99$, so our integration step of $0.01$ orbital periods is not adapted to such high eccentricities.} $e_1=e_2=0.7$. The trends observed for $e_1=e_2=0.65$ are still present: the stability domain for the tadpole and horseshoe configurations continue to shrink, although the neighbourhood of the hyperbolic equilibrium $AL_3$ harbour stable orbits.\\

Moreover, a new domain of stable orbits appears: following the $\cF^{sc}_4$ family emanating from $L_4$ as $\Dv$ increases, we encounter a new separatrix $g=0$ (see figure~\ref{fig:freqge5}, bottom) before the unstable domain is reached. Above this separatrix lies a new stable domain that we call the $G$ configuration. An example of the $G$ trajectories is identified by the $4$ markers $+$ in the figure~\ref{fig:glob_e6} (bottom) and is plotted in figure~\ref{fig:orbe7}. Each of the $G$ trajectories passes near both the trojan configuration librating around $L_4$ and $L_5$. These trajectories hence librate around the families $\cF^{s\!f}_k$ with $k \in \{3,4,5\}$, where $\cF^{s\!f}_4$ and $\cF^{s\!f}_5$ are stable (elliptic), and $\cF^{s\!f}_3$ is unstable (hyperbolic), outside the separatrix $g=0$ in a similar way to the blue trajectories in the figure~\ref{fig:ppH0b}. The domain of $G$ splits as well in orbits that librate around $\Dv=0^\circ$ and orbits librating around $\Dv=180^\circ$. Note that the eccentricities of the orbits in this domain have a huge amplitude of variation, therefore these orbits may not exist when the mass of one co-orbital is significantly smaller than the mass of the other. \\

Until now, most of the integrations were performed with $\eps=10^{-5}$. We recall that the method we used to determine the position of the $\Fsc$ in the averaged problem (section \ref{sec:recon}) is independent of the value of $\eps$. To illustrate the effect of $\eps$ on the position of the $\Fsf$ and on the whole phase space, we integrate trajectories emanating from $\cV$ with $\eps=10^{-6}$ (see Fig. \ref{fig:glob_e7_m6}). The trajectories in this figure are also integrated over $10^6$ orbital periods\footnote{Note that in this case the integration time is too short to properly account for the effect of the secular dynamics (which is also of the order of $10^6$ orbital periods).}. The color code for the non-ejected orbits displays an indicator of the value of the total energy of the system at a given position on $\cV$. Trajectories that were found in the neighbourhood of the $\cF$ families are also displayed. Comparing this figure with the bottom graphs in figure~\ref{fig:glob_e6}, we can see that the intersection of the $\cF$ families (and the manifolds $g=0$ and $\nu=0$) with the reference manifold $\cV$ appears to not depend on the value of $\eps$ ($=m_1/m_0=m_2/m_0$). Consequently, the reconnection of the $\cF$ families and the topology of the phase space seem to be independent from the value of $\eps$, as long as we have $m_1=m_2$. The size of the stability domains, however, is impacted by $\eps$.

In order to identify the origin of the different unstable areas of the phase space in figure~\ref{fig:glob_e7_m6}, we took three initial conditions ($+$, $\times$ and $*$) in the top left quadrant with respect to $L_4$ (the quadrants are delimited by the families $\Fsc$ and $\Fsf$). These initial conditions are taken very close to the collision manifold. By integrating these trajectories over a few periods of $g$, we can identify for each of these orbits the three other points of $\cV$ near which they pass. For a given trajectory, each point is represented by the same symbol. The three trajectories pass near the instability border in each quadrant, these borders thus seem to have the same origin: they emanate from the collision manifold.
 
 \subsubsection{Stability}
 
In this work, the trajectories emanating from a given reference manifold were generally integrated over $10/\eps$ orbital periods. Although this is enough to take into account the secular dynamics (time scale of the order of $1/\eps$), it is not enough to infer the long term stability of a given orbit.

The long term stability of the new orbital configurations that are discussed in this work was studied for various values of the masses and eccentricities \citep[][section 2.5.2]{these}: asymmetric horseshoe for $e_1=e_2>0.5$ and $m_1=m_2=10^{-5} m_0$, continuous path between $L_4$, $L_5$ and the horseshoe configuration for $e_1=e_2=0.7$ and $m_1=m_2=10^{-6} m_0$, stable orbits near $AL_3$ for $e_1=e_2=0.7$ and  $m_1=m_2=10^{-5} m_0$, and $G$ configuration for $e_1=e_2=0.7$ and  $m_1=m_2=10^{-5} m_0$. In these cases, these configurations were stable for duration long with respect to their secular period (over $100/\eps$ orbital period).  \\

 The stability was checked by studying the diffusion of the mean mean-motion of the planet $m_1$ during long-term integrations \citep{RoLa2001}. When required by the high values of the eccentricity, we used the variable-step integrator DOPRI (Runge-Kutta (7)8). The agreement between integrators (SABA4 and DOPRI) was also checked.




 
 

\section{Phase space of eccentric co-orbitals in the case $m_1 \neq m_2$}

\begin{figure}[h!]
\begin{center}
\includegraphics[width=0.5\linewidth]{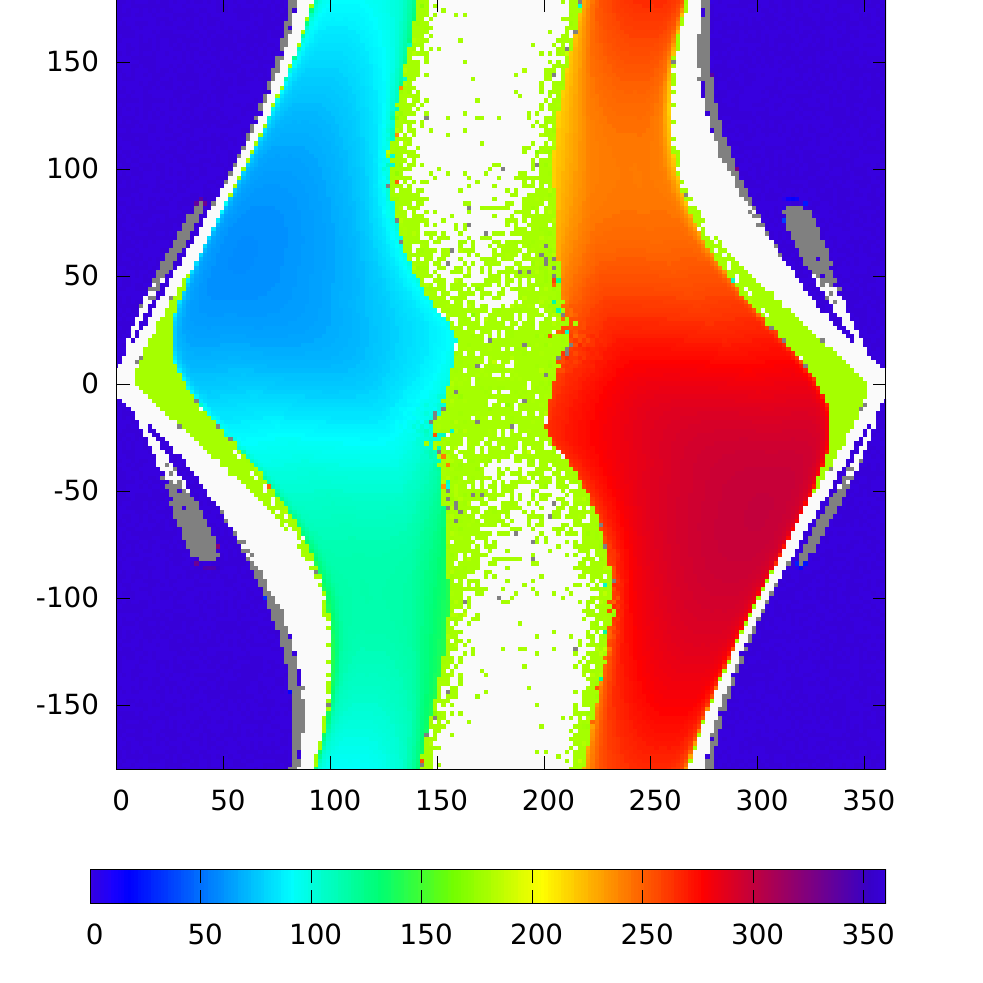}\includegraphics[width=0.5\linewidth]{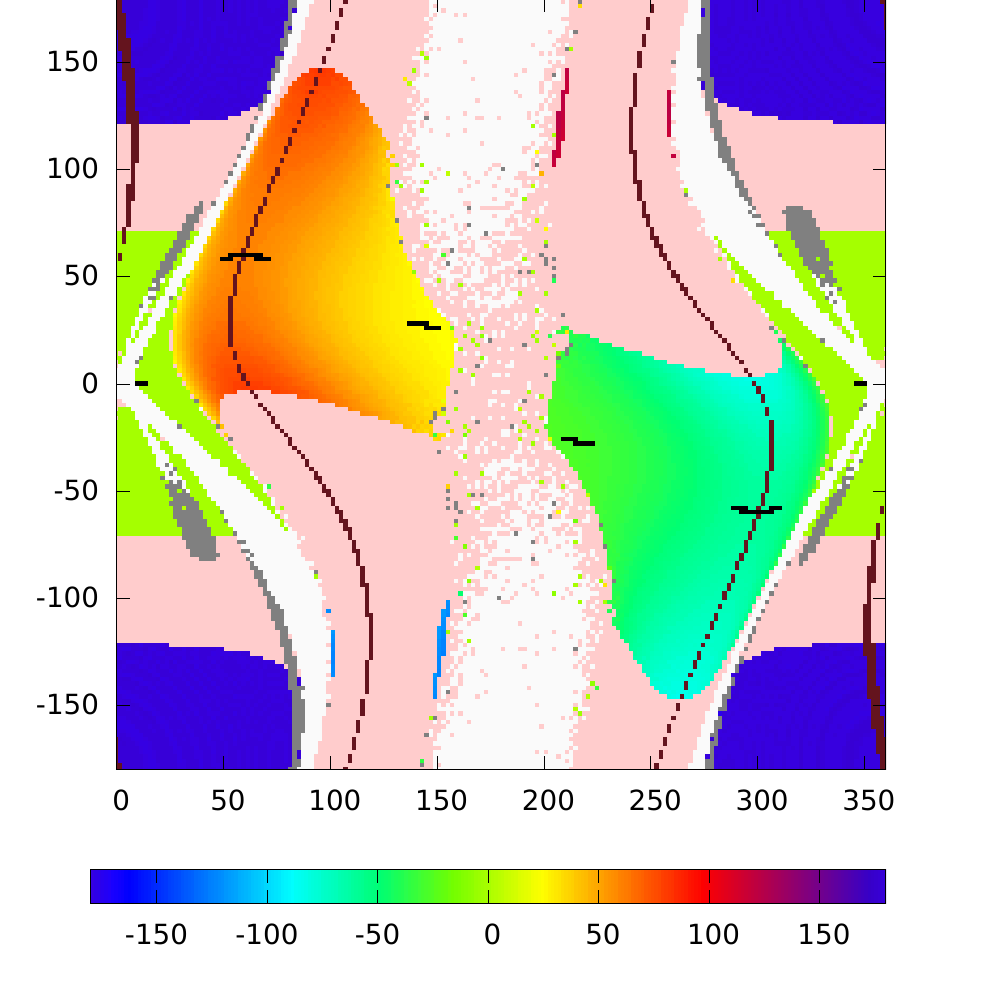}\\
  \setlength{\unitlength}{0.067\linewidth}
\begin{picture}(.001,0.001)
\put(-7.5,4.5){\rotatebox{90}{$\Dv$}}
\put(0,4.5){\rotatebox{90}{$\Dv$}}
\put(-3.8,1.5){{$\zeta$}}
\put(3.8,1.5){{$\zeta$}}
\put(-3.8,0.5){{moy($\zeta$)}}
\put(3.3,0.5){{moy($\Dv$)}}
\end{picture}\\
\includegraphics[width=0.5\linewidth]{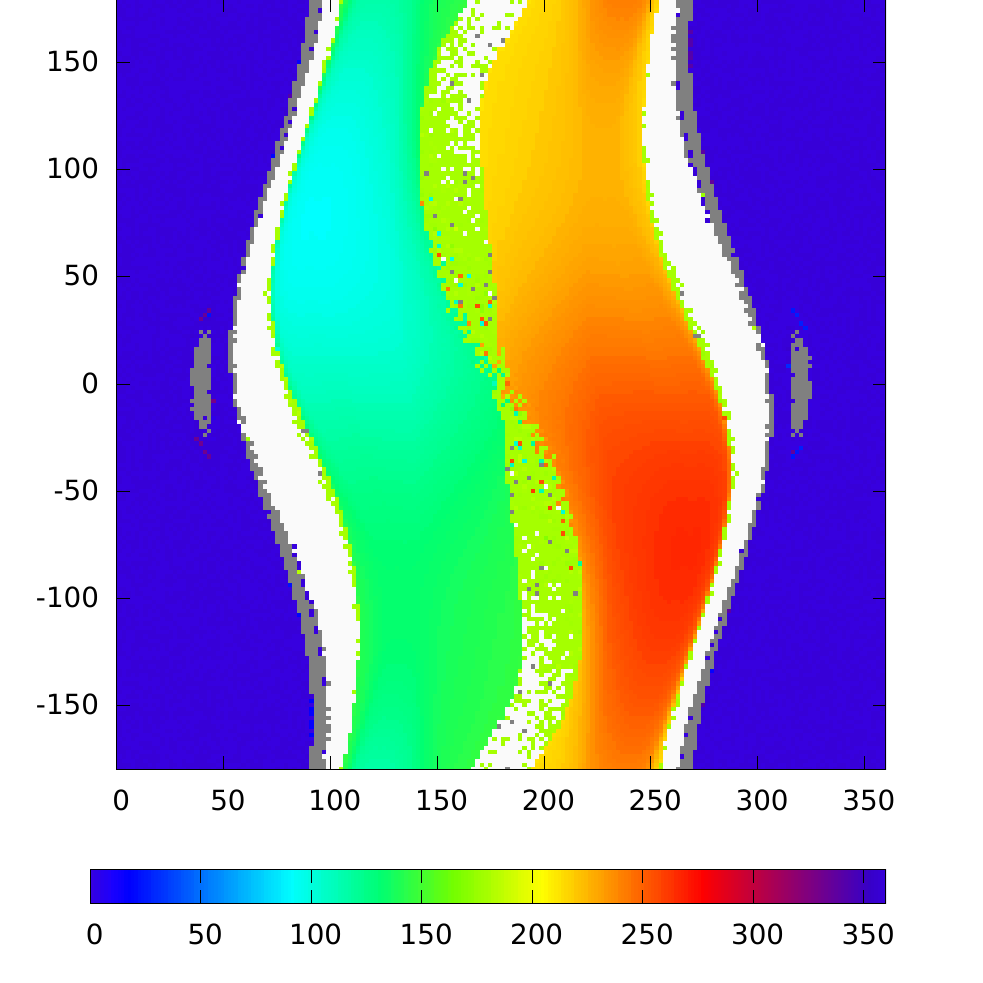}\includegraphics[width=0.5\linewidth]{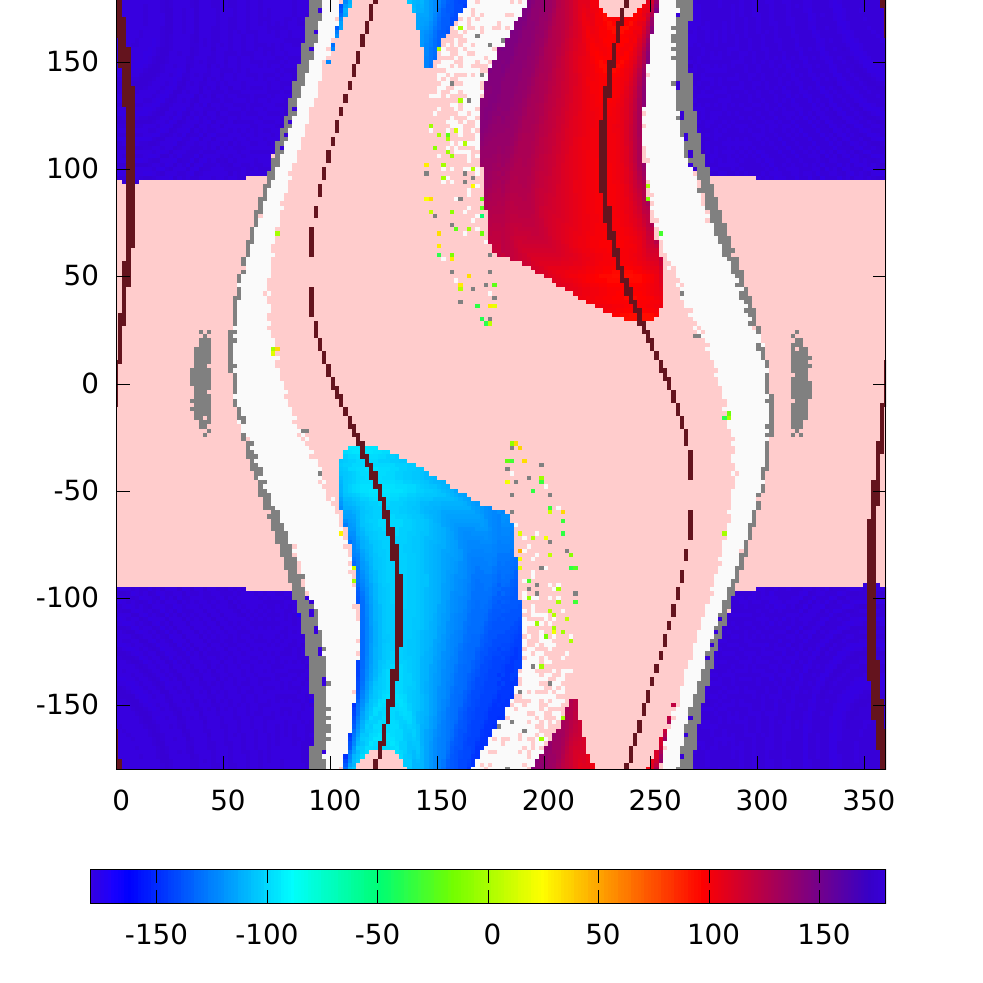}\\
  \setlength{\unitlength}{0.067\linewidth}
\begin{picture}(.001,0.001)
\put(-7.5,4.5){\rotatebox{90}{$\Dv$}}
\put(0,4.5){\rotatebox{90}{$\Dv$}}
\put(-3.8,1.5){{$\zeta$}}
\put(3.8,1.5){{$\zeta$}}
\put(-3.8,0.5){{moy($\zeta$)}}
\put(3.3,0.5){{moy($\Dv$)}}
\end{picture}
\caption{\label{fig:glob_e4_md} Grid of initial condition with $a_1=a_2=1\,au$, $m_2=3m_1=1.5\, 10^{-5}$ and $e_1=e_2=0.4$ (top), and  $e_1=0.7$ and $e_2\approx0.18$ (bottom, same value of the angular momentum than the case $e_1=e_2=0.4$). The color code gives the mean value of $\zeta$ (left) and the mean value of $\Dv$ (right). If the angle $\Dv$ circulate, the salmon color is displayed instead of the color code. The orbits in the neighbourhood of $\Fsc$, hence those verifying (\ref{eq:condFb02}) with $\epsilon_\nu=10^{-3.5}$, are represented by brown pixels. The orbits close to  $\Fsf$ hence those verifying (\ref{eq:condFb1}) with $\epsilon_g= 3^\circ$ are represented by black pixels. The initial conditions that lead to a diffusion of the mean semi-major axis over $\epsilon_a=10^{-5.5}$ are displayed in grey.
}
\end{center}
\end{figure}
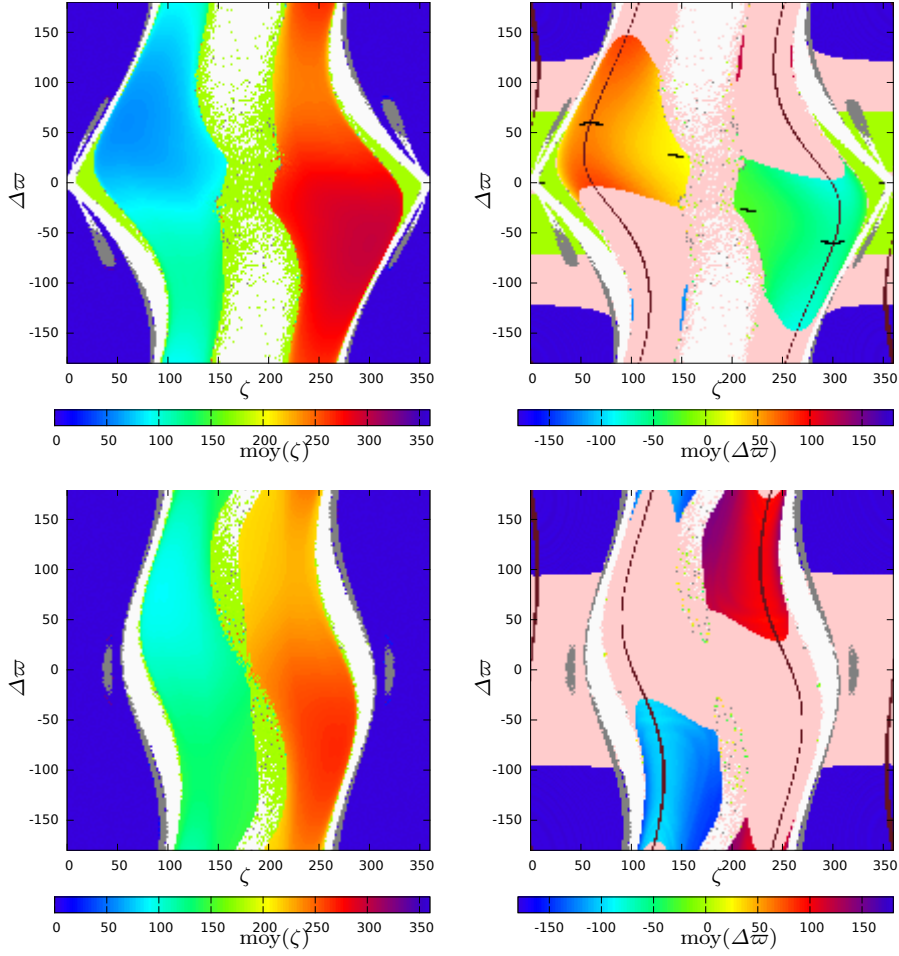

In this section we check if the modifications in the phase space observed for the case $m_1=m_2$ still occur for different mass ratios. We take $m_2=3m_1=1.5 \times 10^{-5} m_0$. It is important to remember that in this case, the manifolds of initial conditions that we consider are no longer reference manifolds as in section \ref{sec:RM}, they are just sections of the phase space that can miss part of, or entire, co-orbital configurations.

\subsection{Moderate eccentricities}

In figure~\ref{fig:glob_e4_md} we show the same information as in section \ref{sec:Vmeq}. In addition, on the right graphs the salmon color represents the initial conditions of the trajectories for which the angle $\Dv$ circulates. 

On the top graphs, the initial conditions are taken across the plane $e_1=e_2=0.4$, $a_1=a_2$ (with $m_2=3m_1=1.5 \times 10^{-5} m_0$). 
On the left hand side (evolution of the mean value of $\zeta$), the dynamics of the pair ($Z,\zeta$) seems to not change much from the case $m_1=m_2$ for the same value of the total angular momentum (compare with the figure~\ref{fig:glob_e01} - bottom). 
On the right hand graph, we can see that the dynamics of the pair ($\varPi,\Dv$) is different from the case $m_1=m_2$: $\Dv$ circulates for a large amount of the integrated trajectories (salmon color). 
Since some trajectories satisfy the criterion (\ref{eq:condFb02}), the manifold $\Fsc$ seems to be close to this plane of initial conditions. However, the $\Fsf$ families depart from it as soon as we quit the neighbourhood of the $L_4$ equilibrium. This is consistent with the analytic estimation of the position of $\Fsf$ \citep[see][section 2.7.2]{these}. \\

The bottom graphs represent another section of the same phase space, with $e_1=0.7$ and $e_2 \approx 0.18$. This plane intersects the phase space closer to the trojan domain librating around the $AL_4$ equilibrium ( $e_1 \approx 0.67$, $e_2\approx0.22$ in the linear approximation eq. \ref{eq:relat_AL}). Some of the trajectories that take initial conditions on this plane librate around $AL_4$ (domain centred on $\zeta=130^\circ$, $\Dv=-100^\circ$), but none librate around $L_4$. Note that for $m_1 \neq m_2$ it may be impossible to pass directly from orbits librating around $L_4$ to orbits librating around $AL_4$ as it seems that these areas are separated by a region where $\Dv$ circulates (checked for $e_1 \in \{0, 0.15, 0.3 0.55 ,0.7\}$ and $e_2$ such that $J_1=J_1(e_1=e_2=0.4)$).\\

Interestingly, although the case $m_2=3m_1$ is far from the restricted case, the phase space of both cases possess similar features. One can compare for example figures 7 and 8 in \cite{NeThoFeMo02} with the figure \ref{fig:glob_e4_md} in this paper, which represents different sections of a similar phase space.

 \subsection{High eccentricities}
\begin{figure}[h!]
\begin{center}
\includegraphics[width=0.5\linewidth]{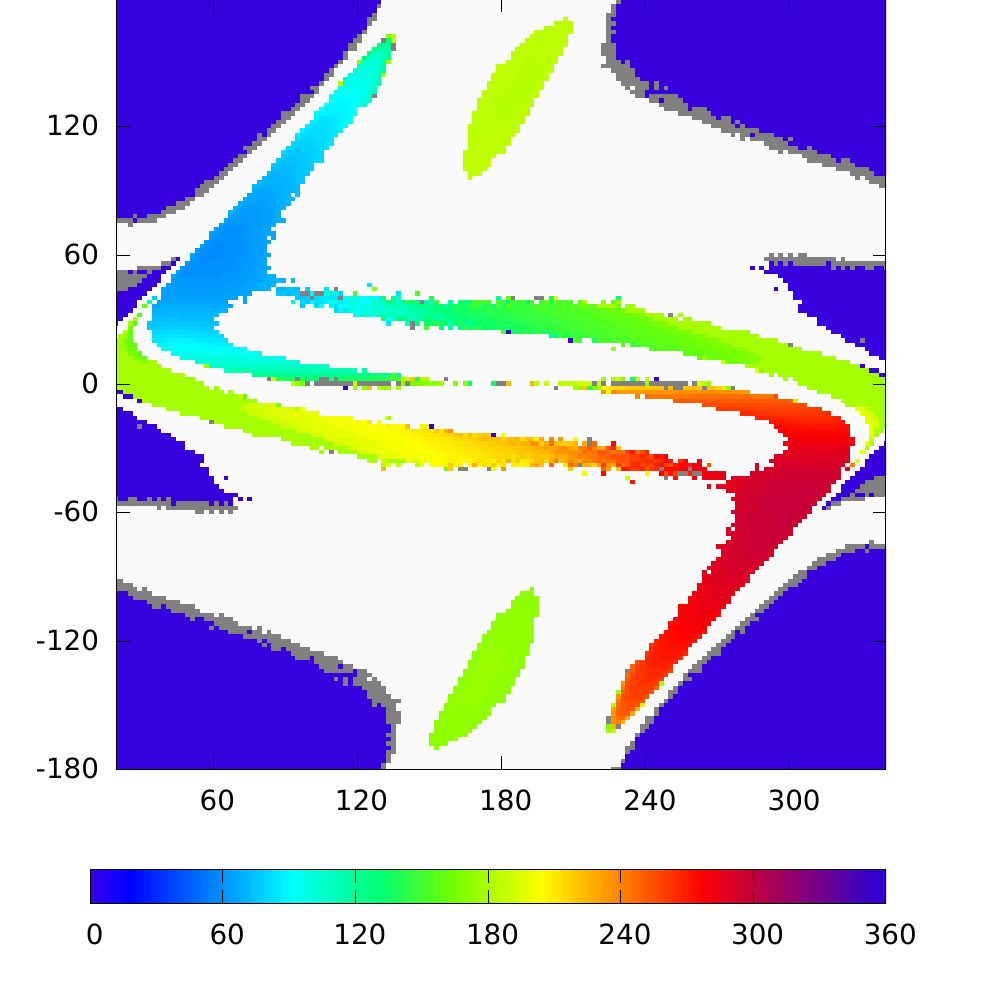}\includegraphics[width=0.5\linewidth]{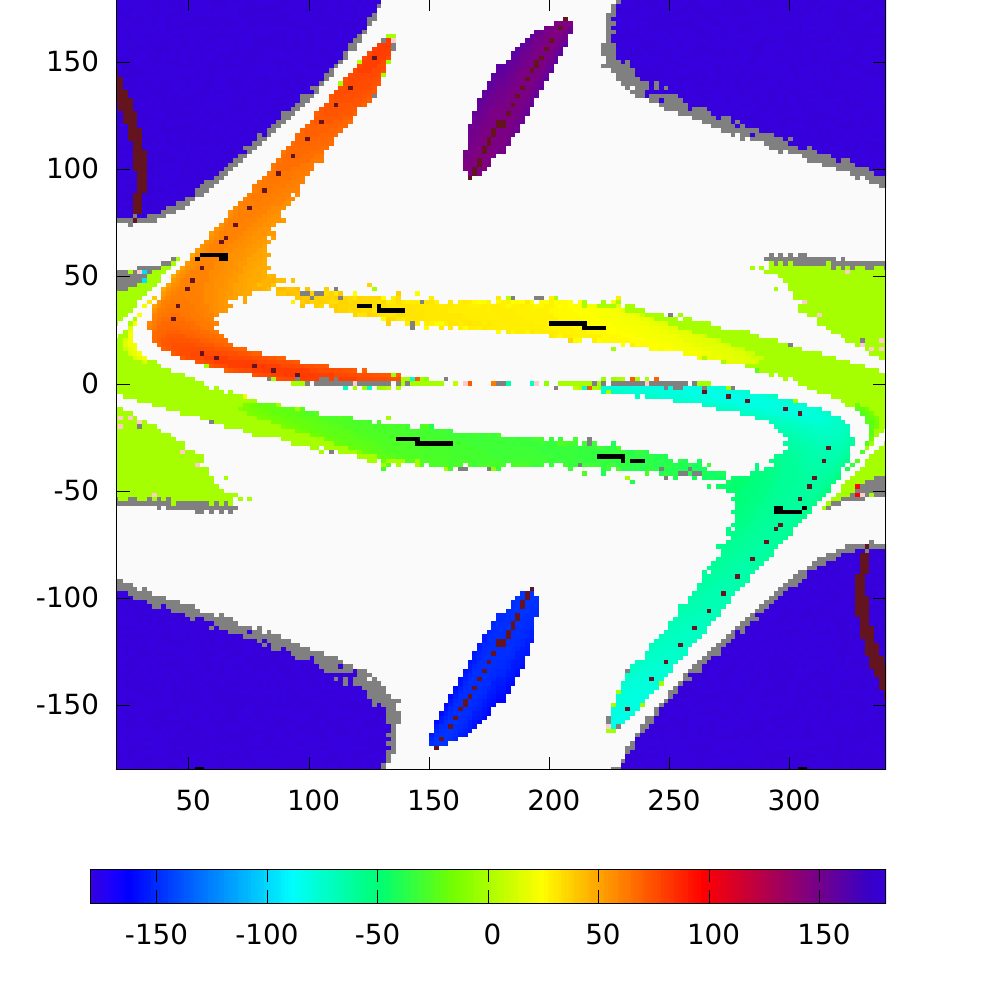}\\
  \setlength{\unitlength}{0.067\linewidth}
\begin{picture}(.001,0.001)
\put(-7.5,4.5){\rotatebox{90}{$\Dv$}}
\put(0,4.5){\rotatebox{90}{$\Dv$}}
\put(-3.8,1.5){{$\zeta$}}
\put(3.8,1.5){{$\zeta$}}
\put(-3.8,0.5){{moy($\zeta$)}}
\put(3.3,0.5){{moy($\Dv$)}}
\end{picture}
\caption{\label{fig:glob_e7_md}  Grid of initial conditions with $a_1=a_2=1\,au$, $m_2=3m_1=1.5\, 10^{-5}$,  $e_1=e_2=0.7$. The color code gives the mean value of $\zeta$ (left) and the mean value of $\Dv$ (right). The orbits in the neighbourhood of $\Fsc$, i.e. those verifying (\ref{eq:condFb02}) with $\epsilon_\nu=10^{-3.5}$, are represented by brown pixels. The orbits close to  $\Fsf$, i.e. those verifying (\ref{eq:condFb1}) with $\epsilon_g= 3^\circ$, are represented by black pixels. The equilibrium $AL_4$ and $AL_5$ are not in this plane of initial conditions \citep[for the chosen value of angular momentum, they are located in the manifold $\e_1\approx 0.8$, $e_2\approx0.6$, see][]{HaVo2011}. The initial conditions that lead to a diffusion of the mean semi-major axis over $\epsilon_a=10^{-5.5}$ are displayed in grey. These integrations were performed with a time step of $0.001$ orbital period. The initial conditions for $\zeta \in [0^\circ:20^\circ]$ were not integrated to save computer time.
}
\end{center}
\end{figure}
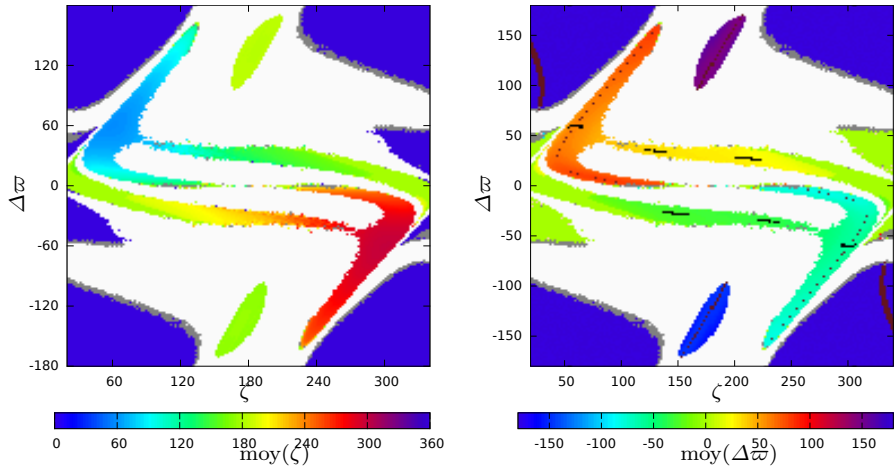

%

In figure~\ref{fig:glob_e7_md} we show the mean value of the angles $\zeta$ and $\Dv$ when the initial conditions are taken across the plane $e_1=e_2=0.65$, $a_1=a_2$ (with $m_2=3m_1=1.5 \times 10^{-5} m_0$). In this case, since the integration time step of $0.01$ orbital periods ejects too many stable trajectories, we adopt $0.001$ orbital periods as time step and slightly reduced the span of initial conditions to save computer time.\\

 The topological change that we described in the case $m_1=m_2$ occurs in this case as well (compare figure~\ref{fig:glob_e7_md} with the top graphs in figure~\ref{fig:glob_e6}): the stable trojan area around $L_4$ and $L_5$ are well separated from those around $AL_4$ and $AL_5$, while asymmetric horseshoe domains emerge, linking the $L_4$ and $L_5$ equilibriums. Note that this plane of initial conditions does not intersect the stability domain of the $G$ configuration. In addition, unstable area splits the quasi satellite domain between the orbit which librates around $0^\circ$ and those librating around $180^\circ$. However, part of this instability may be due to numerical issues, since the eccentricity of the smaller body tends to one at the boundary between the two domains.
Finally, no orbit for which $\Dv$ circulates crosses this plane.

\section{Conclusion}

We studied the dynamics and stability of eccentric coplanar co-orbitals in the planetary case. We observed the topological changes occurring in the phase space as the eccentricity of the co-orbitals increase, and we linked these changes to the evolution of the position of families of quasi-periodic orbits of non-maximal dimension. These changes where mainly quantified in the case $m_1=m_2$ since those families are easier to find, but we checked that the evolution of the phase space is qualitatively the same when $m_1 \neq m_2$. 

In the case $m_1=m_2$, we showed that the orbits emanating from the manifold of initial conditions $\cV=\{e_1=e_2,a_1=a_2\}$ represents a significant part of the orbits of the reduced averaged phase space for a fixed value of the total angular momentum. We hence only need to integrate orbits emanating from this manifold to explore most of the orbital behavior of this phase space. From $e_1=e_2=0$ to $e_1=e_2 \lesssim 0.5$ no major modifications were observed in the phase space with respect to the quasi-circular case: trojan and hoseshoe orbits are separated by a separatrix along which $\nu=0$, and the collision manifold separates the horseshoe orbit from the quasi-satellite ones. As $e_1=e_2$ increases, the position of these separatrix evolves in the phase space, the stable quasi-satellite area get larger, while the size of the trojan and horseshoe stable domains decrease.

Around $e_1=e_2\approx0.55$, a first significant modification occurs: the secular frequency $g$ vanishes within the horseshoe domain, splitting it in three domains: The symmetric horseshoes, which are the same that existed in the circular case, and two domains of asymmetric horseshoe, located between the separatrices $g=0$ and $\nu=0$. These asymmetric horseshoes blur the difference between horseshoe and tadpole.

Between $0.605\leq e_1=e_2 \leq 0.61$, a second major change occurs: the family of quasi-periodic of non-maximal dimension $\Fsc$ reconnects, forming a single family going through the eccentric Lagragian equilibrium $L_k$ and anti-Lagrangian equilibrium $AL_k$ for $k \in \{3,4,5\}$. This reconnection  leads to an unstable area appearing between the tadpole that were orbiting around $L_4$ (resp. $L_5$), and those that are orbiting around $AL_4$ (resp. $AL_5$), creating two distinct stable areas. The reconnection of the $\Fsc$ opens the way to the reconnection of another family: the $\Fsf$. This family has members in each stable domain, and for masses small enough, there is a path of stable quasi-periodic orbits of non-maximal dimension that link continuously the trojan domain librating around the $L_4$ and $L_5$ equilibrium, to the asymmetric and symmetric horseshoe domains. Finally, we note the presence of a new separatrix $g=0$ in the trojan domain, beyond which a new stable configuration, that we called $G$, appears. In this configuration, the difference of the mean longitudes librates around $180^\circ$ with a significant amplitude ($\sim 100^\circ$), while $\Dv$ librates around $0^\circ$ or $180^\circ$ on a secular time scale with large variations of the quantity $e_1-e_2$.

\begin{acknowledgements}
The authors acknowledge financial support from the Observatoire de
Paris Scientific Council, CIDMA strategic project UID/MAT/04106/2013,
ENGAGE SKA POCI-01- 0145-FEDER-022217 (funded by COMPETE 2020 and FCT,
Portugal), and the Marie Curie Actions of the European Commission
(FP7-COFUND). Parts of this work have been carried out within the frame of the National Centre for Competence in Research PlanetS supported by the SNSF. This work was granted access to the HPC resources of MesoPSL financed by the Region Ile de France and the project Equip@Meso (reference
ANR-10-EQPX-29-01) of the programme Investissements d’Avenir supervised
by the Agence Nationale pour la Recherche. The authors thank the referees for useful suggestions that greatly improved the description of the results.
\end{acknowledgements}

\bibliographystyle{apalike}


\newpage
\appendix

\section{Time scales} 
\label{sec:tsai}

The planar 3-body co-orbital problem has 3 time scales: the fast time scale, associated with the mean mean-motion $\eta=\gO(1)$, the semi-fast time scale of fundamental frequency $\nu=\gO(\sqrt{\eps})$,  associated with the evolution of the resonant angle $\zeta$, and the secular time scale of fundamental frequency $g_1$ and $g_2$, of order $\gO(\eps)$, associated with the evolution of the eccentricities and the arguments of perihelia. The separation of these time scales is a classical approach for the study of mean motion resonances \citep{HeCa1989,Morbidelli02,BaMo2013,Delisle2012,Delisle2014}. 



In theory, this separation allows for two averaging of the Hamiltonian: a first averaging over the fast angle $\lambda_2$, that we already considered in section \ref{sec:Dyncp}, and a second one over the semi-fast angle $\zeta$, in order to obtain the secular Hamiltonian. In the double averaged reduced case, we would obtain a 1 degree of freedom Hamiltonian which would describe the secular dynamics of the resonance. It would add an additional parameter $J_0$ (the action variable associated with the degree of freedom $Z$,$\zeta$). The canonical transformation for the variables $\varPi$ and $\Dv$ associated with this second averaging differs from the identity only with coefficients of the order of $\gO(\sqrt \eps)$ \citep{Morbidelli02}. 
%

\subsection{Adiabatic invariants}

In practice, this second averaging is rather difficult because the variables $Z$ and $\zeta$ are not close to action-angle variables \citep{Morais1999,Morais2001,BeaugeR01,PaEf2015}. However, the possibility to do it gives us important information on the dynamics of the system: in the averaged reduced problem (2 degree of freedom), the evolution of the variables $\varPi$ and $\Dv$ is of size $\sqrt{\eps}$ over durations of the order of $1/\nu$, these variables can be considered as constant on a time scale short with respect to $1/g$. For sufficiently low-mass co-orbitals, we can hence consider that the variables $\varPi$ and $\Dv$ are adiabatic invariants.

\subsection{Interpretation of numerical simulations}
\label{sec:ins}

The change of coordinate (eq. \ref{eq:Hbtrans}) from the variables of the planar 3-body problem to the variables of the averaged problem is $\eps$ close from identity for all variables except $\zeta_2$. Similarly, the perturbations of the semi-fast time-scale on the secular variables are of size $\sqrt{\eps}$ \citep{Morbidelli02}. 
%
Thus, if we integrate numerically the full 3-body problem for co-orbital with low enough masses, for quasi-periodic orbits we can consider on the one hand the evolution of the variables ($Z,\zeta$) as their evolution in the averaged problem (they are $\eps$ close), and on the other hand the evolution of the variables $e_j$ and $\varpi_j$ as their evolution in the secular problem (they are $\sqrt\eps$ close).

\section{Reference manifold}

In this section we aim to verify that all the trajectories of the phase space pass as close as we want from the reference manifold $\cV$ defined by the equation (\ref{eq:RM}).

\subsection{At first order in $e_j$}
\label{sec:foe}

The equation (\ref{eq:eqerdi}) holds at first order in $e_j$. Hence, for $m_1=m_2$, all trajectories go through the plane $a_1=a_2$ twice per period $2\pi/\nu$. On the other hand, near a solution of the circular coplanar case $\zeta(t)$, the equation of variation in the direction ($x_j,\xt_j$), where the $x_j$ are the canonical Poincar\'e variables defined eq. (\ref{eq:poincvar}), is given by the matrix \citep{RoPo2013}:
\begin{equation}
X=
   \begin{pmatrix}
 x_1\\
 x_2
   \end{pmatrix} 
   \hspace{0.3cm}
   \text{et}
   \hspace{0.3cm}
 M(t)= i\eps \eta \frac{m'_1 m'_2}{m_0}
   \begin{pmatrix}
\frac{A(\zeta(t))}{m'_1} & \frac{\bar{B}(\zeta(t))}{\sqrt{m'_1 m'_2}}\\
\frac{B(\zeta(t))}{\sqrt{m'_1 m'_2}} & \frac{A(\zeta(t))}{m'_2} 
   \end{pmatrix}  \, , 
\label{eq:eqvar2}
\end{equation}
where $A$ and $B$ depend on the considered trajectory and on the time. For a given trajectory, since $\nu \gg g$, we can obtain an approximation of the secular dynamics in the direction ($x_j,\xt_j$) by averaging the expression of this matrix over a period $2\pi/\nu$ with respect to the time $t$. For equal mass co-orbitals, the symmetries of this matrix give relations of the form:
\begin{equation}
\begin{aligned}
x_1 & =\alpha \sqrt{m_2} \e^{i (\frac{\pi}{3}+g t)} + \beta \sqrt{m_1} \e^{i\frac{\pi}{3}}\, , \\
x_2 & =- \alpha \sqrt{m_1} \e^{i g t} + \beta \sqrt{m_2} \, ,
\end{aligned}
\label{eq:x1x2L}
\end{equation}
with $\alpha$ and $\beta$ complexes. Replacing these expression in the one of $\varPi$ (eq. \ref{eq:CIred}), and noting $\alpha \bar \beta= C \operatorname{e}^{ic}$, we obtain: 
\begin{equation}
\begin{aligned}
\varPi & = (\alpha\bar{\alpha}-\beta\bar{\beta})(m_1-m_2)-2C\sqrt{m_1 m_2} \cos (g t + c)\, .
\end{aligned}
\label{eq:PiL4}
\end{equation}

On the other hand, we have:
\begin{equation}
\begin{aligned}
\Dv & =\arg(x_1\bar{x}_2) \, , \ \ \text{where}\ \\
x_1\bar{x}_2 &=[\sqrt{m_1m_2}(\beta\bar{\beta}-\alpha\bar{\alpha})+Cm_2\e^{i(gt+c)}-Cm_1\e^{-i(gt+c)}]\e^{i\pi/3}\, .
\end{aligned}
\label{eq:DpiL4}
\end{equation}
When $m_1=m_2$, $\varPi$ librates around $0$ with a frequency $g$. Using once more the expression (\ref{eq:CIred}), we obtain that the quantity $e_1^2-e_2^2$ behaves like an harmonic oscillator, librating around $0$ with the frequency $g=2|c|=\gO(\eps)$. All the trajectories of the phase space hence goes through the plane $e_1=e_2$ twice per period $2\pi/g$.\\

 As long as $\nu$ and $g$ are non-resonant, all trajectory get as close as we want to the manifold $\cV$ in a finite time. 

\subsection{Large eccentricities}
\label{sec:fle}
\begin{figure}[h!]
\begin{center}
 \includegraphics[width=0.24\linewidth]{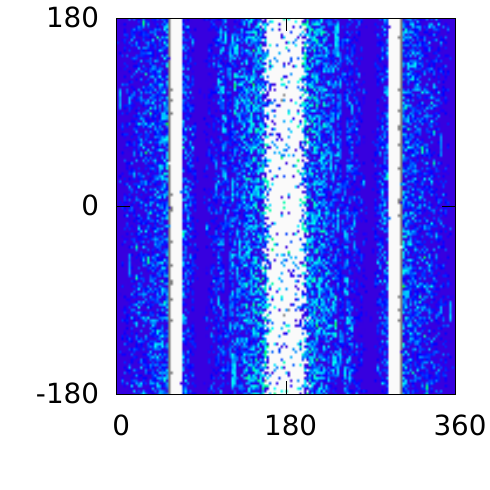}  \includegraphics[width=0.24\linewidth]{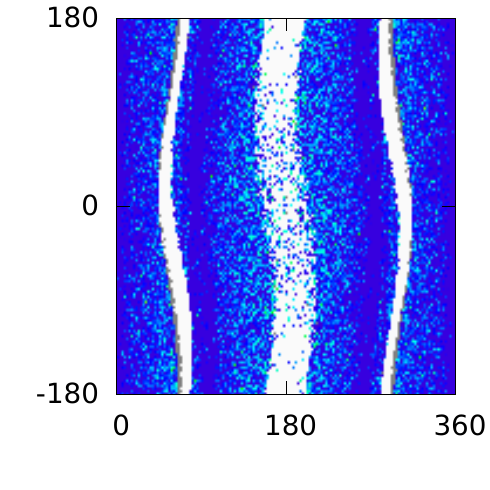}
  \includegraphics[width=0.24\linewidth]{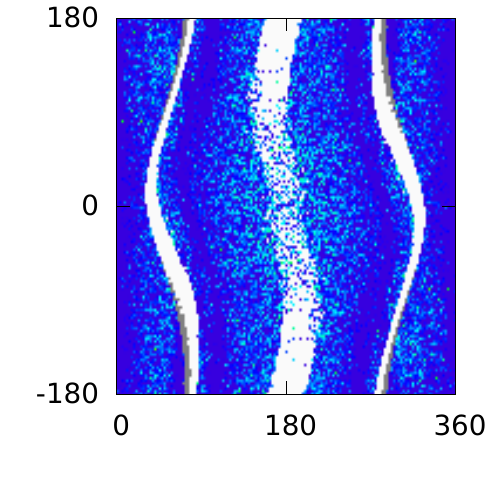}   \includegraphics[width=0.24\linewidth]{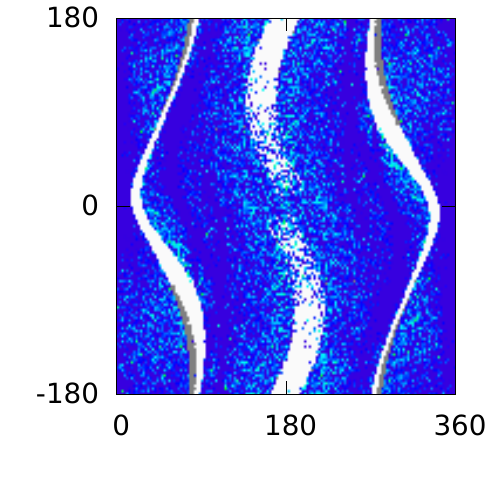}\\
  \setlength{\unitlength}{0.1\linewidth}
\begin{picture}(.001,0.001)
\put(0.2,3.6){\rotatebox{90}{$\Dv$}}
\put(0.2,1.2){\rotatebox{90}{$\Dv$}}
\put(1.4,4.9){(a)}
\put(3.8,4.9){(b)}
\put(6.3,4.9){(c)}
\put(8.8,4.9){(d)}
\put(1.45,0){{$\zeta$}}
\put(3.9,0){{$\zeta$}}
\put(6.4,0){{$\zeta$}}
\put(8.9,0){{$\zeta$}}
\put(3,-0.7){{$\log[\min{}((a_1/\bar{a}-a_2/\bar{a})^2+(e_1-e_2)^2)]$}}
\end{picture}
%
  \includegraphics[width=0.24\linewidth]{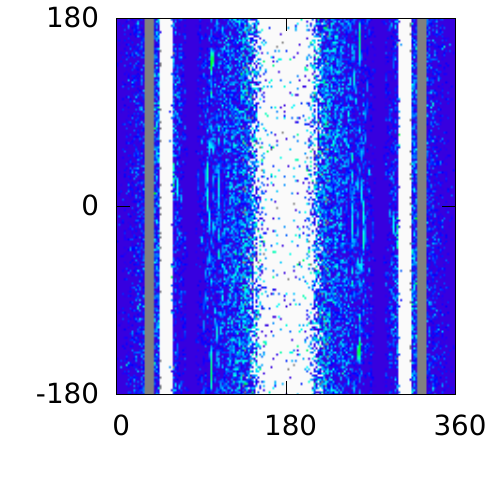}
   \includegraphics[width=0.24\linewidth]{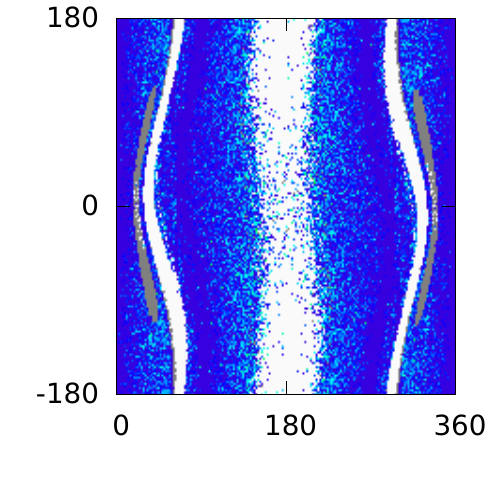}
  \includegraphics[width=0.24\linewidth]{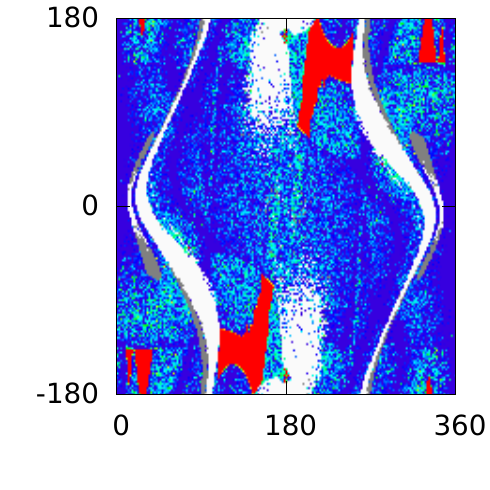}
   \includegraphics[width=0.24\linewidth]{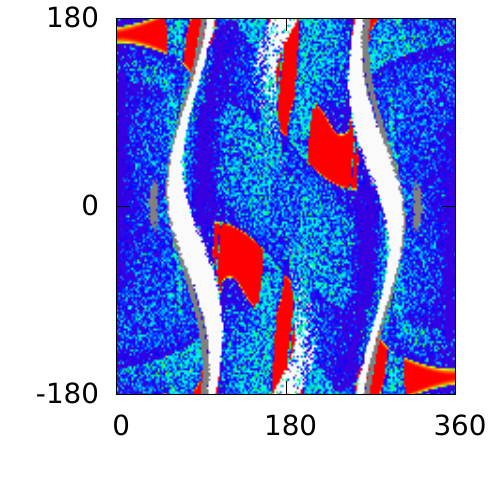}\\
     \setlength{\unitlength}{0.067\linewidth}
\begin{picture}(.001,0.001)
\end{picture}
   \includegraphics[width=0.5\linewidth]{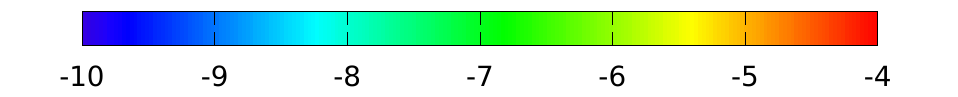}\\
\caption{\label{fig:varefmeqe07} minimal value of the quantity $\log \left( (a_1/\bar a-a_2/\bar a)^2+(e_1-e_2)^2 \right)$ over $10\times 10^{5}$ orbital periods with a step of $0.01$~orbital period and a fixed value of the angular momentum $J_1(e_1=e_2=0.4)$, with the following initial conditions: $a_1=a_2=1$, $m_1=m_2=10^{-5} m_0$ for the top row and $m_2=3m_1=1.5\, 10^{-5}$ for the bottom one. (a) $e_1=0.00$ and $e_2 \approx 0.55$; (b) $e_1=0.10$ and $e_2 \approx 0.54$; (c) $e_1=0.20$ and $e_2 \approx 0.52$; (d) $e_1=0.30$ and $e_2 \approx 0.47$. the trajectories ejected before the end of the integration are identified by white pixels. See section \ref{sec:ECC} for more details about the integrations.} 
\end{center}
\end{figure}
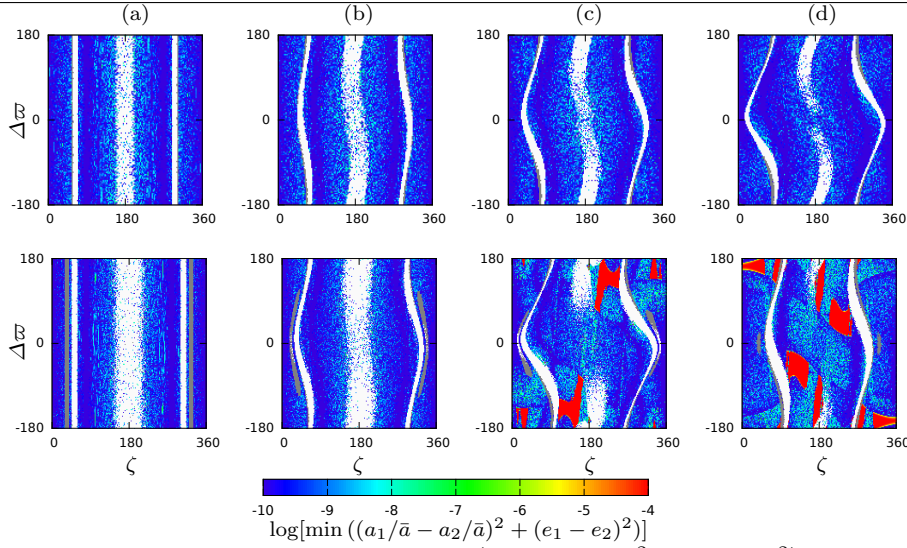

We check numerically if the definition $\cV=\{a_1=a_2,e_1=e_2\}$ holds for higher eccentricities. Note that we consider only trajectories that reach $a_1=a_2$ on their orbit. 
To perform this check, we take grids of initial conditions for $\zeta \in [ 0^\circ : 360^\circ ] $ and $\Dv \in [-180^\circ:180^\circ]$ and several values of $\varPi$ for a fixed value of $J_1$ such that $J_1=J_1(e_1=e_2=0.4)$. The corresponding values of the eccentricities are given by:
\begin{equation}
\begin{aligned}
e_2=\sqrt{1-\frac{2J_1}{\Lambda^0_1}-\sqrt{1-e_1^2}}
\end{aligned}
\label{eq:e2J1cst}
\end{equation}
Figure~\ref{fig:varefmeqe07} shows the value of $(a_1/\bar{a}-a_2/\bar{a})^2+(e_1-e_2)^2$ for several values of $\varPi$ for $m_1=m_2$ (top line) and $m_1 \neq m_2$ (bottom line). The integrations are conducted over $10/\eps$ orbital periods, hence only a few times $2\pi/g$ at best. For all initial conditions when $m_1=m_2$, the criterion
\begin{equation}
\frac{(a_1-a_2)^2}{\bar{a}^2}+(e_1-e_2)^2 < \epsilon_\Sigma
\label{eq:critVaRemeq}
\end{equation}
 is met for $\epsilon_\Sigma \approx 10^{-8}$. Although this verification is not exhaustive, it suggests that the chosen reference manifold represents a significant part of the phase space of the averaged reduced problem. However it is possible that, especially at high eccentricities, stable domains appear for which the orbits never reach $e_1=e_2$ even in the case $m_1=m_2$, but none was discovered during this study. A study performed in the case $e_1=e_2=0.7$ yielded similar results.\\ 
 
 In order to compare with the case $m_1 \neq m_2$, the bottom line of the figure~\ref{fig:varefmeqe07} shows that there are areas of the phase space where the criterion (\ref{eq:critVaRemeq}) is not verified for $\epsilon'_\Sigma=10^{-4}$. There are hence orbits in the phase space that are not represented by the trajectories taking their initial conditions on the manifold $\cV=\{a_1=a_2,e_1=e_2\}$. This is not surprising: we know for example that, at least for moderate eccentricities, the position of the $AL_4$ equilibrium is approximated by $m_1e_1=m_2e_2$. In the case $m_1=m_2$, any trajectory librating sufficiently close to this equilibrium would never cross the $e_1=e_2$ manifold.

\section{Identification of the $\cF$ families}
\label{sec:SAIF}
We show here how the separation of the time scales allows us to identify the position of the $\cF$ anywhere in the phase space.

\subsection{Identification of the $\Fsc$ families}
\label{sec:IdFsc}
The $\Fsc$ families are families of periodic orbit of the reduced averaged problem, whose period is associated with the secular time scale. The position of the $\Fsc$ families can be identified by studying the critical points of the averaged Hamiltonian. Let us use the hypothesis of adiabatic invariant for the variables $\varPi$ and $\Dv$ (see appendix \ref{sec:tsai}): 
on a short time scale with respect to $1/g$, $\Fsc$ is made of orbits that behave as fixed points of the reduced averaged problem. The orbits belonging to $\Fsc$ are thus orbits which satisfy:
\begin{equation}
\frac{\partial }{\partial Z}\gH_{\cR\cM} = \frac{\partial }{\partial \zeta} \gH_{\cR\cM}= 0\, .
\label{eq:condFb0}
\end{equation}
where $Z$ and $\zeta$ are conjugated canonical variables. This is equivalent to:
\begin{equation}
\dot \zeta = \dot Z= 0\, .
\label{eq:condFb02}
\end{equation}
Starting from the reduced Hamiltonian (section \ref{sec:RoP}), we can estimate the value of the averaged Hamiltonian at any point of the phase space by doing a numerical averaging over the fast angle $Q$. We can identify the orbits belonging to $\Fsc$ by finding the orbits for which $\dot Z = 0$ on the manifold $\dot{\zeta}=0$. For a given value of the constants $\varPi$ and $\Dv$, we take a grid of values for $\zeta$ and estimate the averaged Hamiltonian at each point. We can then have the approximate position of the points where $\dot Z = 0$ by finding the positions on the grid where the equation
\begin{equation}
\frac{\partial }{\partial \zeta} \gH_{\cR\cM}|_{\zeta=\zeta_k} \times \frac{\partial }{\partial \zeta} \gH_{\cR\cM}|_{\zeta=\zeta_{k+1}} < 0\, .
\label{eq:condFb0n}
\end{equation}
is satisfied, with 
\begin{equation}
\frac{\partial }{\partial \zeta} \gH_{\cR\cM}|_{\zeta=\zeta_k} = \frac{\gH_{\cR\cM}(Z,\zeta_{k+1},\Dv,\varPi) -\gH_{\cR\cM}(Z,\zeta_{k-1},\Dv,\varPi) }{|\zeta_{k+1}-\zeta_{k-1}|}\, . 
\label{eq:expZp}
\end{equation}
Note that it is not guaranteed that the associated trajectory in the full 3-body problem is quasi-periodic.\\

Alternatively, numerical integrations allow us to determine an empiric criterion for a numerical determination of the position of $\Fsc$. In the various integrations that we computed through this study, we noted that the amplitude of variation of $Z$ (hence $a_1-a_2$) seems not to be impacted much by the frequency $g$. We hence make the hypothesis that if an orbit in a regular area of the phase space verifies the condition  %
\begin{equation}
(\max{(Z)}-\min{(Z)} ) < \epsilon_\nu\, , 
\label{eq:condFb02}
\end{equation}
 with $\epsilon_\nu \propto \sqrt{\eps}$, this orbit is in the neighbourhood of the manifold $\Fsc$. One can check in figures \ref{fig:glob_e01} to \ref{fig:glob_e7_m6}, where the quasi-periodic orbits that verify equation (\ref{eq:condFb02}) are identified by brown pixels, and the point of the phase space satisfying the equation (\ref{eq:condFb0}) are identified by purple dots, that both methods yield very similar results in the regular area of the phase space.

  \subsection{Application in the case $m_1=m_2$}
\label{sec:appmm}

We can apply the research of the critical points of the Hamiltonian to identify the position of $\Fsc$ in the case $m_1=m_2$. We assume that, as it is the case for circular co-orbitals, the manifold $\dot{\zeta}=0$ is located at $Z=0$. We know that the equilibriums $L_k$ and $AL_k$ are all located in the plane $\varPi =0$ ($e_1=e_2$). We can hence explore the manifold $\cV =\{Z,\zeta , \varPi ,\Dv / Z=\varPi =0 \}$. We chose a grid of initial condition for $\zeta$ and $\Dv$ with a step of $0.5^\circ$, and we compute numerically the averaged Hamiltonian at each point of the grid. In figure \ref{fig:Fb0meq} we show all the points of $\cV$ that verify the condition (\ref{eq:condFb0n}). Each graph corresponds to a different value of the total angular momentum (different value of $e_1=e_2$).

 \subsection{Identification of the $\Fsf$ families}

The $\Fsf$ families are families of periodic orbit of the reduced averaged problem, whose period is associated with the semi-fast (resonant) time scale.The method developed in section \ref{sec:IdFsc} cannot be used directly to determine the position of the $\Fsf$ manifold because it requires to numerically average the Hamiltonian over the semi-fast angle $\zeta$, which is somehow laborious, see section \ref{sec:tsai}. 

However, the evolution of the variables $\Dv$ and $\varPi$ during the numerical integrations of the 3-body problem are $\gO(\sqrt \eps)$ close to their evolution in the secular problem (see section \ref{sec:tsai}). Since the orbits belonging to $\Fsf$ are fixed points of the 1-degree of freedom secular problem, we make the following hypothesis: all orbits in a regular area of the phase space (far from the separatrix, the chaotic and the unstable areas) that verify 
\begin{equation}
(\max{(\Dv)}-\min{(\Dv)} ) < \epsilon_g\, , 
\label{eq:condFb1}
\end{equation}
 with $\epsilon_g \propto \sqrt{\eps}$ are in the neighbourhood of $\Fsf$. One can check that such orbits (represented by black pixel in the figures \ref{fig:glob_e01} to \ref{fig:glob_e7_m6}) are indeed in the neighbourhood of the analytical approximation of the positions of the $\Fsf$ families, see \cite{these}, section 2.7.2.

\end{document}